\documentclass[letterpaper,11pt, nofootinbib]{article}
\pdfoutput=1

\usepackage{silence}
\usepackage{comment}
\usepackage{jheppub}
\usepackage{epsfig}
\usepackage{amsmath}
\usepackage{bbm}
\usepackage{graphicx}
\usepackage{appendix} 
\usepackage{capt-of}
\usepackage{tikz-cd}
\usepackage[cal=boondoxo]{mathalfa}
\usetikzlibrary{arrows}

\newcommand{\midarrow}{\tikz \draw[-triangle 90] (0,0) -- +(.1,0);}

\usepackage{hyperref}
\usepackage{braket}
\usepackage{array}
\newcolumntype{P}[1]{>{\centering\arraybackslash}p{#1}}

\newcommand*{\rom}[1]{\expandafter\@slowromancap\romannumeral #1@}

\usepackage[missing=]{gitinfo2}

\DeclareSymbolFont{AMSa}{U}{msa}{m}{n}
\DeclareSymbolFont{AMSb}{U}{msb}{m}{n}
\let\Box\relax
\DeclareMathSymbol{\Box}{\mathord}{AMSa}{"03}

\newcommand{\Sum}{\displaystyle\sum\limits}

\allowdisplaybreaks

\newcommand{\cH}{{\mathcal H}}

\newcommand{\cB}{{\mathcal B}}

\newcommand{\cO}{{\mathcal O}}

\newcommand{\bbZ}{{\mathbb Z}}

\newcommand{\Z}{{\bbZ}}

\newcommand{\uR}{v}
\newcommand{\uP}{u}
\newcommand{\idTDL}{I}

\newcommand{\insfigsvg}[3]{

\medskip
\noindent
\begin{minipage}{\linewidth}

\makebox[\linewidth]{\includegraphics[keepaspectratio=true,scale=#2]{figures/#1.pdf}}

\captionof{figure}{#3}

\label{fig:#1}
\end{minipage}
\medskip

}

\title{Lattice Realizations of Topological Defects in the critical (1+1)-d Three-State Potts Model}

\author[1]{Madhav Sinha,}
\author[1,5]{Fei Yan,}
\author[2]{Linnea Grans-Samuelsson,}
\author[1]{Ananda Roy,}
\author[3,4]{and Hubert Saleur}
\affiliation[1]{Department of Physics and Astronomy, Rutgers University, Piscataway, NJ 08854-8019 USA}
\affiliation[2]{Microsoft Station Q, Santa Barbara, CA 93106-6105 USA}
\affiliation[3]{Institut de physique théorique, CEA, CNRS, Université Paris-Saclay, France}
\affiliation[4]{Physics Department, University of Southern California, Los Angeles, USA}
\affiliation[5]
{Brookhaven National Laboratory, Upton, NY 11973, USA
}
\emailAdd{ms3066@physics.rutgers.edu, fyan.hepth@gmail.com, linneag@microsoft.com, ananda.roy@physics.rutgers.edu, hubert.saleur@ipht.fr}

\abstract{Topological/perfectly-transmissive defects   play a fundamental role in the analysis of the symmetries of two dimensional conformal field theories (CFTs). In the present  work, spin chain regularizations for these defects are proposed and analyzed in the case of  the three-state Potts CFT. In particular, lattice versions for all the primitive defects are presented, with the remaining defects obtained from the fusion of the primitive ones. The defects are obtained by introducing modified interactions around two given sites of an otherwise homogeneous spin chain with periodic boundary condition. The various primitive defects are topological on the lattice except for one, which is topological only in the scaling limit. The lattice models are analyzed using a combination of exact diagonalization and density matrix renormalization group techniques. Low-lying energy spectra for different defect Hamiltonians as well as entanglement entropy of blocks located symmetrically around the defects are computed. The latter provides a convenient way to compute the $g$-function which characterizes various defects.  Finally, the  eigenvalues of the line operators in the ``crossed channel'' and fusion of different defect lines are also analyzed. The results are all in agreement with expectations from conformal field theory. }

\begin{document}

%\tracingall

\noindent{{\tiny \color{gray} \tt \gitAuthorIsoDate \gitAbbrevHash}}

\maketitle
\flushbottom

%%%%%%%

\section{Introduction}

Line operators play many important roles in quantum field theory. In 4d gauge theories for example, the expectation values of Wilson lines and 't Hooft lines can be used to distinguish different phases  \cite{PhysRevD.10.2445,THOOFT19781}, while  line operators and their correlation functions capture many aspects of the global structure \cite{Kapustin:2005py,Gaiotto:2010be,Aharony:2013hda}. In condensed matter physics, line defects (i.e. extending along the time direction) also have interesting applications, such as modeling, at low energies, impurities in physical systems \cite{Vojta_1999,Vojta_2000}. In statistical physics, line defects were used very early on to understand the operator content of critical points \cite{KadanoffCeva}. Moreover, dynamical properties of line defects and defect renormalization flows can be studied using various techniques \cite{Cuomo_2022,Cuomo:2021kfm,Cuomo:2022xgw,Aharony:2022ntz,Beccaria:2022bcr,Gabai:2022vri,Popov:2022nfq,Rodriguez-Gomez:2022xwm,Gimenez-Grau:2022czc,Bianchi:2022sbz,Pannell:2023pwz,Hu:2023ghk,Aharony:2023amq}.

An area where line defects play a particularly crucial  role is the study of quantum entanglement. Entanglement entropy (EE) has become a powerful tool to analyze quantum critical phenomena \cite{Holzhey1994,Vidal2002,Jin_2004,Calabrese2004}, to study the dynamics of strongly coupled systems \cite{Calabrese:2005in,Eisler_2007}, and to characterize quantum phases of matter \cite{Kitaev:2005dm,Levin2006}. Physical information is often encoded within the scaling behavior of EE. For example, in (1+1)-d critical systems EE exhibits a logarithmic dependence on the subsystem size \cite{Holzhey1994,Calabrese2004,Calabrese2009}, with a   coefficient  that is determined by the universality class of the critical point.\footnote{For critical points that are described by conformal field theories, the coefficient is proportional to the central charge.} With insertions of line defects (or point-like impurities), the scaling behavior of EE is more complicated. It depends not only on the size of the subsystem, but also on the type  of defects, and their insertion location relative to the subsystem. This kind of problems have been explored both from condensed matter and high energy physics perspectives frequently in (1+1)-d (see e.g. \cite{Oshikawa1997,Bachas2001,Sakai2008,Affleck2009,Eisler2010,Peschel2012e,Bachas2013,Brehm2015,Gutperle2015,Erdmenger_2016,RoySaleur2022}).\footnote{A related setup is the study of EE for (1+1)-d systems with boundaries. The simplest case where the subsystem ends on the boundary has been studied extensively, see for example \cite{Calabrese2004,Calabrese2009,Affleck2009,Roy2020a}. The more general case of subsystem located away from the boundary is recently explored in \cite{Sully_2021,Estienne_2022,Estienne_2023,Karch:2023evr,Estienne:2023ekf}.}

Conversely, EE can also be used to investigate the  physical properties of defects. An example of this is the so-called {\it{g-function}} associated to line defects, which is a monotonically decreasing quantity along the defect renormalization group flow. In (1+1)-d this property was conjectured by \cite{Affleck1991} and proven in \cite{Friedan_2004,Kutasov_2000,Casini_2016}. Generalization of the $g$-function and  $g$-theorem to line defects in arbitrary dimensions was recently achieved by \cite{Cuomo_2022}. In (1+1)-d, the g-function of a line defect can be extracted from the sub-leading term of EE, with the defect inserted at the center of subsystem \cite{Oshikawa1997}.\footnote{In higher dimensions, the relation between the g-function and EE is more complicated \cite{Lewkowycz_2014,Cuomo_2022,Casini:2022bsu,Casini:2023kyj}.}

An important class of line operators/defects is the so-called topological lines \cite{Frohlich2006,Davydov:2010rm,Kapustin:2014gua,Gaiotto:2014kfa,Chang:2018iay} -  lines which can be deformed without  changing the flat space-time partition function as long as they do not pass over each other or cross local  operators insertions. Well-known examples in (1+1)-d include the Verlinde lines \cite{Verlinde:1988sn,Petkova:2000ip,Drukker:2010jp,Gaiotto:2014lma} in diagonal rational CFTs. More recently, various tools have been developed to study topological lines in (1+1)-d QFTs, see e.g. \cite{Frohlich:2009gb,Carqueville:2012dk,Brunner:2013xna,Bhardwaj:2017xup,Chang:2018iay,Lin:2019hks,Ji:2019ugf,Ji:2019jhk,Thorngren:2019iar,Komargodski:2020mxz,Gaiotto:2020iye,Thorngren:2021yso,Lin:2022dhv,Freed:2022qnc,Kaidi:2022cpf,Lin:2023uvm}.

While topological lines have been studied mostly in the continuum, there is also a growing interest in tackling their properties starting from lattice discretizations, where  various techniques such as integrability or numerical simulations can be applied \cite{Aasen2016,Aasen2020,Belletete2018,Belletete2020,Roy2020a, Rogerson_2022}.  An interesting question is then the following: given a certain lattice model which flows to some continuum field theory in the infrared, what are the lattice counterparts for the topological lines in the continuum? The answer to this question can, for instance, make possible calculations of EE using DMRG techniques: this is not a vain exercise, since comparison with field theory calculations  reveals unexpected discrepancies  already in the case of the Ising model \cite{Gutperle2015,Roy2020a}. Of course, one generally expects  to have  more topological lines in the continuum than on the lattice due to the possibility of emergent symmetries. Identifying candidates on the lattice that become topological in the continuum limit is therefore challenging, and many results in this area remain conjectural \cite{Belletete2020}.

While the case of the  Ising model has been studied before \cite{Oshikawa1997,Grimm2001,Frohlich2004,Aasen2016,Hauru2016,Aasen2020,RoySaleur2022,Li:2023mmw,Seiberg:2023cdc,Seifnashri:2023dpa}, we turn our attention here to the (1+1)-d critical three-state Potts model, described by the continuum three-state Potts CFT in the infrared. Topological lines in the three-state Potts CFT have been studied before \cite{Petkova:2000ip,Chui_2003,Chang:2018iay}. Our main goals is  to construct their lattice counterpart. For the majority of topological lines (exceptions have been discussed in \cite{Aasen2020,Grimm1992}) such lattice counterparts were previously unknown. We distinguish between {\it{defects}} and {\it{line operators}}, where the former corresponds to local defects around two sites on the spatial lattice (the one-dimensional spin chain), and the latter is an operator acting on the Hilbert space of the (periodic) spin chain. We note that, when going from the (1+1)-d to the 2d statistical model point of view, these two versions correspond to having a defect line extending in the (imaginary) time resp. space dimension. Accordingly, in what follows, we will often use the concepts of ``direct'' and ``crossed'' channels to refer to lines extending in time and space, respectively.  
	
The main tools we will use to justify our constructions are the numerical study (using both direct diagonalization and Density Matrix Renormalization Group (DMRG)) of spectra in the presence of defects, eigenvalues of line operators, and EE and defect g-function. 

\bigskip

For the reader's convenience,  we now summarize our main results. Details can be found in the rest of the paper. Additional technical aspects are discussed in appendices.

The three-state Potts model  is a natural generalization of the transverse field Ising (TFI) model \cite{Pfeuty1970,Kogut1979}. We choose the following critical periodic-chain Hamiltonian \footnote{Our convention here differs from the usual convention in \cite{Mong:2014ova} by an overall scaling and shift.}
\begin{equation}\label{eq:Potts_Ham_Summary}
H = -\frac{1}{3\sqrt{3}} \Sum\limits_{i=1}^{L} \left(1 + \sigma_i^\dagger \sigma_{i+1} +\sigma_{i+1}^\dagger \sigma_i \right)-\frac{1}{3\sqrt{3}} \Sum\limits_{i=1}^L \left(1 + \tau_i + \tau_i^\dagger \right),
\end{equation}
where $\sigma_i$, $\tau_i$ act on the three-dimensional Hibert space on the $i$-th spin site, and $\sigma_{L+1}$ ($\tau_{L+1}$) is identified with $\sigma_1$ ($\tau_1$). Our convention here is 
\begin{equation}
\sigma_i=\begin{pmatrix}
0 & 1 & 0 \\
0 & 0 & 1 \\
1 & 0 & 0
\end{pmatrix}, ~ 
\tau_i=\begin{pmatrix}
1 & 0 & 0\\
0 & \omega & 0\\
0 & 0 & \omega^2
\end{pmatrix}, ~  \omega =\text{e}^{2\pi \text{i}/3} \, .
\end{equation}
This lattice model flows to the continuum three-state Potts CFT in the infrared.

This CFT has in total 16 simple topological lines~\footnote{A simple topological line is a line that can not be written as a direct sum of other simple topological lines.}. These topological lines are generated by fusion of the following primitive lines: 
\begin{equation}
\idTDL, ~ \eta, ~  C, ~ N, ~ W \, .
\end{equation}
Here, $\eta$ and $C$ generate the $\Z_3$ symmetry\footnote{In the following, by {\it symmetries} we mean ordinary 0-form symmetries.} and $\Z_2^C$ charge-conjugation symmetry of the Potts CFT respectively. The $N$ line is a duality-type line \cite{Frohlich2004} associated with the Kramers-Wannier duality of the model \cite{Krammers1941}. The $\Z_3$ symmetry in Potts CFT is non-anomalous; moreover the $\Z_3$-orbifold of Potts CFT is isomorphic to itself. The $N$ line can be constructed by gauging the $\Z_3$ symmetry on one side of the line and imposing Dirichlet boundary condition for the $\Z_3$ gauge field along the line.\footnote{Recently, Kramers-Wannier-like duality defect has also been constructed in higher dimensions \cite{Choi:2021kmx,Kaidi:2021xfk,Choi:2022zal}.} Finally, the $W$ line is a topological line obeying the Fibonacci fusion relation : $W^2= \idTDL + W $.

In the following, we list the lattice counterpart for the $\eta, C, N, W$ lines. We distinguish between {\it defects} and {\it line operators}. Concretely a defect corresponding to $D$ amounts to local modifications of the 1d lattice Hamiltonian which gives rise to the defect Hamiltonian $H_{D}$ , while a line operator $\widehat{D}$ is supported on the whole 1d quantum spin chain. In both cases, we found that the lattice  realizations of $\eta, C, N$ are topological {\it on the lattice}. This is as expected from the perspectives of symmetries. The $\Z_3$ and $\Z_2^C$ symmetries are manifest on the lattice model: correspondingly $\eta$ and $C$ can be constructed based on the symmetry action and should be  topological even in finite size. Addtionally, gauging the $\Z_3$ symmetry can also be performed directly on the lattice (see \autoref{sec:KWgauging}), which implies that the lattice realizations of $N$ would also be topological.
\insfigsvg{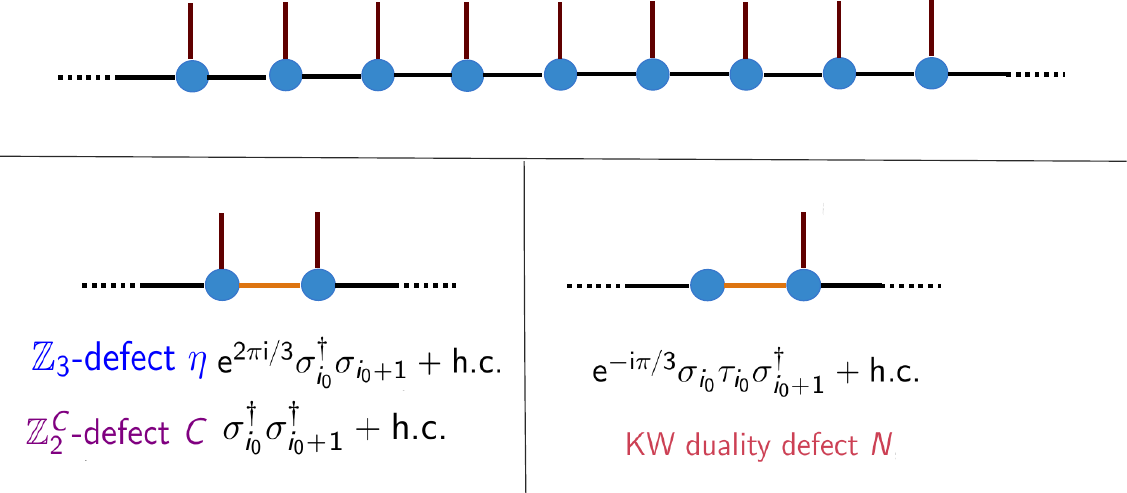}{0.56}{{\it Above}: A graphical illustration of the three-state Potts model, where each blue dot represents a spin site. Nearest-neighbor interactions are represented by horizontal links connecting the spins, while vertical links represent the transverse field. {\it Below}: Local modifications involved in the $\eta$, $C$ and $N$ defect Hamiltonians. The $\eta$ and $C$ defects effectively introduce a twist by their appropriate symmetry actions. The N defect Hamiltonian involves removing the transverse field at one site of the defect link, while modifying the nearest-neighbor interaction to mimic the coupling between the original spin and the dual spin.}

\begin{enumerate}
\item
The defect Hamiltonians involve local modifications to two spin sites, say site $i_0$ and site $i_0+1$. For symmetry defects, such as $\eta$ and $C$, these local modifications follow from the corresponding symmetry actions. The duality defect $N$ (and similarly $N' = CN$) can be constructed in two ways, either from an integrable lattice construction, or  from $\Z_3$ gauging on half the chain: the results of  these two methods agree. We illustrate the modifications for the $\eta$, $C$ and $N$ defect Hamiltonians in \autoref{fig:PottsTLD}.

\item The $N$ defect and $W$ defect can be obtained using an integrable lattice construction based on the  Temperley-Lieb (TL) algebra, as developed in \cite{Belletete2018,Belletete2020,Grans-Samuelsson:2022zyv} (for earlier albeit quite implicit results, see \cite{Chui_2003}), in the $\mathbf{D}_4$ RSOS model. The corresponding three-state Potts representation for the TL generators is given  by:
\begin{equation}
e_{2i-1}=\frac{1}{\sqrt{3}}\left( 1+\sigma_i^\dagger \sigma_{i+1} +\sigma_{i+1}^\dagger \sigma_i \right), ~ e_{2i}=\frac{1}{\sqrt{3}} \left( 1+\tau_i +\tau_i^\dagger \right).
\end{equation}
The critical Potts model Hamiltonian is then given as follows, with two TL sites representing one physical spin site:
\begin{equation}\label{eq:no_def_Ham}
H=-\frac{\gamma}{\pi\text{sin}\gamma}\Sum\limits_{i=1}^{2L} e_i, ~~ \gamma=\frac{\pi}{6} \, .
\end{equation}
In the general integrable model  context, a defect is inserted by shifting the spectral parameter at the defect location. The technical starting point is a Transfer Matrix,  which one can use to obtain a Hamiltonian with/without defect at a particular site. The no-defect case corresponds to a system with homogeneous spectral parameter, and leads to the Hamiltonian in equation \eqref{eq:no_def_Ham}. The $N$ and $W$ defects are then obtained by  shifting  the spectral parameter at one site by $\text{i}\infty$ and $-\pi/2$ respectively (here we use the conventions of \cite{Klumper:1992vt} for spectral parameter dependent face weights).

The resulting  $W$ defect lattice Hamiltonian is then given by:
\begin{equation}
H_W=-\frac{\gamma}{\pi\text{sin}\gamma}\Sum\limits_{i=1}^{2L} e_i+\frac{\gamma}{\pi\text{sin}\gamma\text{cos}\gamma}\left(e_{2i_0-1}e_{2i_0}+e_{2i_0}e_{2i_0-1}\right).
\end{equation}
This defect is not topological on the lattice. We provide numerical evidence that it flows to the $W$ line in the continuum Potts CFT - in particular, by considering the spectrum of $H_W$ in the scaling limit.

\item We numerically compute (using DMRG) the ground state {\it symmetric EE} of the  system with defect - specifically, the ground state von-Neumann entropy where the subsystem is an interval centered on the defect.\footnote{By folding, this setup gives rise to a  system  where the defect becomes a boundary condition \cite{Oshikawa1997,Saleur1998,Saleur2000}.} The defect g-function \cite{Affleck1991} can then be extracted from the sub-leading term of EE \cite{Calabrese2004,Calabrese2009}. The numerical values are consistent with expectations from the Potts CFT \cite{Petkova:2000ip,Chang:2018iay}, namely
\begin{equation}
g_\eta = g_C =1, ~ g_N = \sqrt{3}, ~ g_W = \frac{1+\sqrt{5}}{2} \, .
\end{equation}

\item  Like their defect counterparts, the line operators $\widehat{\eta}$, $\widehat{C}$, and $\widehat{N}$ are topological on the lattice. $\widehat{\eta}$ and $\widehat{C}$ are the symmetry operators for $\Z_3$ and $\Z_2^C$ respectively, 
\begin{equation}
\widehat{\eta}=\prod_{i=1}^L \tau_i^\dagger, ~ \widehat{C}=\prod_{i=1}^L c_i, ~ c_i=\begin{pmatrix}
1 & 0 & 0\\ 0 & 0& 1 \\ 0 & 1 & 0\end{pmatrix}  ,
\end{equation}

The $\widehat{N}$ and $\widehat{W}$ operators are again constructed using integrability techniques, where the spectral parameter for every site in a row is now shifted by $\text{ i}\infty$ and $-\pi/2$ respectively. The $\widehat{N}$ operator is central in the TL algebra and therefore topological on the lattice \cite{Belletete2018,Belletete2020}. Concretely, it can be written as 
\begin{equation}
\widehat{N}=(-q)^{\frac{1}{2}} \uR g_{2L-1}...g_2 g_1+(-q)^{-\frac{1}{2}}g_1^{-1}g_2^{-1}... g_{2L-1}^{-1} \uR^{-1}  
\end{equation}
where $q=\text{e}^{\text{i}\pi/6}$, $\uR$ is the shift operator of the Affine TL algebra, and $g_i^{\pm 1}$ are the braid operators (not to be confused with g-functions) given by
\begin{equation}
g_i^{\pm 1} = (-q)^{\pm 1/2} 1 + (-q)^{\mp 1/2} e_i.
\end{equation} 
We also verify that the $\widehat{N}$ operator indeed implements the Kramers-Wannier duality transformation on the 3-state Potts chain. The $\widehat{W}$ operator, on the other hand, is  not topological on the lattice. Again, we provide numerical evidence that it flows to the $\widehat{W}$  line in the continuum Potts CFT. We further remark that, the Kramers-Wannier duality operators $\widehat{N}$ and $\widehat{N'}$ constructed here map from the Hilbert space associated with the sites of the 1d quantum Potts chain to a dual Hilbert space associated with the links, similar to the situation of transverse field Ising model considered in \cite{Aasen2016,Aasen2020,Tan:2022vaz,Li:2023mmw}. This is related to, albeit different from recent constructions in \cite{Cheng:2022sgb,Seiberg:2023cdc, Shao:2023gho,Seifnashri:2023dpa} of non-invertible operators which act on a single Hilbert space.

\end{enumerate}

Finally, a word on notations. We use the same symbol $\widehat{D}$ for line operator in all the cases we study. We also use the same symbol for lattice defect Hamiltonians - $H_D$. On the basis of what space these operators are acting on, it is clear what model we are referring to. 
For simplicity, we also do not use separate notations for the lattice realizations of the defects, and for their  continuum CFT versions.

\section{Topological lines in the three-state Potts CFT}
\label{sec:CFT}
The critical three-state Potts model is described by the $c=4/5$ three-state Potts CFT in the continuum limit. Topological  lines in the Potts CFT have been well studied before \cite{Petkova:2000ip,Chui_2003, Chang:2018iay}. In this section we will review {\it simple} topological lines, namely topological lines that can not be written as a direct sum of other simple topological lines, and write down  their defect Hilbert space spectra. 

The $c=4/5$ three-state Potts CFT is a non-diagonal Virasoro minimal model, which can also be viewed as a diagonal RCFT with respect to the extended $W_3$ algebra. It has in total 12 Virasoro primary fields, listed in Table \ref{table:PottsPrimary}.

\begin{table}[h]
	\centering
	\begin{tabular}{|c|c|c|c|c|c|} \hline
		      & left Kac label    & right Kac label   & $(h,\bar{h})$     & scaling dim. & spin \\
		\hline 
		$\mathbbm{1}$ & $(1,1)$ & $(1,1)$ & $(0,0)$ & $0$ & $0$\\
		\hline
		$\epsilon$ & $(2,1)$ & $(2,1)$ & $(\frac{2}{5},\frac{2}{5})$ & $\frac{4}{5}$ & $0$\\
		\hline 
		$\sigma^{(1)}$, $\sigma^{(2)}$ & $(2,3)$ & $(2,3)$ & $(\frac{1}{15},\frac{1}{15})$ & $\frac{2}{15}$ & 0\\
		\hline
		$Z^{(1)}$, $Z^{(2)}$ & $(1,3)$ & $(1,3)$ & $(\frac{2}{3},\frac{2}{3})$ & $\frac{4}{3}$ & $0$\\
		\hline
		$X$ & $(3,1)$ & $(3,1)$ & $(\frac{7}{5},\frac{7}{5})$ & $\frac{14}{5}$ & $0$\\
		\hline
		$Y$ & $(1,5)$ & $(1,5)$ & $(3,3)$ & $6$ & $0$\\
		\hline
		$\Phi$ & $(3,1)$ & $(2,1)$ & $(\frac{7}{5},\frac{2}{5})$ & $\frac{9}{5}$ & $1$\\
		\hline
		$\overline{\Phi}$ & $(2,1)$ & $(3,1)$ & $(\frac{2}{5},\frac{7}{5})$ & $\frac{9}{5}$ & $1$\\
		\hline
		$\Omega$ & $(1,5)$ & $(1,1)$ & $(3,0)$ & $3$ & $3$\\
		\hline
		$\overline{\Omega}$ & $(1,1)$ & $(1,5)$ & $(0,3)$ & $3$ & $3$\\
		 \hline
	\end{tabular}
\caption{The primary fields in the $c=4/5$ three-state Potts CFT.}
\label{table:PottsPrimary}
\end{table}

The theory has a $S_3$ $0$-form symmetry, generated by an order-$3$ element denoted as $\eta$, and a charge conjugation element denoted as $C$. Correspondingly we have six invertible topological line defects implementing the $S_3$ symmetry. Three of them: $\idTDL$, $\eta$, $\bar{\eta}=\eta^2$ are elements of a $\Z_3$ subgroup and are Verlinde lines, when we view the Potts CFT  as a diagonal RCFT with $W_3$ algebra as the chiral algebra. On the other hand charge conjugation does not commute with the $W_3$ algebra. 

Under gauging the $\Z_3\subset S_3$ symmetry of the Potts CFT, the resulting orbifold theory is in fact isomorphic to the Potts CFT itself. One can then construct a Kramers-Wannier duality defect denoted as $N$, by gauging the $\Z_3$ symmetry on a half space and imposing Dirichlet boundary condition for the $\Z_3$ gauge field. There is also a closely related topological line $N'=CN$. This gives rise to  two sets of the $\Z_3$ Tambara-Yamagami  fusion rules \cite{Tambara1998TensorCW}
\begin{equation}\label{eq:fusion_of_lines}
\begin{split}
&N\times N =N'\times N' =\idTDL+\eta+\bar{\eta},\\
&N \times \eta= \eta \times N=N, ~ N'\times \eta =\eta \times N' =N' \, .
\end{split}
\end{equation}

Both $N$ and $N'$ are examples of non-invertible topological lines. Another elementary non-invertible topological line, denoted as $W$, is a Verlinde line when considering the Potts CFT as a diagonal RCFT with respect to the $W_3$ algebra. It obeys the following fusion relation:
\begin{equation}
W\times W=\idTDL+W \, .
\end{equation}
Other simple topological lines in the model can be realized as fusion products of the above lines. 

In Section \ref{sec:TDLlatticedirect}, we will describe the realization for the $\eta$, $C$, $N$ ($N'$), and $W$ line defects in the critical three-state Potts lattice model. To confirm  that we have gotten the correct lattice description for the corresponding lines, we shall numerically extract the defect spectrum for the 1d periodic lattice system with insertion of the defect, and compare with the CFT computation of the spectrum in the defect Hilbert space on a circle. 

Given a topological line $D$, there is a space $\cH_{D}$ of point operators that could live at its end. By the state/operator correspondence, $\cH_{D}$ is the Hilbert space of the theory on a circle where $D$ sits at a point of the circle. States in $\cH_{D}$ are encoded by the torus partition function $Z_{D}$ where the line $D$ wraps the temporal cycle
\begin{equation}
Z_{D} (\tau,\bar{\tau}):= \text{Tr}_{\cH_{D}}\left[ q^{L_0-\frac{c}{24}}\bar{q}^{\bar{L}_0-\frac{c}{24}}\right], ~ q=\text{e}^{2\pi \text{i} \tau} \, . 
\end{equation}
The spectrum of $\cH_{D}$ organizes itself into product of representations of the left- and right-moving Virasoro algebras. We then have
\begin{equation}
Z_{D}(\tau,\bar{\tau})=\Sum\limits_{ij} n_{ij} \chi_i(\tau) \chi_j(\bar{\tau}), ~ n_{ij}\in \Z_{\geq 0}\, , 
\label{partfint}
\end{equation}
where $\chi_i$ is the Virasoro character for an irreducible representation, and $n_{ij}\in \Z_{\geq 0}$.

By a modular S-transformation this is related to the torus partition function $Z^{D}$ where the topological line now  goes along the spatial cycle, so  the line is now an operator $\widehat{D}$ acting on the bulk Hilbert space $\cH$ 
\begin{equation}
Z^{D}(\tau,\bar{\tau}):=\text{Tr}_{\cH} \left[\widehat{D} q^{L_0-\frac{c}{24}}\bar{q}^{\bar{L}_0-\frac{c}{24}}\right]=Z_{D}(-1/\tau,-1/\bar{\tau}) \, . 
\end{equation}
As the line $D$ is topological, $\widehat{D}$ commutes with the left- and right-moving Virasoro algebras.  The condition that $n_{ij}\in \Z_{\geq 0}$ in (\ref{partfint}) puts strong constraints on the action of $\widehat{D}$ as an operator.

In the three-state Potts CFT, the $\widehat{\eta}$, $\widehat{C}$, $\widehat{N}$ and $\widehat{W}$ line operators act on the primaries as shown in Table \ref{table:PottsTDLaction}. In Section \ref{sec:TDLlatticecrossed}, we will describe the lattice realizations of these line operators, and compare their expectation value with continuum expectation values given in Table \ref{table:PottsTDLaction}.
\begin{table}[h]
	\centering
	\begin{tabular}{|c|c|c|c|c|c|c|c|c|c|c|c|c|} \hline
		& $\mathbbm{1}$   & $\epsilon$   &  $\sigma^{(1)}$    & $\sigma^{(2)}$ & $Z^{(1)}$ & $Z^{(2)}$ & $X$ & $Y$ & $\Phi$ & $\overline{\Phi}$ & $\Omega$ & $\overline{\Omega}$ \\
		\hline 
		$\widehat{\eta}$ & $1$ & $1$ & $w$ & $w^2$ & $w$ & $w^2$ & $1$ & $1$ & $1$ & $1$ & $1$ & $1$\\
		\hline
		$\widehat{C}$ & $1$ & $1$ & \multicolumn{2}{|c|}{$\sigma^{(1)}\leftrightarrow\sigma^{(2)}$} & \multicolumn{2}{|c|}{$Z^{(1)}\leftrightarrow Z^{(2)}$} & $1$ & $1$ & $-1$ & $-1$ & $-1$ & $-1$ \\
		\hline
		$\widehat{N}$ & $\sqrt{3}$ & $-\sqrt{3}$ & $0$ & $0$ & $0$ & $0$ & $\sqrt{3}$ & $-\sqrt{3}$ & $\sqrt{3}$ & $-\sqrt{3}$ & $-\sqrt{3}$ & $\sqrt{3}$ \\
		\hline
		$\widehat{W}$ & $x$ & $-x^{-1}$ & $-x^{-1}$ & $-x^{-1}$ & $x$ & $x$ & $-x^{-1}$ & $x$ & $-x^{-1}$ & $-x^{-1}$ & $x$ & $x$\\
		\hline
	\end{tabular}
	\caption{The action of the $\widehat{\eta}$, $\widehat{C}$, $\widehat{N}$ and $\widehat{W}$ lines on the primary fields of the Potts CFT. Here $w=\text{e}^{2\pi\text{i}/3}$ and $x=(1+\sqrt{5})/2$.}
	\label{table:PottsTDLaction}
\end{table}

By performing the modular S-transformation, we can then read off the spectra in the defect Hilbert space. The results are summarized below - to make comparison with  lattice results easier, we labelled  the Virasoro representations using their conformal weights $(h,\bar{h})$.
\begin{equation}
\begin{split}
\label{eq:eta_Part}
\cH_{\eta}:&\left(\frac{1}{15},\frac{1}{15}\right)\oplus \left(\frac{2}{5},\frac{1}{15}\right)\oplus \left(\frac{1}{15},\frac{2}{5}\right)\oplus \left(\frac{2}{3},0\right)\oplus \left(0,\frac{2}{3}\right)\oplus \left(\frac{2}{3},\frac{2}{3}\right)\\
&\oplus\left(\frac{7}{5},\frac{1}{15}\right)\oplus \left(\frac{1}{15},\frac{7}{5}\right)\oplus \left(3,\frac{2}{3}\right)\oplus \left(\frac{2}{3},3\right)
\end{split}
\end{equation}

\begin{equation}
\begin{split}
\label{eq:C_Part}
\cH_C: &\left(\frac{1}{40},\frac{1}{40}\right)\oplus \left(\frac{1}{8},\frac{1}{8}\right)\oplus \left(\frac{21}{40},\frac{1}{40}\right)\oplus \left(\frac{1}{40},\frac{21}{40}\right)\oplus \left(\frac{21}{40},\frac{21}{40}\right) \oplus \left(\frac{13}{8},\frac{1}{8}\right)\\
&\oplus \left(\frac{1}{8},\frac{13}{8}\right)\oplus \left(\frac{13}{8},\frac{13}{8}\right)
\end{split}
\end{equation}

\begin{equation}
\begin{split}
\label{eq:N_Part}
\cH_N: & 2\left(\frac{1}{40},\frac{1}{15}\right)\oplus \left(\frac{1}{8},0\right)\oplus \left(\frac{1}{40},\frac{2}{5}\right)\oplus 2\left(\frac{21}{40},\frac{1}{15}\right)\oplus 2\left(\frac{1}{8},\frac{2}{3}\right)\oplus \left(\frac{21}{40},\frac{2}{5}\right)\\
&\oplus \left(\frac{1}{40},\frac{7}{5}\right)\oplus \left(\frac{13}{8},0\right)\oplus \left(\frac{21}{40},\frac{7}{5}\right)\oplus 2\left(\frac{13}{8},\frac{2}{3}\right)\oplus \left(\frac{1}{8},3\right)\oplus \left(\frac{13}{8},3\right)
\end{split}
\end{equation}

\begin{equation}
\begin{split}
\label{eq:W_Part}
\cH_W:& 2\left(\frac{1}{15},\frac{1}{15}\right)\oplus \left(\frac{2}{5},0\right)\oplus \left(0,\frac{2}{5}\right)\oplus 2\left(\frac{2}{3},\frac{1}{15}\right)\oplus 2\left(\frac{1}{15},\frac{2}{3}\right)\oplus \left(\frac{2}{5},\frac{2}{5}\right)\\
&\oplus \left(\frac{7}{5},0\right) \oplus \left(0,\frac{7}{5}\right)\oplus \left(\frac{7}{5},\frac{2}{5}\right)\oplus \left(\frac{2}{5},\frac{7}{5}\right)\oplus \left(\frac{7}{5},\frac{7}{5}\right)\oplus \oplus \left(3,\frac{2}{5}\right)\\
&\oplus \left(\frac{2}{5},3\right)\oplus \left(3,\frac{7}{5}\right) \oplus \left(\frac{7}{5},3\right)
\end{split}
\end{equation}

Finally, we point out that in the general RSOS construction of \cite{Sinha2023a}, it is important to label the topological lines via their Kac labels. In \autoref{sec:2d3d}, we give such labelings of topological lines in the Potts CFT, viewed as a non-diagonal $M(6,5)$ minimal model.

\section{Lattice Hamiltonians for the three-state Potts model}

\subsection{The spin model} 

The critical three-state Potts quantum spin chain with  periodic boundary condition has the following Hamiltonian \cite{Mong:2014ova} 
\begin{equation}
\label{eq:D_I}
H_{\idTDL} =-\frac{1}{3\sqrt{3}}\Sum\limits_{i=1}^{L} \left(1+\sigma_i^\dagger \sigma_{i+1} +\sigma_{i+1}^\dagger \sigma_i \right)-\frac{1}{3\sqrt{3}}\Sum\limits_{i=1}^L\left(1+ \tau_i +\tau_i^\dagger \right) \, , 
\end{equation}
where $\sigma_{L+1} = \sigma_1, \tau_{L+1}= \tau_1$, and we adopt the convention that
\begin{equation}\label{eq:Potts_convention_1}
\sigma_i=\begin{pmatrix}
0 & 1 & 0 \\
0 & 0 & 1 \\
1 & 0 & 0
\end{pmatrix}, ~ 
\tau_i=\begin{pmatrix}
1 & 0 & 0\\
0 & \omega & 0\\
0 & 0 & \omega^2
\end{pmatrix}, ~  \omega =\text{e}^{2\pi \text{i}/3} \, .    
\end{equation}

These operators obey the following  relations:
\begin{equation}\label{eq:pottsgenalgebra}
\sigma_i^3=1, ~ \tau_i^3=1, ~ \sigma_i\tau_i =\omega \tau_i \sigma_i, ~ \sigma_i\tau_j =\tau_j \sigma_i ~(i\neq j).
\end{equation}

The Hamiltonian in \eqref{eq:D_I} can be viewed as a natural generalization of the quantum Ising chain~\cite{Pfeuty1970, Kogut1979}, where the first term describes the ferromagnetic interaction between nearest neighbors and the last term is the transverse field. Depending on the relative strength of the two terms, the spin chain can be in the ordered~(ferromagnetic) phase or the disordered~(paramagnetic) phase. When the two competing interactions balance each other~[as is the case in Eq.~\eqref{eq:D_I}], the spin chain is critical and is described by the Potts CFT in the infrared. 

The lattice model, just like the continuum CFT, has $S_3$ symmetry, generated by the $\Z_3$ symmetry of cyclical permutation of the 3-state spins and a charge conjugation. In the spin chain realization, the $\Z_3$ charge operator is
\begin{equation}
\label{eq:Q_def}
\widehat{\eta}=Q_{\Z_3}=\prod_{i=1}^L \tau_i^\dagger,
\end{equation}
where \begin{equation}\label{eqn:Z3action}
Q_{\Z_3} \sigma_i Q_{\Z_3}^\dagger = \omega \sigma_i , ~ Q_{\Z_3} \tau_i Q_{\Z_3}^\dagger = \tau_i.
\end{equation}
The charge conjugation operator is 
\begin{equation}\label{eqn:chargeconj}
\widehat{C}=\prod_{i=1}^{L} c_i, ~ 
c_i=\begin{pmatrix}
1& 0 & 0\\
0 & 0 & 1\\
0 & 1 &0
\end{pmatrix},
\end{equation}
with 
\begin{equation}
\label{eq:C_act}
\widehat{C}^2=\mathbbm{1}, ~ \widehat{C} \sigma_i \widehat{C} =\sigma_i^\dagger, ~ \widehat{C}\tau_i \widehat{C} =\tau_i^\dagger.
\end{equation}

It is useful in what follows to  write the Hamiltonian (\ref{eq:D_I}) in terms of the Potts representation of the  Temperley-Lieb algebra, following the conventions of Refs.~\cite{Koo_1994}:
\begin{equation}
\label{eq:H_I_TL}
H_{\idTDL} = -\frac{\gamma}{\pi\sin\gamma}\sum_{i  = 1}^{2L}e_{i},
\end{equation}
with $\gamma = \pi/6$. Here the Temperley-Lieb generators are given by
\begin{equation}
\label{eq:TL_Potts}
e_{2i-1} = \frac{1}{\sqrt{3}} \left(1+\sigma_i^\dagger \sigma_{i+1} +\sigma_{i+1}^\dagger \sigma_i \right),\ e_{2i} = \frac{1}{\sqrt{3}}\left(1+ \tau_i +\tau_i^\dagger \right),
\end{equation}
Recall that the $e_i$ satisfy the relations
\begin{subequations}\label{templiebdef}
\begin{equation}\label{middlelooptempremove}
    e_i ^2 = \left(q + q^{-1}\right) \ e_i \, ,
\end{equation}
\begin{equation}
    e_i  e_{i \pm 1}  e_i = e_i \, ,
\end{equation}
\begin{equation}
    e_i  e_j = e_j  e_i \text{ if } \mid i - j \mid  \ \geq 2 \, ,
\end{equation}
\end{subequations}
where $q = e^{i \gamma }$, $\gamma = \frac{\pi}{6}$, and the label $i$ is identified with $i+2L$.

There are additional ways to realize the three-state Potts Model on a lattice. One of them is to consider a lattice built using the $\mathbb{Z}_3$ Tambara-Yamagami fusion category \cite{Aasen2020}. Another way is to use a reformulation as a  Restricted Solid-on-Solid (RSOS) model \cite{Pasquier:1986jc}. This turns out to be more convenient for certain aspects of the problem, and further has the advantage to allow generalizations to other values of the central charge \cite{Bazhanov:1987zu}.\footnote{This  we will discuss in a subsequent paper \cite{Sinha2023a}.}
\subsection{The RSOS  model}

The input data for a RSOS model on a square lattice is a graph, which for us will be a Dynkin diagram denoted as $\mathbf{G}$. The choice of $\mathbf{G}$ as $\mathbf{A}_3$ gives rise to the Ising model and $\mathbf{G}=\mathbf{D}_4$ corresponds to the three-State Potts model. The spins (called {\it heights} for RSOS models) lie on the vertices of the lattice, and their values are labels on $\mathbf{G}$. The heights for nearest neighbour sites must be adjacent  on the Dynkin diagram. 

\begin{figure}
    \centering
    \includegraphics[scale = 0.5]{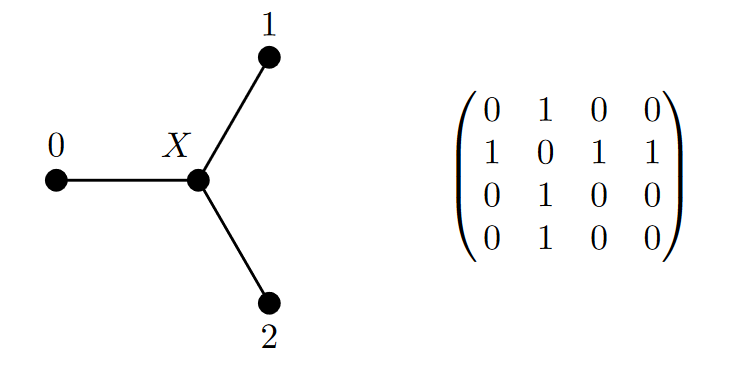}
    \caption{The $\mathbf{D}_4$ dynkin diagram and its adjacency matrix.}
    \label{fig:enter-label}
\end{figure}

 A row of   $N $ sites on the square lattice  can be thought of as a walk of length $N$, formed from the labels on $\mathbf{G}$:  0,$X$,1,2 for $\mathbf{D}_4$, and respecting the rule that only adjacent labels can follow each other along the walk. Let us further impose periodic boundary conditions (PBC) on these walks, which forces  $N$ to be even, i.e. $N = 2 L$, for $L \in \mathbb{Z}_{>0}$. The linear span of the walks with PBC makes up the Hilbert space - $\mathcal{H}_{2L}$. Further, the labels $x$ on $\mathbf{G}$ are given mod $N$, i.e. $x_{N+i} = x_i$. In the following, we adopt the convention that a product state in $\cH_{2L}$ is expressed as $|x_0, x_1,\ldots,x_{2L-1}\rangle$.

 For a given Dynkin diagram $\mathbf{G}$
, the adjacency matrix $G$ has entries $G_{ij}$ - the number of links connecting nodes $i$ and $j$. $G$ is a non-negative matrix and it can be shown (using the  Perron-Frobenius Theorem) that  the eigenvector  $\psi^{(1)}$  associated with the largest eigenvalue is completely positive\cite{mehta1977elements}. For $\mathbf{D}_4$, this vector  is $[  1 , \ \sqrt{3}, \ 1, \ 1 ]$. Now we define the operators
\begin{equation}
   \bra{....,x_{i-1}',x_{i}',x_{i+1}',....} e_i \ket{....,x_{i-1}, x_{i}, x_{i+1}....} = \left( \prod_{j \neq i}\delta_{x_j, x_j'} \right)\frac{\left(\psi^{(1)}_{x_i}\psi^{(1)}_{x_i'}\right)^{\frac{1}{2}}}{\psi^{(1)}_{x_{i-1}}}\delta_{x_{i-1}, x_{i+1}} \, ,
\end{equation}
where $i \in \{ 1,2, \ldots, N \}$. At a site $i$, there are two cases $x_i = 0 / 1 / 2$ or $x_i = X$. For these two cases, we write the action of the $e_i$ operators below in terms of their non-trivial matrix elements
\begin{subequations}\label{TLgenPotts}
    \begin{equation}
         \bra{....,x_{i-1}',x_{i}',x_{i+1}',....} e_i \ket{....,x_{i-1}, X, x_{i+1}....} = \left( \prod_{j }\delta_{x_j, x_j'} \right)\sqrt{3}  \ \delta_{x_{i-1}, x_{i+1}} \, ,
    \end{equation}
        \begin{equation}
         \bra{....,x_{i-1}',x_{i}',x_{i+1}',....} e_i \ket{....,X, a, X,....} = \left( \prod_{j \neq i}\delta_{x_j, x_j'} \right)\frac{1}{\sqrt{3}} \, ,
    \end{equation}
\end{subequations}
where $a \in \{ 0,1,2 \}$.\footnote{This discussion was presented in \cite{Saleur:1990uz}.  Equation (\ref{TLgenPotts}) exactly matches equation (3.64) in \cite{Aasen2020}, where it was derived in terms of projection operators in a Tambara-Yamagami Category. In \autoref{sec:TYtoRSOS}, we describe how Tambara-Yamagami Category can be used to study the 3-state Potts model.} In the action of the operators $e_{2L-1}$ and $e_{2L}$ we require $x_{2L}$ and $x_{2L+1}$, these are equal to $x_0$ and $x_1$ respectively, as $x_{2L+i} = x_{i}$. These operators obey the Temperley-Lieb algebra relations (\ref{templiebdef}).

 On the Hilbert space $\mathcal{H}_{2L}$, we can also define an operator $\uR$  called the shift  operator, which acts as 
\begin{equation}
 \uR\ket{x_0,x_1,x_2, ......,x_{2L-1}} = \ket{x_{2L-1},x_0,x_1, ......,x_{2L-2}} \, .  
\end{equation}
 The following relations are then satisfied by $\uR$ and  the $e_i$'s
\begin{subequations}
\begin{equation}
    \uR e_i = e_{i+1} \uR \, ,
\end{equation}
\begin{equation}
\uR^2 e_{N-1} = e_1  e_2.....e_{N-1} \ .
\end{equation}
\end{subequations}
$\uR$ and the TL generators form the affine TL algebra - $aTL_{N}(q)$ (see \cite{Belletete:2018eua} for details and more careful statements).

For  later use, we also define the braid operators ($g_j$) 
\begin{subequations}
\begin{equation}
\label{eq:g_def}
g_j = (-q)^{1/2}\mathbf{1} + (-q)^{-1/2}e_j,
\end{equation}
\end{subequations}
where $\mathbf{1} $ is the identity operator on the Hilbert space and  $j = 1, \ldots, N$. The braid operators satisfy  the relations
\begin{subequations}
\begin{equation}
g_i g_j = g_j g_i \text{ if }  \lvert \ i - j \ \rvert \geq 2 \, ,
\end{equation}
\begin{equation}
 g_i g_{i+1} g_i = g_{i+1} g_i g_{i+1} \, ,
\end{equation}
\end{subequations}
and form the braid group on $2L$ strands.

If we simulate the quantum Hamiltonian for some finite $N=2L$, the $N$-site $\mathbf{D}_4$ RSOS model exactly gives us the same ground state energy and low energy eigenvalues as the $L$-site three-state Potts model, provided we choose the correct normalizations. The only difference between these two models is that all the eigenvalues in the RSOS model appear with twice the multiplicity of what is seen in three-state Potts. (The mapping between the two bases of the Hilbert space is easy to write down, and is discussed further in \autoref{sec:spinchaintorsos}). This can be remedied by working in an even/odd sector in the RSOS model, where by even (odd) sector we mean all the walks described above must start and end with $X$ (0,1 or 2). It is not hard to check that both these subspaces are in fact modules for the Temperley-Lieb algebra, but not the affine-TL algebra, as $\uR$ moves us from the even to the odd sector, and  vice versa.

\section{Line Operators in the crossed channel} \label{sec:TDLlatticecrossed}

In this section and section \autoref{sec:TDLlatticedirect}, we provide the lattice realizations of topological lines in the Potts CFT on the three-state Potts quantum spin chain (as well as the  $D_4$ RSOS model). We first consider  the crossed channel, where the topological lines are realized as extended line operators supported on the whole spin chain. We denote the operator corresponding to the topological line $D$ by $\widehat{D}$.

\subsection{The symmetry operators $\widehat{\eta},\widehat{\overline{\eta}},\widehat{C}$}

The simplest examples are the invertible symmetry lines, whose lattice realizations  in the crossed channel are given by the corresponding symmetry charge operators. In particular, the operator corresponding to the $\Z_3$ line $\eta$ is given by (\ref{eq:Q_def}), namely
\begin{equation}\label{eq:eta_line_crossed}
\widehat{\eta} = Q_{\Z_3} = \prod_{i=1}^L \tau_i^\dagger .
\end{equation}
Similarly, the operator corresponding to the $\overline{\eta}$ line is
\begin{equation}
\widehat{\overline{\eta}} = \prod_{i=1}^L \tau_i \, ,
\end{equation}
and the operator corresponding to the charge conjugation line $C$ is given by (\ref{eqn:chargeconj}).  As symmetry charges these operators  commute with the Hamiltonian. 

We now use the RSOS formulation of the three-state Potts model to study the line operators in the crossed channel. In what follows, we construct RSOS operators $\widehat{\eta}_{\text{RSOS}}$ and $\widehat{C}_{\text{RSOS}}$, which are the same as $\widehat{\eta}$ and $\widehat{C}$ operator when one maps the $D_4$ RSOS model to the three-state Potts chain, as done in \autoref{sec:spinchaintorsos}. We will now drop the subscript RSOS from these operators since, from the Hilbert space on which these line operators act, it is clear whether we are talking about $\widehat{\eta}_{\text{RSOS}}$  ($\widehat{C}_{\text{RSOS}}$) or $\widehat{\eta}$  ($\widehat{C}$). Further, in the rest of this subsection and in \autoref{sec:spinchaintorsos}, we use the convention that 
\begin{equation}\label{eqn:convention2}
\sigma =\begin{pmatrix}
1 & 0 & 0\\
0 & \omega & 0\\
0 & 0 & \omega^2
\end{pmatrix}, ~ 
\tau=\begin{pmatrix}
0 & 0 & 1\\
1 & 0 & 0\\
0 & 1 & 0
\end{pmatrix}, ~ \omega=\text{e}^{2\pi\text{i}/3} \, .
\end{equation}    
The $\mathbb{Z}_3$ charge operator $\widehat{\eta}$ and the charge conjugation operator $\widehat{C}$, have the same definitions as in equations \eqref{eq:Q_def} and \eqref{eqn:chargeconj}. This convention and the convention we use in the rest of this paper are related by a unitary transformation, which we discuss in \autoref{sec:spinchaintorsos}. In particular, under this convention the relations between operators given in equations \eqref{eq:pottsgenalgebra}, \eqref{eqn:Z3action}, and \eqref{eq:C_act}  still hold.
It is useful in what follows to also consider the action of these symmetry operators in the RSOS representation of the Potts model. A bijection $\sigma : \{0,1,2 \} \to \{0,1,2 \}$, can also be considered  as an element of $S_3$. Corresponding to $\sigma$, we have a map, denoted as $D(\sigma)$ in what follows, which acts by permuting the Potts spins\footnote{In the RSOS Potts Hilbert space, we call 0,1, and 2 ``physical spins'' - these  can be mapped to states $\uparrow$, $\searrow$, and $\swarrow$ in the  Potts spin chain. $X$ acts like a wall, separating these spins. See \autoref{sec:spinchaintorsos} for more details.},  for example
\begin{equation}
    D(\sigma) \ket{X,1,X,0,X,2,X,2} = \ket{X, \sigma(1), X, \sigma(0),X,\sigma(2),X, \sigma(2)} \, .
\end{equation}
This operator can in general be written as 
\begin{equation}
D(\sigma) \ket{\ldots , X, x_{i}, X, x_{i + 1}, X, x_{i + 2}, X, \ldots} = \ket{\ldots , X, \sigma(x_{i}), X, \sigma(x_{i + 1}), X, \sigma(x_{i + 2}), X, \ldots} \ .
\end{equation}
It can be shown that this operator commutes with $e_i$. Indeed, recall that the action of $e_i$ is 
\begin{subequations}
    \begin{equation}
            e_i \ket{...,x_{i-1},X,x_{i+1},...} = \sqrt{3} \ \delta_{x_{i-1},x_{i + 1}}\ket{...,x_{i - 1},X,x_{i + 1},...}  \, ,
    \end{equation}
    \begin{equation}
            e_i \ket{...,X, x_i ,X,...} = \frac{1}{\sqrt{3}}\sum_{b \in \{0,1,2 \} } \ket{...,X,b,X,...} \, .
    \end{equation}
\end{subequations}
Let us first consider the action of $D(\sigma) \, e_i$ on  the two possible kinds of basis elements
\begin{equation}
\begin{split}
    D(\sigma) \, e_i \ket{\ldots, x_{i - 1}, X,  x_{i + 1}, \ldots } &  =  \sqrt{3} \ \delta_{x_{i-1},x_{i + 1}} \, D(\sigma) \ket{...,x_{i - 1},X,x_{i + 1},...} \, ,   \\
     =  \sqrt{3} \ \delta_{x_{i-1},x_{i + 1}} \, &  \ket{...,\sigma(x_{i - 1}),X, \sigma(x_{i + 1}),...} \, . \\
     D(\sigma) \, e_i \ket{\ldots, X,  x_{i}, X, \ldots } = \frac{1}{\sqrt{3}} & \, D(\sigma) \sum_{b \in \{0,1,2 \} }  \ket{...,X,b,X,...} \, ,   \\
     = \frac{1}{\sqrt{3}} \, \sum_{b \in \{0,1,2 \} } &  \ket{..\sigma(x_{i-1}),X,b,X,\sigma(x_{i+1})...}  \, .
\end{split}
\end{equation}
Applying $e_i D(\sigma)$ instead, for the first type of basis elements we get 
\begin{equation}
\begin{split}
e_i  \,  D(\sigma) \ket{\ldots, x_{i - 1}, X,  x_{i + 1}, \ldots } = e_{i} \ket{\ldots, \sigma(x_{i - 1}), X , \sigma(x_{i + 1}), \ldots} \, ,  \\
= \sqrt{3} \, \delta_{\sigma(x_{i-1}),\sigma(x_{i + 1})}  \ket{\ldots, \sigma(x_{i - 1}), X , \sigma(x_{i + 1}), \ldots} \, ,  \\ 
= \sqrt{3} \, \delta_{x_{i-1},x_{i+1}} \ket{\ldots, \sigma(x_{i - 1}), X , \sigma(x_{i + 1}), \ldots} \, , 
\end{split}
\end{equation}
where $\delta_{\sigma(x_{i-1}), \sigma(x_{i+1})} = \delta_{x_{i-1},x_{i+1}} $, since $\sigma$ is a bijection. For the second type of basis elements we have
\begin{equation}
    \begin{split}
        e_i  \,  D(\sigma) \ket{\ldots, X,  x_{i}, X, \ldots } = e_{i} \ket{\ldots \sigma(x_{i - 1}), X,\sigma(x_{i}),X,\sigma(x_{i + 1}) \ldots} \, , \\
        =  \frac{1}{\sqrt{3}} \, \sum_{b \in \{0,1,2 \} }   \ket{..\sigma(x_{i-1}),X,b,X,\sigma(x_{i+1})...} \, .
    \end{split}
\end{equation}
Hence, we have shown that 
\begin{equation}
    D(\sigma) \, e_{i} = e_{i} \, D(\sigma) \, .
\end{equation}
Further, it is not hard to see that 
\begin{equation}
    D(\sigma) \, \uR = \uR \, D(\sigma) \, .
\end{equation}
Therefore, $D(\sigma)$ commutes with all generators of the affine Temperley Lieb algebra. This can be used to argue for topological invariance on the lattice following \cite{Belletete2018}.

For any $\sigma \in S_3$,  it can be shown that $D(\sigma^{-1}) \circ D(\sigma) = D(\mathbf{1}) = \textbf{Id}$, where $\mathbf{1}$ is the identity of $S_3$ and $\textbf{Id}$ is the identity defect.  We will now call $\widehat{\eta}$ and $\widehat{C}$ the operators on the lattice corresponding  specifically to the choices $(\sigma(0),\sigma(1),\sigma(2)) = (1,2,0)$ and  $(\sigma(0),\sigma(1),\sigma(2)) = (0,2,1)$ respectively. Using exact diagonalization, we list the action of these operators in Table \ref{invertible-lines-expectation}. Comparing with the second and third column in Table \ref{table:PottsTDLaction} allows us to identify our lattice operators $\widehat{\eta}$ and $\widehat{C}$ with the space-like invertible topological lines $\widehat{\eta}$ and $\widehat{C}$ in the continuum. As $\widehat{\eta}$ and $\widehat{C}$ generate the full $S_3$ symmetry, we have found RSOS lattice operators corresponding to the six invertible lines in the Potts CFT.

We note that, although these six invertible lines commute with all elements of the affine TL algebra, they cannot be written in terms of the $Y$ and $\overline{Y}$ operators introduced in \cite{Belletete2018}, which are central in the affine TL algebra. Indeed from Table \ref{invertible-lines-expectation}, one can see that $Y$ and $\overline{Y}$ act in the same way on  $\ket{\frac{1}{15},\frac{1}{15}}^{(1)}$ and  $\ket{\frac{1}{15},\frac{1}{15}}^{(2)}$. It is therefore not possible to use these operators alone to construct $\widehat{\eta}$ and $\widehat{C}$, since the latter  act  differently on these two states. This happens because the results of \cite{Belletete2018} are only valid for generic $q$ and the whole affine TL algebra. The Potts model does not provide a faithful representation, since it involves only a subset of all representations: within this subset, the center of the algebra can be larger, as it indeed is. 

\subsection{The duality operators $\widehat{N}$ and $\widehat{N'}$}\label{subsec:duality}

The $\widehat{N}$ and $\widehat{N'}$ operators are expected to implement the Kramers-Wannier duality transformation on the three-state Potts quantum chain. On the other hand, they are a special instance of the $Y$ and $\overline{Y}$ operators discussed in \cite{Belletete:2018eua,Belletete2020}. The $Y$ and $\overline{Y}$ operators, also called the hoop operators, are defined in general using braid generators 

\begin{subequations}\label{Yoperator}
    \begin{equation}
        Y = (-q)^{-1/2} \,   \, g^{-1}_{1} \ldots g_{2L - 1}^{-1}  \uR^{-1} 
        +  (-q)^{1/2} \, \uR \, g_{2L -1}  \ldots g_1    \,  ,
    \end{equation}
    \begin{equation}
        \overline{Y} =  (-q)^{-1/2} \,  \uR  \, g^{-1}_{2L - 1} \ldots g_{1}^{-1} +  (-q)^{1/2} \, g_1 \ldots g_{2L -1}  \, \uR^{-1}    \,  .
    \end{equation}
\end{subequations}    
These  operators lie in the center of $aTL_{N}(q)$ with $N=2L$, in fact they generate  (for $q$ generic) the center as shown in \cite{Belletete:2018eua}. 
In this reference one can also find nice diagrammatic representations of the operators we have defined in this section and the relations satisfied by them.  We will show elsewhere that the $Y$ ($\overline{Y}$) operators can also be realized (up to normalization) as a transfer matrix with the spectral parameters along the entire row set to ${\rm i} \infty$ ($ - {\rm i} \infty$), see \cite{Finch:2014nxa} for a similar discussion but for a different model. We also note that although the form of $Y$ and $\overline{Y}$ operators in \eqref{Yoperator} seems to depend on the choice of a starting site, the actual operators do not have such a dependence. If we consider the operator $\uR^i Y \uR^{-i}$, then all the indices in \eqref{Yoperator} will increase by $i$ (mod $2L$). However, $\uR^i Y \uR^{-i} = Y$ as $Y$ commutes with $\uR$ by integrability. (The same is true for the $\overline{Y}$ operator.) This is consistent with the periodic boundary condition that we chose to use.

As the $Y$ and $\overline{Y}$ operators are central in the affine Temperley-Lieb algebra, they are expected to be topological on the lattice: this can be proven e.g. using Reidemeister moves as in \cite{Belletete2018}. We note that an alternative characterization of the topological nature of the lines is obtained by demanding commutation with the Virasoro (chiral and antichiral)  generators. Since the combinations of the  Temperley-Lieb generators provide regularizations of the Virasoro algebra~\cite{Koo_1994}, the fact  $Y$ and $\overline{Y}$ lie in the center of affine TL guarantees, in this approach, that their continuum limit commutes with Virasoro, and therefore is topological.

We computed numerically the expectation values of $Y$ and $\overline{Y}$ on   various low energy states of the Potts model, as detailed in Table \ref{invertible-lines-expectation}. These  expectation values match perfectly the continuum limit predictions in Table \ref{table:PottsTDLaction}. In fact, using representation theory of affine TL, it can be shown that the expectation values in finite size  are exactly those of the continuum limit \cite{Belletete2018}, which agrees with the fact that  these operators are topological on the lattice. In the following, we will show $Y$ and $\overline{Y}$, while obtained initially using algebraic techniques, realize indeed  the Kramers-Wannier duality transformation, and can be identified with   the $\widehat{N}$ and $\widehat{N'}$ operators. 

Extended operators realizing the Kramers-Wannier duality transformation have been studied before, see e.g. \cite{Aasen2016,Aasen2020,Tan:2022vaz,Li:2023mmw,Seiberg:2023cdc}. Here we will construct operators mapping between states on a given lattice and states on its dual lattice, where we adopt the convention of Eq. (\ref{eqn:convention2}). For convenience, we denote the Hilbert spaces for such states as $\cH_{\text{even}}$ and $\cH_{\text{odd}}$ respectively, where we adopt the periodic boundary condition for both $\cH_{\text{even}}$ and $\cH_{\text{odd}}$, namely $h_{L+1}$ is identified with $h_1$.\footnote{This is different from the setup of \cite{Li:2023mmw} and related setup of \cite{Seiberg:2023cdc}. We thank Shu-Heng Shao and Yunqin Zheng for helpful discussions regarding boundary conditions, lattice translation and the translational invariance.} In $\cH_{\text{even}}$, physical spins are placed on even sites and the states take the form of $| X h_1 X h_2 ... X h_L \rangle$ where $h_i \in \{ 0,1,2 \}$ labels the possible single-site spin configurations in the Potts model and we have used $X$ to label the auxiliary sites. In $\cH_{\text{odd}}$, the physical spins are placed on odd sites and the states take the form of $| h_1 X h_2 X ... h_L X \rangle$. The actions of $\widehat{N}$ and $\widehat{N'}$ line operators are then given by
\begin{equation}\label{eqn: KWop}
\begin{aligned}
\widehat{N} | X h_1 X h_2 \dots X h_L\rangle & = \bigotimes_{r=1}^L \frac{1}{\sqrt{3}} \left( |0 X\rangle  + w^{h_{r-1}+2h_r} |1 X\rangle +w^{2(h_{r-1}+2h_r)} |2 X\rangle \right) \, ,\\
\widehat{N} |  h_1 X h_2 \dots X h_L X\rangle & = \bigotimes_{r=1}^L \frac{1}{\sqrt{3}} \left( |X 0 \rangle  + w^{h_{r}+2h_{r+1}} | X 1 \rangle +w^{2(h_r+2h_{r+1})} |X 2 \rangle \right) \, , \\
\widehat{N'} | X h_1 X h_2 \dots X h_L\rangle & = \bigotimes_{r=1}^L \frac{1}{\sqrt{3}} \left( |0 X\rangle  + w^{2h_{r-1}+h_r} |1 X\rangle +w^{2(2h_{r-1}+h_r)} |2 X\rangle \right) \, , \\
\widehat{N'} |  h_1 X h_2 \dots X h_L X\rangle & = \bigotimes_{r=1}^L \frac{1}{\sqrt{3}} \left( |X 0 \rangle  + w^{2h_{r}+h_{r+1}} | X 1 \rangle +w^{2(2h_r+h_{r+1})} |X 2 \rangle \right) \, , \\
\end{aligned}
\end{equation}
where we consider states on a periodic chain of $L$ sites and $w=\text{e}^{2\pi\text{i}/3}$ as usual. Equation (\ref{eqn: KWop}) is consistent with the fact that the Kramers-Wannier duality transform interchanges the Potts ferromagnetic and paramagnetic phases. As an example, the states $|X0X0...X0\rangle$, $|X1X1...X1\rangle$ and $|X2X2...X2\rangle$ all get mapped to $|+X+X...+X\rangle$ where $|+\rangle=1/\sqrt{3}(|0\rangle +|1\rangle+|2\rangle)$. At the level of operators, we have $\sigma^\dagger_i\sigma_{i+1} \widehat{N} = \widehat{N} \tau_i$ and $\widehat{N}\sigma_i^\dagger\sigma_{i+1}=\tau_{i+1}\widehat{N}$ on $\cH_{\text{even}}$ and similar relations (up to labelling) for $\cH_{odd}$, where the operators on the LHS and RHS act on the physical spins in $\cH_{\text{odd}}$ (or $\cH_{\text{even}}$) and $\cH_{\text{even}}$ (or $\cH_{\text{odd}}$) respectively.

Using (\ref{eqn: KWop}) we can also check the $\Z_3$ TY fusion relations on the Potts lattice. Recall that, the $\Z_3$ symmetry operator $\hat{\eta}$ acts on the physical spins as a shift operation, namely
\begin{equation}
\begin{split}
\widehat{\eta} | X h_1 X h_2 \dots X h_L\rangle & = | X (h_1+1) X (h_2+1) \dots X (h_L+1)\rangle\, ,\\
\widehat{\eta} |  h_1 X h_2 X \dots  h_L X\rangle &= |  (h_1+1) X (h_2+1)X \dots  (h_L+1) X\rangle \, .
\end{split}
\end{equation} 
Since simultaneously shifting all $h_i$ by $1$ does not change the weights appearing in (\ref{eqn: KWop}), we confirm the following fusion relations of line operators on the lattice:
\begin{equation}
\widehat{N}\widehat{\eta}=\widehat{N}, ~ \widehat{\overline{N}}\widehat{\eta}=\widehat{\overline{N}} \, .
\end{equation}
Additionally, since the charge conjugation operator $\widehat{C}$ exchanges the spin value $h_i=1$ with $h_i=2$ and vice versa, it effectively exchanges the weights $w^{h_{r-1}+2h_r}$ with $w^{2h_{r-1}+h_r}$ etc in (\ref{eqn: KWop}). Therefore we have
\begin{equation}
\widehat{N}\widehat{C} = \widehat{N'}, ~ \widehat{N'}\widehat{C} = \widehat{N} \, .
\end{equation}
To check the fusion of two $\widehat{N}$ (or $\widehat{N'}$) operators, on $\cH_{\text{even}}$ we consider the amplitude 
\begin{equation}
\langle Xh''_1\ldots Xh''_L|\widehat{N}\widehat{N} |Xh_1\ldots Xh_L\rangle .
\end{equation}
It can be shown that this amplitude vanishes unless $h_i'' = h_i$, or $h_i'' = h_i+1$, or $h_i''=h_i+2$ for all $i$. Similar statements hold for $\cH_{\text{odd}}$. This represents the fusion rule
\begin{equation}
\widehat{N}\widehat{N}=\widehat{\idTDL} +\widehat{\eta}+\widehat{\overline{\eta}} \, .
\end{equation} 

We will now show that actions of the $\widehat{N}$ and $\widehat{N'}$ operators, as written in (\ref{eqn: KWop}), match actions of the $Y$ and $\overline{Y}$ operators. In the following, we demonstrate this for the $Y$ operator, while the case of $\overline{Y}$ operator works out analogously. The $Y$ operator maps states between $\cH_{\text{even}}$ and $\cH_{\text{odd}}$ as illustrated in \autoref{fig:Yoperator}. 
\insfigsvg{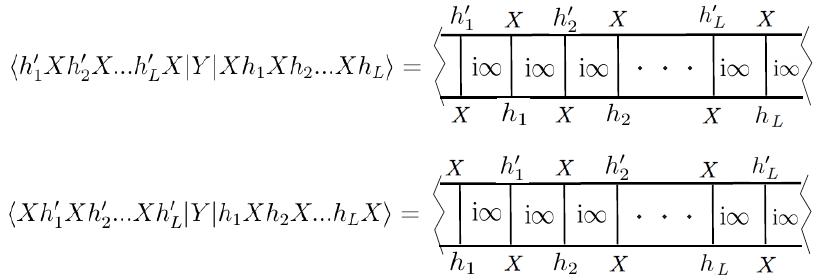}{0.8}{The $Y$ operator maps between $\cH_{\text{even}}$ and $\cH_{\text{odd}}$.}

In \autoref{fig:Yoperator}, each face with the spectral parameter set to  $\text{i}\infty$ contributes the factors shown in \autoref{fig:localweights}.

\insfigsvg{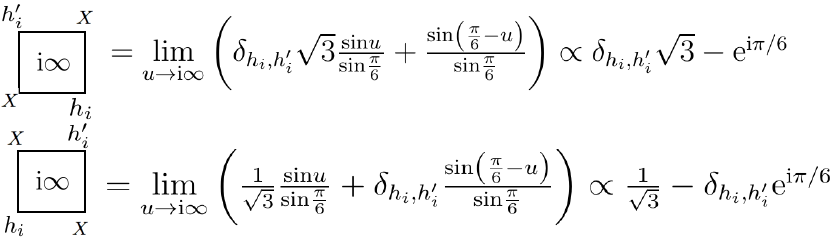}{0.8}{Local weights associated with a face with the spectral parameter set to $\text{i}\infty$.}

Thanks to the local structure of (\ref{eqn: KWop}), it suffices to check the weights associated with two neighboring faces, which should produce a single factor inside the product in (\ref{eqn: KWop}). For example, to check the action of $Y$ on states in $\cH_{\text{even}}$, we need to consider the local weights associated with the following two faces:
\begin{figure}[h]
\centering
\includegraphics[scale=0.8]{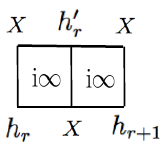}
\end{figure}

Denoting the corresponding weight as 
\begin{equation}
A_{h_r,h_r',h_{r+1}}:= \delta_{h_r',h_{r+1}}+\delta_{h_r,h_r'}\text{e}^{\text{i}\frac{\pi}{3}}-\delta_{h_r,h_r'}\delta_{h_r',h_{r+1}}\sqrt{3}\text{e}^{\text{i}\frac{\pi}{6}}-\frac{1}{\sqrt{3}}\text{e}^{\text{i}\frac{\pi}{6}},
\end{equation}
to produce the local structure in (\ref{eqn: KWop}), we must have
\begin{equation}
A_{h_r, 1, h_{r+1}} = w^{h_r+2h_{r+1}}A_{h_r,0,h_{r+1}}, ~ A_{h_r, 2, h_{r+1}} = w^{2(h_r+2h_{r+1})}A_{h_r,0,h_{r+1}}\, ,
\end{equation}
which we can check is satisfied by enumerating all possible cases.

\begin{table}[h!]
\renewcommand{\arraystretch}{1.4}
    \centering
\begin{tabular}{ |P{3cm}| P{3cm}|P{3cm}|P{3cm}|P{3cm}| }
\hline
 \multicolumn{5}{|c|}{States and Expectation values for lattice operators - 16 RSOS sites } \\
 \hline
 State & $\widehat{\eta}$ & $\widehat{C}$ &  $T({\rm i } \, \infty) = Y$  &  $T ({-\rm i } \, \infty) = \overline{Y}$  \\ [2ex]
 \hline
  $\ket{0,0} $   & 1     & 1  & 1.73205081 & 1.73205081    \\

 $\ket{\frac{1}{15},\frac{1}{15}}^{(1)}$    & $\omega$  & $\ket{\frac{1}{15},\frac{1}{15}}^{(2)}$   &  0   & 0\\
  $\ket{\frac{1}{15},\frac{1}{15}}^{(2)}$   & $\omega^2$   & $\ket{\frac{1}{15},\frac{1}{15}}^{(1)}$    &   0 & 0     \\
 $\ket{\frac{2}{5},\frac{2}{5}}$      & 1    & 1   &    -1.73205081 & -1.73205081     \\
 $L_{-1}\ket{\frac{1}{15},\frac{1}{15}}^{(1)}$    & $\omega$    & $L_{-1}\ket{\frac{1}{15},\frac{1}{15}}^{(2)} $  & 0 &  0          \\
 $L_{-1}\ket{\frac{1}{15},\frac{1}{15}}^{(2)}$    & $\omega^2$   & $L_{-1}\ket{\frac{1}{15},\frac{1}{15}}^{(1)} $  & 0 & 0   \\
 $\bar{L}_{-1}\ket{\frac{1}{15},\frac{1}{15}}^{(1)}$    & $\omega$     & $\bar{L}_{-1}\ket{\frac{1}{15},\frac{1}{15}}^{(2)}$   & 0  &  0          \\
 $\bar{L}_{-1}\ket{\frac{1}{15},\frac{1}{15}}^{(2)}$    & $\omega^2$   & $\bar{L}_{-1}\ket{\frac{1}{15},\frac{1}{15}}^{(1)}$  &  0 & 0   \\
 $\ket{\frac{2}{3},\frac{2}{3}}^{(1)}$     & $\omega$ &  $\ket{\frac{2}{3},\frac{2}{3}}^{(2)} $    & 
0 & 0 \\  
 $\ket{\frac{2}{3},\frac{2}{3}}^{(2)} $      & $\omega^2$ &  $\ket{\frac{2}{3},\frac{2}{3}}^{(1)}$   &   0 & 0    \\
 \hline
\end{tabular}
\caption{In this table we list the actions of the lattice operators $\widehat{\eta},\widehat{C}, Y$ and $\overline{Y}$. Numerical values are given when the states are eigenstates of the lattice operators, otherwise we list the state we get after the action of the operator.}
\label{invertible-lines-expectation}
\end{table}

\subsection{The operator $\widehat{W}$}
We now provide numerical evidence that a lattice operator - $\widehat{W}$, which acts on the RSOS Hilbert space, is the lattice realization of the $\widehat{W}$ topological line operator in the Potts CFT. The form of this operator on the lattice is  
\begin{equation}\label{eq:(2,1)defect}
    \widehat{W} = T\left(-\frac{\pi}{2}\right) \uR^{-1} , 
\end{equation}
where $T$ denotes the transfer matrix with the spectral parameters set to $-\pi/2$.

Now, since this operator can be written in terms of products of transfer matrices and their inverses at different spectral parameters (recall $\uR = T(0)$), it  commutes with the transfer matrix at any other spectral parameter, as a consequence of the Yang Baxter Equation. In particular, it commutes with $Y$ ($= T( {\rm i} \infty)$), $\overline{Y}$ (= $T( -{\rm i} \infty) $), and the Hamiltonian ($ = T^{-1}(0)  \dot{T}(0)$), which is something we would also demand from the $\widehat{W}$ line operator in the continuum. 

\begin{figure}[h!]
    \centering
    \includegraphics[scale = 0.5]{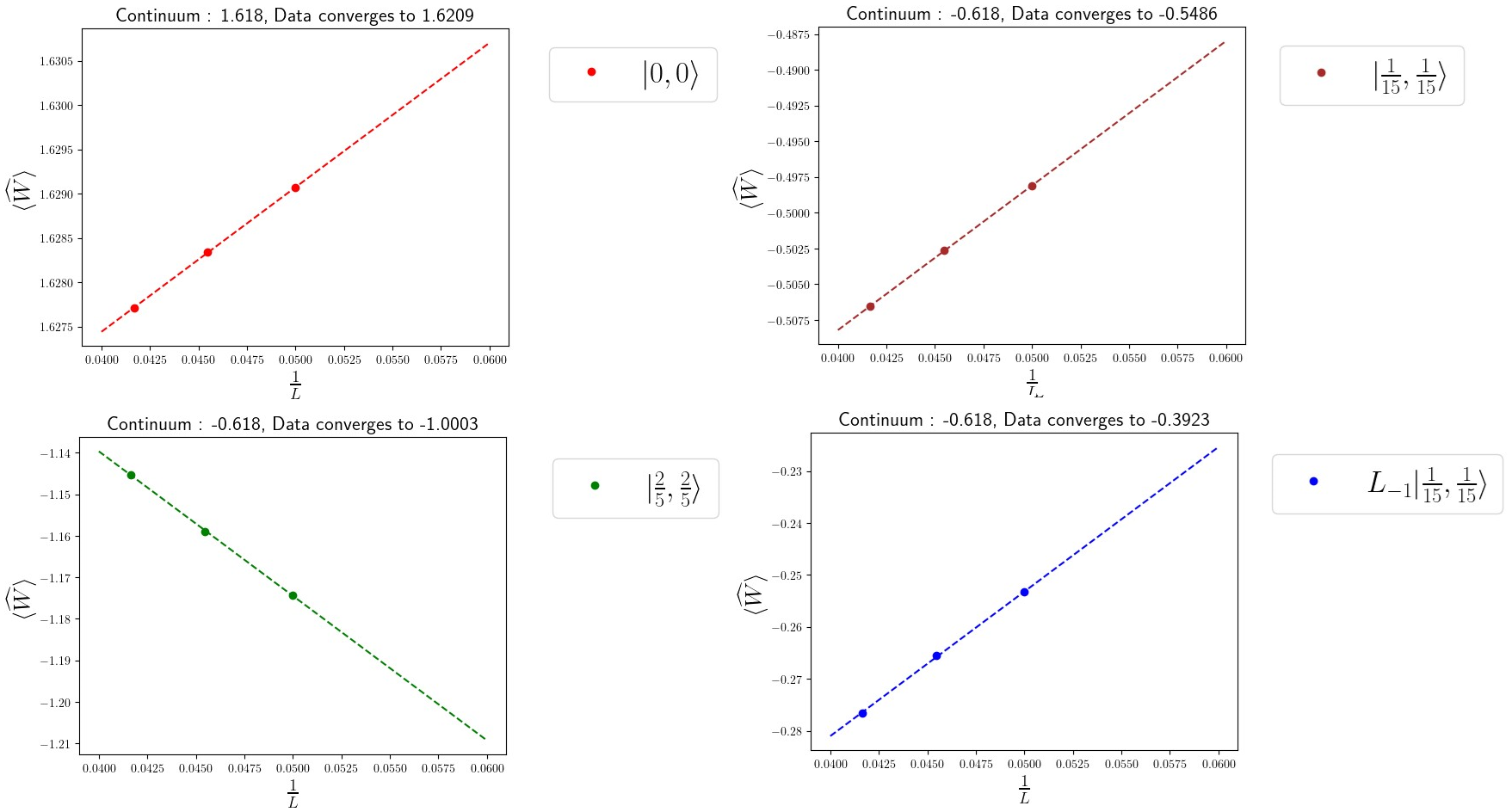}
    \caption{Expectation values of the $\widehat{W}$ lattice operator for several sets of states. The results for the two groups should converge to the golden ratio $x$ (resp. $- x^{-1}$) as $L\to\infty$, see Table \ref{table:PottsTDLaction}.}
    \label{fig:W_expectation_value}
\end{figure}

Since $\widehat{W}$ commutes with the transfer matrix, and therefore the Hamiltonian, it is diagonal in the eigenbasis of $H_{\idTDL}$. In the scaling limit, the corresponding expectation values should be constant within a given Virasoro module, and take different values in different sectors as indicated in Table \ref{table:PottsTDLaction}. The expectation values $\langle\widehat{W}\rangle$ are shown in \autoref{fig:W_expectation_value} for low-lying eigenstates of $H_{\idTDL}$ obtained using exact, numerical diagonalization of the corresponding RSOS Hamiltonian. Since the constructed line operator approaches the continuum counterpart only in the scaling limit, we perform a finite-size scaling analysis to obtain the expectation value in the thermodynamic limit. While the obtained results are in reasonable agreement with the CFT predictions for the two lowest energy eigenstates, the same is not true for higher energy states. We believe this is due to the limited set of system-sizes probed in the exact computation and could be improved using larger scale numerical analysis or using Bethe Ansatz. To further substantiate the claim that the proposed line operator approaches the CFT counterpart in the scaling limit,  we also analyze the commutators of the line operator with the TL generators for different system sizes. Since the TL generators can be used to build  lattice regularizations of the  Virasoro generators \cite{Koo_1994}, the vanishing of  these commutators  is an important way to check the topological nature of the line operator in the scaling limit. 

\begin{figure}
    \centering
    \includegraphics[scale = 0.75]{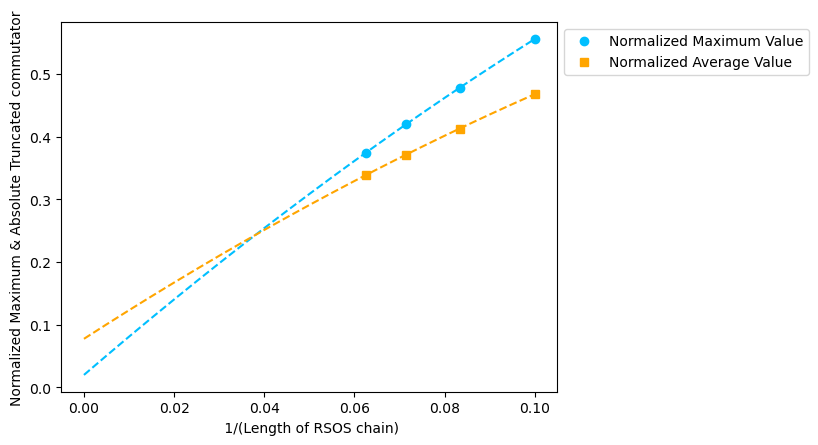}
    \caption{Maximum and Average absolute values of elements in the truncated commutator $[\widehat{W}, e_i]$ as a function of inverse system-size, normalized using the operator $\widehat{W}e_i$. The lines are constructed by fitting the data points with a polynomial fit, further we only study the lowest 24 energy eigensates. Normalized Maximum Value and Normalized Average value converge to 0.01988233 and 0.07748208 respectively. The computed quantities tending to zero, as the system-size is increased is indicative of the topological nature of the constructed line operator in the scaling limit. }
    \label{fig:W_commutativity}
\end{figure}

To investigate the topological nature of the $\widehat{W}$ operator, we look at the commutator of $\widehat{W}$ with the $e_i$'s. Since we are interested in a property emerging in the continuum limit, we can restrict  to  low-lying eigenstates of the RSOS Hamiltonian: in practice we study properties of $\left[\widehat{W},e_i\right]$ restricted to the first 24 eigenstates. To obtain meaningful results we need to set a scale, and normalize the commutator using the $\widehat{W} e_i$ operator. Figure \ref{fig:W_commutativity} shows the maximum and average absolute value of elements in the truncated commutator matrix $[\widehat{W}, e_i]$, divided by the maximum and average value of the truncated $\widehat{W} e_i$ matrix respectively. Both the normalized maximum and average absolute value approaches zero as the system size is increased, indicating that the $\widehat{W}$-line becomes topological in the scaling limit indeed.

\section{Defect Hamiltonians in the direct channel }
\label{sec:TDLlatticedirect}
In this section, we turn to the study of lattice realizations of  the different topological defect lines in the direct channel. 

It is interesting first  to pause briefly to discuss the relationship between the crossed and the direct channels. In the latter case,  the defects are associated with local modifications of the interactions in an otherwise periodic quantum chain, giving rise to some sort of twisted boundary conditions. While in the crossed channel, the topological nature was characterized by the simple condition that the defect operator commutes with the left and right copies of the Virasoro algebra, asserting that a defect is topological in the direct channel does not seem so obvious. Of course, one can always start by requiring that the defect be topological from the (1+1)-d quantum system point of view. This corresponds to demanding 
invariance of expectation values of observables under a shift of the defect in the space direction, as long as this shift does not cross the locations where the observables are measured. This has been discussed recently e.g. in \cite{Seifnashri:2023dpa,Seiberg:2023cdc,Shao:2023gho}. A   necessary condition for this property to hold  is the existence of a local unitary operator which shifts the location of the defect. For instance, let  $H_D^{[j, j+1]}$ be the Hamiltonian with a defect bond $D$ between sites $j$ and $j+1$. The local unitary operator $U_D[j]$,\footnote{As shown in equation \eqref{eq:translation_op}, the two defect Hamiltonians,$H_D^{[j, j+1]}$ and $H_D^{[j-1,j]}$ are related  by a unitary transformation, therefore the energy spectra remain the same. Hence, we will usually not write the location of the defect in the Hamiltonian.} which we call the defect-shift operator, would then have to satisfy 
\begin{equation}\label{eq:translation_op}
   U_D[j] \, H_D^{[j, j+1]} \, U_D[j]^{-1} =   H_D^{[j-1,j]} \, , 
\end{equation}
to shift the defect location to the left by 1 latice unit.
 We note that corresponding to local {\it invertible} defects ($H_D^{[j, j+1]}$) with local shift operators ($U_D[j]$), such as the ones we discuss in this section, a prescription for generating a unitary operator commuting with the no-defect Hamiltonian has been discussed in \cite{Seifnashri:2023dpa}. In this work, we observe that if $H_D^{[j, j+1]}$ is the defect Hamiltonian corresponding to an invertible TDL $D$, then the unitary operator constructed is exactly the line operator $\widehat{D}$.

Going back to our main goal, we can check that the lattice realization of a certain  defect  has been properly identified in the direct channel by studying  the energy and momenta of eigenstates of the corresponding defect Hamiltonians, and comparing them with predictions from CFT. 

Technically, the momentum $P$  is obtained by computing the eigenvalues,~$e^{i2\pi Pa}$, of the translation operator,~$T$, where  $a$ is the lattice spacing (set to one in numerical computations). $T$ itself involves several ingredients. The translation operator for a periodic chain with no defect is 
\begin{equation}
\begin{split}
    T_I = \uP \, , 
\end{split}
\end{equation}
where $\uP$ is the shift operator,  moving the Potts states to the right by one site (for the RSOS model, the corresponding operator is $\uR$). 
The translation operator for the defect Hamiltonian is then related with the defect-shift operator, $U_{D}[j]$,  by 
\begin{equation}
    T_D[j] = \uP \, U_D[j] =  U_D[j+1] \,  \uP \, .
\end{equation}
Defect translation operator commutes with the corresponding defect Hamiltonian, i.e. $T_D[j]$ commutes with $H_D^{[j, j+1]}$. We also note that for the defect Hamiltonians obtained using integrability ($H_N$, $H_{N'}$, and $H_W$), the translation operators can be  constructed using the transfer matrix with the same spectral parameter as the one used to obtain the defect Hamiltonian. Only at the special values of the  spectral parameters corresponding to $N, N',$ and $W$ lines are  these translation operators unitary. 

Now, standard finite-size scaling relations~\cite{Cardy:1986ie}:
\begin{subequations}
\begin{equation}
    E_{h, \bar{h}}(L_{\rm eff}) = \frac{2 \pi }{L_{\rm eff}}\left(-\frac{c}{12} + h + \bar{h} \right) + O\left(\frac{1}{L_{\rm eff}^2}\right),
\end{equation}
\begin{equation}
    P_{h, \bar{h}}(L_{\rm eff}) = \frac{2 \pi }{L_{\rm eff}}\left(h - \bar{h} \right) + O\left(\frac{1}{L_{\rm eff}^2}\right) \, ,
\end{equation}
\end{subequations}
enable determination of the conformal dimensions for the different energy eigenstates. Here, $L_{\rm eff}$ is the `effective' length of the quantum spin chain. As explained below for the $N, N'$ defects, the effective length can differ from the actual length of the chain~(see Refs.~\cite{Grimm2001, RoySaleur2022, Tan:2022vaz} for similar analysis for the Ising case). We emphasize finally that, despite their sometimes unusual appearance, the Hamiltonians that we construct here are hermitian, as they should be.

\subsection{The $\idTDL$ defect}
The~$\idTDL$ defect corresponds to the usual Potts spin chain with periodic boundary condition. The relevant Hamiltonian, $H_{\idTDL}$, is given in~\eqref{eq:D_I}. In  this case (and this case only), the relevant translation operator is given by $T_I = \uP$, where $\uP$ shifts each site of the lattice to the right by one site. As explained earlier, the conformal dimensions of the low-lying states are obtained by determining the energies and the momenta of the latter. Results obtained using the DMRG technique~\cite{DMRG_TeNPy} are presented in Table~\ref{tab:Idefect} below. They are certainly not new, see Refs.~\cite{vonGehlen_1987, Mong:2014ova, Hauru2016}, but we include them here  for completeness, and  to benchmark our numerical technique. 

\begin{table}[h]
\renewcommand{\arraystretch}{1.2}

    \centering
\begin{tabular}{ |P{3cm}| P{2.5cm}|P{2.5cm}|P{2.5cm}|P{2.5cm} | }

\hline
 \multicolumn{5}{|c|}{States and Conformal dimensions $\idTDL$-defect - Up to 60 Potts site - Scaling done with L} \\
 \hline
 State & $h + \bar{h}$ & Theoretical value   & $h - \bar{h}$& Theoretical value  \\ [2ex]
 \hline

 $\ket{\frac{1}{15},\frac{1}{15}}        $    & 0.1335  & 0.1333       &  0            & 0 \\
 $\ket{\frac{1}{15},\frac{1}{15}}        $    & 0.1335  & 0.1333       &  0            & 0 \\
$\ket{\frac{2}{5},\frac{2}{5}}          $    & 0.807   & 0.8  &   0& 0       \\
 $L_{-1} \ket{\frac{1}{15},\frac{1}{15}} $    & 1.135       & 1.133    &  1   & 1       \\
 $L_{-1} \ket{\frac{1}{15},\frac{1}{15}} $    & 1.135       & 1.133    &  1   & 1       \\
 $\bar{L}_{-1} \ket{\frac{1}{15},\frac{1}{15}} $    & 1.135       & 1.133    &  -1   & -1       \\
 $\bar{L}_{-1} \ket{\frac{1}{15},\frac{1}{15}} $    & 1.135       & 1.133    &  -1   & -1       \\
 
 $\ket{\frac{2}{3},\frac{2}{3}}$      & 1.338        & 1.333    &  0   & 0        \\
 $\ket{\frac{2}{3},\frac{2}{3}}$      & 1.338        & 1.333    &  0   & 0        \\

$\ket{\frac{7}{5},\frac{2}{5}}          $    & 1.798  & 1.8  &   1& 1      \\

$\ket{\frac{2}{5},\frac{7}{5}}          $    & 1.798   & 1.8  &   -1& -1       \\

 \hline
\end{tabular}
    \caption{DMRG results for the values of $h+\bar{h}$ and $h-\bar{h}$ found using the Hamiltonian and translation operator for Potts model with the identity defect.}
    \label{tab:Idefect}

\end{table}

\subsection{The symmetry defects $\eta$, $\overline{\eta}$, and $C$}\label{sec: D_eta_C}

Next, we discuss the defect Hamiltonians corresponding to the $\eta$, $\overline{\eta}$, and $C$ defects. The latter are based upon the $\Z_3$ and $\Z_2^C$ symmetry actions on the model. These  defects are manifestly topological on the lattice.  

The Hamiltonians for the $\eta$ and $\overline{\eta}$ defects are $\Z_3$ generalizations of the antiperiodic Ising chain. As such, the corresponding defect Hamiltonians can be obtained by starting with the Hamiltonian of the periodic Potts chain and multiplying the strength of the ferromagnetic interaction between two neighboring sites, $i_0$ and $i_0 + 1$, by  $\omega$ and $\omega^{-1}$ respectively. The transverse field terms are kept unchanged. Explicitly, the~$\eta$ defect Hamiltonian reads
\begin{equation}\label{eq:H_eta}
H_\eta = H_{\idTDL} + \frac{1}{3\sqrt{3}}\left(\sigma_{i_0}^\dagger \sigma_{i_0+1} +{\rm h.c.} \right) -\frac{1}{3\sqrt{3}}\left(\omega\sigma_{i_0}^\dagger \sigma_{i_0+1} +{\rm h.c.} \right),
\end{equation}
where $H_{\idTDL}$ is defined in (\ref{eq:D_I}).
The defect Hamiltonian for $H_{\overline{\eta}}$ can be obtained by interchanging $\omega$ and $\omega^{-1}$ in the last term of~\eqref{eq:H_eta}.

In  presence of the defect, the Hamiltonian is clearly not translation-invariant. The translation operator is not simply the shift operator $u$ anymore. The effective translation operator can be constructed by composing $u$ with the local unitary which shifts the topological defect back by one site. In particular, the effective translation operators in presence of a $\eta$ defect or a $\overline{\eta}$ defect are given by
\begin{equation}\label{eq:T_eta} 
T_\eta = \tau_{i_0+1}^\dagger \uP, \ T_{\overline{\eta}} = \tau_{i_0 + 1} \uP.
\end{equation}

Numerical results obtained using DMRG for the~$\eta$ defect is shown in Table \ref{tab:etadefect}. We note that this case has been also analyzed in Ref.~\cite{Hauru2016}.

The Hamiltonian for the charge conjugation defect is constructed similarly. Recalling the action of the charge conjugation symmetry on the lattice operators in equation \eqref{eq:C_act}, the corresponding defect Hamiltonian can be shown to be:
\begin{equation}\label{eq:H_C}
H_C = H_{\idTDL} + \frac{1}{3\sqrt{3}}\left(\sigma_{i_0}^\dagger \sigma_{i_0+1} +{\rm h.c.} \right) - \frac{1}{3\sqrt{3}}\left(\sigma_{i_0}^\dagger \sigma_{i_0+1}^\dagger+{\rm h.c.} \right).
\end{equation}
Conjugation by $c_{i_0}$ (as defined in equation \eqref{eqn:chargeconj}) shifts the defect location by one unit while keeping other terms in $H_C$ unchanged, which is again consistent with the topological nature of the lattice $C$ defect. 
The modified translation operator in presence of a $C$ defect is given by 
\begin{equation}\label{eq:T_C} 
T_C = c_{i_0+1}\uP.
\end{equation}
Note that the charge conjugation defect does not conserve the characteristic $\Z_3$ charge defined in~\eqref{eq:Q_def}, of the Potts model.\footnote{Additionally, one can also realize these defect Hamiltonians (\eqref{eq:H_eta}, \eqref{eq:H_C}) in the $D_4$ RSOS model, using the mapping between the spin chain model and the RSOS model. We present the RSOS Hamiltonians corresponding to these defects in Appendix \ref{subsec:C_and_eta_RSOS_Direct}.} The DMRG results for the charge conjugation defect is shown in Table~\ref{tab:Cdefect}.
\begin{table}[h]
\renewcommand{\arraystretch}{1.2}

    \centering
\begin{tabular}{ |P{3cm}| P{2.5cm}|P{2.5cm}|P{2.5cm}|P{2.5cm} | }

\hline
 \multicolumn{5}{|c|}{States and Conformal dimensions $\eta$-defect - Up to 60 Potts site - Scaling done with L} \\
 \hline
 State & $h + \bar{h}$ & Theoretical value   & $h - \bar{h}$& Theoretical value  \\ [2ex]
 \hline

 $\ket{\frac{1}{15},\frac{1}{15}}        $    & 0.1334  & 0.13333       &  0            & 0 \\
  $\ket{\frac{2}{5},\frac{1}{15}}       $    & 0.4663   & 0.466667   &   0.333333 & 0.333333     \\
  $\ket{\frac{1}{15},\frac{2}{5}}       $    & 0.4665   & 0.466667   &   -0.333333& -0.333333       \\
 $\ket{\frac{2}{3},0}                   $    & 0.6664        & 0.666667    &  0.666667   & 0.666667       \\
 $\ket{0,\frac{2}{3}}                   $    & 0.6664        & 0.666667    &  -0.666667   & -0.666667       \\
  $L_{-1}\ket{\frac{1}{15},\frac{1}{15}}        $    & 1.1341  & 1.13333       &  1            & 1 \\
 $\bar{L}_{-1}\ket{\frac{1}{15},\frac{1}{15}}   $    & 1.1342  & 1.13333       &  -1            & -1 \\
$\ket{\frac{2}{3},\frac{2}{3}}                   $    & 1.3297       & 1.33333    &  0   & 0       \\

\hline
\end{tabular}
    \caption{\label{tab:etadefect}Comparison of the $\eta$ defect spectrum obtained using DMRG with theoretical values.   }
\end{table}

\begin{table}[h]
\renewcommand{\arraystretch}{1.2}

    \centering
\begin{tabular}{ |P{3cm}| P{2.5cm}|P{2.5cm}|P{2.5cm}|P{2.5cm} | }
\hline
 \multicolumn{5}{|c|}{States and Conformal dimensions $C$-defect - Up to 60 Potts site - Scaling done with L} \\
 \hline
 State & $h + \bar{h}$ & Theoretical value   & $h - \bar{h}$& Theoretical value  \\ [2ex]
 \hline

 $\ket{\frac{1}{40},\frac{1}{40}}        $    & 0.05023349  & 0.05   &  0    & 0                \\
  $\ket{\frac{1}{8},\frac{1}{8}}       $    & 0.249984029   & 0.25   &   0 & 0     \\
 $\ket{\frac{21}{40},\frac{1}{40}}          $    & 0.5501037    & 0.55   &    0.5 & 0.5     \\
 $\ket{\frac{1}{40},\frac{21}{40}}          $    & 0.5501037    & 0.55   &    0.5 & -0.5     \\
 
 \hline
\end{tabular}
    \caption{\label{tab:Cdefect}Comparison of the $C$ defect spectrum obtained using DMRG with theoretical values.}
\end{table}

As noted in Sec.\ref{sec:CFT}, the $\bar{\eta}$ defect line may be obtained by fusion of two $\eta$ lines. This can be also demonstrated for the corresponding defect Hamiltonians in the direct channel. Explicitly, the Hamiltonian with the $\eta$ defect between bonds $i_0$ and $i_0 + 1$ and another $\eta$ defect between bonds $i_0 + 1$ and $i_0 + 2$
\begin{equation}
\begin{split}
    H_{\eta, \eta}^{[i_0, i_0 + 1], [i_0 + 1, i_0 + 2]} &= -\frac{1}{3\sqrt{3}}\Sum\limits_{i=1, i \neq i_0, i_0 + 1}^{L} \left(1+\sigma_i^\dagger \sigma_{i+1} +\sigma_{i+1}^\dagger \sigma_i \right)-\frac{1}{3\sqrt{3}}\Sum\limits_{i=1}^L\left(1+ \tau_i +\tau_i^\dagger \right) \\
     - & \frac{1}{3\sqrt{3}} \left( 1 + \omega\sigma_{i_0}^{\dagger}\sigma_{i_0  + 1} + \omega^2 \sigma_{i_0 }\sigma_{i_0 + 1}^{\dagger}  \right)    -  \frac{1}{3\sqrt{3}} \left( 1 + \omega \sigma_{i_0 + 1}^{\dagger}\sigma_{i_0 + 2}  + \omega^2 \sigma_{i_0 + 1}\sigma_{i_0 + 2}^{\dagger} \right),
\end{split}
\end{equation}
coincides with the $H_{\bar{\eta}}$ Hamiltonian, when conjugated with the $\eta$-defect shift operator $\tau_{i_0 + 1}^{\dagger}$, i.e.  $\tau_{i_0 + 1}^{\dagger} H_{\eta, \eta}^{[i_0, i_0 + 1], [i_0 + 1, i_0 + 2]}\tau_{i_0 + 1}=H_{\bar{\eta}}^{[i_0,i_0+1]}$.  Fusion of two $\bar{\eta}$ lines can be similarly analyzed. We further note that corresponding to {\it invertible} defects with local shift operators, a prescription for generating unitary operator commuting with the no-defect Hamiltonian has been discussed in \cite{Seifnashri:2023dpa}. This can be demonstrated, for example, for the $\eta$ defect. Consider the Hamiltonian $H_{\eta, \bar{\eta}}^{[i_0,i_0+1],[i_0+1, i_0 + 2]}$, using the $\tau$ operator, which shifts the $\bar{\eta}$ defect to the left by one site, we can show 
\begin{equation} \label{eq:moving_taubar}
\tau_{i_0 + 1} H_{\eta, \bar{\eta}}^{[i_0,i_0+1],[i_0+1, i_0 + 2]} \tau_{i_0 + 1}^{\dagger} = H,  
\end{equation}
which resembles $\eta \times \bar{\eta} = 1$. Now, let us shift the $\eta$ defect, but instead of shifting it to the right to fuse with $\bar{\eta}$, we keep shifting it to the left and bring it to the bond between sites $i_0 + 2, i_0 + 3$  using the periodicity of the chain, i.e.  
\begin{equation}\label{eq:forming_unitary1}
   \tau_{i_0 + 3}^{\dagger} \ldots \tau_{L}^{\dagger} \tau_1^{\dagger}\ldots \tau_{i_0}^{\dagger} H_{\eta, \bar{\eta}}^{[i_0,i_0+1],[i_0+1, i_0 + 2]}\tau_{i_0}  \ldots  \ldots \tau_1  \tau_{L}\ldots \tau_{i_0 + 3}  = H_{\eta, \bar{\eta}}^{[i_0 + 2, i_0 + 3], [i_0 + 1, i_0 + 2]}\, .
\end{equation}
Now, if we again shift the $\eta$ defect one site to the left, as $\eta \times \bar{\eta} = 1$, we have 
\begin{equation}\label{eq:forming_unitary2}
 \tau_{i_0+2}^{\dagger}   H_{\eta, \bar{\eta}}^{[i_0 + 2, i_0 + 3], [i_0 + 1, i_0 + 2]}\tau_{i_0+2} = H \, .
\end{equation}
Combining  \eqref{eq:moving_taubar}, \eqref{eq:forming_unitary1}, and \eqref{eq:forming_unitary2} we get a unitary symmetry operator, i.e. an operator $U$ such that
\begin{equation}
    U H U^{\dagger} = H, \text{ where } U = \prod_{i = 1}^{L} \tau_i^{\dagger} \, .
\end{equation}
This is precisely the $\eta$ line in crossed channel, discussed in equation \eqref{eq:eta_line_crossed}. This process can be repeated for the charge conjugation shift operator, to get the charge conjugation line discussed in equation \eqref{eqn:chargeconj}. Note that this procedure can only be carried out  for invertible defects, as the defects that are introduced on neighboring links are precisely  inverse of each other. \footnote{For TFI model, if one followed the same procedure for the shift operator for Spin-Flip Defect Hamiltonian, one would produce the $\eta$ line - the invertible $\mathbb{Z}_2$ symmetry line.}

\subsection{The duality defects $N$ and $N'$}
\label{sec: D_N_N'}

Next, we consider the Kramers-Wannier duality defects~\cite{Frohlich2004, Frohlich2006}. These defects have been shown to exhibit more exotic properties than $\idTDL$, $\eta$, $\overline{\eta}$ or $C$ and have been analyzed for the Ising, Potts and the XXZ chains~\cite{Grimm1992, Grimm2001, Aasen2016,Hauru2016, Aasen2020}. The Kramers-Wannier duality defects can be constructed by two different methods. The first method utilizes integrability techniques, and relies  on a   general construction scheme for RSOS models~\cite{Belletete2018, Belletete2020}. The second method builds upon the fact that, Kramers-Wannier duality defects can be obtained by gauging the non-anomalous $\Z_3$ symmetry on a half-chain while imposing  Dirichlet boundary condition for the $\Z_3$ gauge field. The concepts underlying this   idea have been explored in \cite{Frohlich:2009gb,Carqueville:2012dk,Brunner:2013xna,Bhardwaj:2017xup,Thorngren:2019iar,Gaiotto:2020iye,Huang:2021zvu,Choi:2021kmx,Li:2023mmw}: here we apply them  directly to the 1d quantum Potts chain. In this section, we mostly discuss  the duality defect Hamiltonians obtained from integrability techniques;  the construction via gauging is presented  in \autoref{sec:KWgauging}. The results from the two constructions agree, in the sense that there exists a local unitary relating defect Hamiltonians obtained from these two approaches.

In the TL formulation of the 3-state Potts model, the $N$ and $N'$ defects are constructed by shifting the spectral parameter at the defect site by $\text{i}\infty$ and $-\text{i}\infty$ respectively. In contrast to the defect Hamiltonians for $\eta, \overline{\eta}$, and $C$, the $N$ and $N'$ defect Hamiltonians turn out to have simple expressions in terms of the TL generators. We find 
\begin{align}\label{eqn:HN}
H_{N} &= H_{\idTDL} + \frac{\gamma}{\pi\sin\gamma}\left(q \, e_{2i_0-1}e_{2i_0} + q^{-1}e_{2i_0}e_{2i_0-1}\right) \, ,
\end{align}
where  again $H_{\idTDL}$ is  the Hamiltonian for the periodic chain with no defect insertions. The $H_{N'}$ Hamiltonian is given by interchanging $q$ and $q^{-1}$ in the last term of equation (\ref{eqn:HN}). In terms of the three-state spins, the relevant Hamiltonians are given by:
\begin{equation}
\begin{split}
\label{eq:H_N}
H_N =& -\frac{1}{3\sqrt{3}}\sum\limits_{i=1,i\neq i_0}^L  (1+\sigma_i^\dagger\sigma_{i+1}+\sigma_i\sigma_{i+1}^\dagger)-\frac{1}{3\sqrt{3}}\sum\limits_{i=1,i\neq i_0}^L  (1+\tau_i +\tau_i^\dagger) \\
&+ \frac{1}{3\sqrt{3}}\left(\text{e}^{-\text{i}\pi/3} \sigma_{i_0}\tau_{i_0}\sigma_{i_0 + 1}^\dagger + \text{e}^{\text{i}\pi/3}
\tau_{i_0}^\dagger\sigma_{i_0}^\dagger\sigma_{i_0+1}\right) \, ,
\end{split}
\end{equation}
and 
\begin{equation}
\label{eq:H_N'}
\begin{split}
H_{N'} =& -\frac{1}{3\sqrt{3}}\sum\limits_{i=1,i\neq i_0}^L  (1+\sigma_i^\dagger\sigma_{i+1}+\sigma_i\sigma_{i+1}^\dagger)-\frac{1}{3\sqrt{3}}\sum\limits_{i=1,i\neq i_0}^L  (1+\tau_i +\tau_i^\dagger) \\ &+ \frac{1}{3\sqrt{3}}\left(\text{e}^{\text{i}\pi/3} \sigma^\dagger_{i_0}\tau_{i_0}\sigma_{i_0 + 1} + \text{e}^{-\text{i}\pi/3} 
\tau_{i_0}^\dagger\sigma_{i_0} \sigma_{i_0+1}^{\dagger}\right).
\end{split}
\end{equation}

Since the $N'$ defect is the fusion product of the $C$ defect and the $N$ defect, we can check the consistency of $H_{N'}$ with such a fusion. We consider the following Hamiltonian with a $C$ defect inserted between sites $i_0-1$ and $i_0$ and a $N$ defect inserted between sites $i_0$ and $i_0+1$:
\begin{equation}
\begin{split}
H_{C,N}^{[i_0 - 1, i_0],[i_0 , i_0 + 1]}&= -\frac{1}{3\sqrt{3}}\Sum\limits_{i=1,i\neq i_0-1,i_0}^L \left(1+\sigma^\dagger_i \sigma_{i+1}+\sigma_i\sigma^\dagger_{i+1}\right)-\frac{1}{3\sqrt{3}}\Sum\limits_{i=1,i\neq i_0}^L \left( 1+\tau_i +\tau_i^\dagger\right)\\
& -\frac{1}{3\sqrt{3}}\left(1+\sigma_{i_0-1}^\dagger\sigma^\dagger_{i_0} +\sigma_{i_0-1}\sigma_{i_0} \right)+\frac{1}{3\sqrt{3}}\left(\text{e}^{-\text{i}\pi/3}\sigma_{i_0} \tau_{i_0}\sigma_{i_0+1}^\dagger +\text{e}^{\text{i}\pi/3}\tau_{i_0}^\dagger\sigma_{i_0}^\dagger\sigma_{i_0+1} \right)
\end{split}
\end{equation}
Recall that conjugation by the local unitary operator $c_{i_0}$ as defined in (\ref{eqn:chargeconj}) moves the $C$ defect to the right by 1 site, this should produce a $N':=CN$ defect between sites $i_0$ and $i_0+1$. We find
\begin{equation}\label{eqn:CNfusion}
\begin{split}
H_{N'=CN} =& -\frac{1}{3\sqrt{3}}\Sum\limits_{i=1,i\neq i_0}^L \left(1+\sigma^\dagger_i \sigma_{i+1}+\sigma_i\sigma^\dagger_{i+1}\right)-\frac{1}{3\sqrt{3}}\Sum\limits_{i=1,i\neq i_0}^L \left( 1+\tau_i +\tau_i^\dagger\right)\\
& +\frac{1}{3\sqrt{3}}\left(\text{e}^{-\text{i}\pi/3}\sigma_{i_0}^\dagger \tau_{i_0}^\dagger\sigma_{i_0+1}^\dagger +\text{e}^{\text{i}\pi/3}\tau_{i_0}\sigma_{i_0}\sigma_{i_0+1} \right)
\end{split}
\end{equation}
The defect Hamiltonian in (\ref{eqn:CNfusion}) is related to the defect Hamiltonian in (\ref{eq:H_N'}) by a local unitary transformation. Concretely, the following unitary
\begin{equation}
V_{i_0} = \frac{1}{3}\begin{pmatrix}
2+\omega^{-1} & \, 2\omega^{-1}+\omega & \,  2\omega^{-1}+\omega\\
2\omega^{-1}+\omega &  \, 2+\omega^{-1} & \, 2\omega^{-1}+\omega\\
2\omega^{-1}+\omega & \,  2\omega^{-1}+\omega & \,  2+\omega^{-1}
\end{pmatrix}, ~ ~ \omega=\text{e}^{2\pi\text{i}/3}, 
\end{equation}
satisfies
\begin{equation}
V_{i_0}^\dagger \left(\sigma_{i_0}^\dagger \tau_{i_0}^\dagger \right) V_{i_0} = \tau_{i_0}^\dagger \sigma_{i_0}, ~ V_{i_0}^\dagger \sigma_{i_0} V_{i_0} = \sigma_{i_0} \, .
\end{equation}
Therefore, conjugating (\ref{eqn:CNfusion}) by $V_{i_0}$ yields (\ref{eq:H_N'}). Similarly, it can be shown that if we take the Hamiltonian with a $\eta$ defect and a $N$ defect at two neighboring bonds, and then conjugate with the $\eta$ defect shift operator, we recover the $H_N$ defect Hamiltonian. 

We can also bring the above defect Hamiltonians into a simpler form, by performing a local unitary transformation at site $i_0$. For example, the following unitary
\begin{equation}
U_{i_0} = \frac{1}{\sqrt{3}}\begin{pmatrix}
\text{e}^{-\frac{2\pi \text{i}}{3}} & 1 & 1\\
1 & \text{e}^{-\frac{2\pi \text{i}}{3}} & 1 \\
1 & 1 & \text{e}^{-\frac{2\pi \text{i}}{3}}
\end{pmatrix} \, , 
\end{equation}
satisfies
\begin{equation}
U_{i_0}^\dagger \sigma_{i_0} U_{i_0} = \sigma_{i_0}, ~ U_{i_0}^\dagger \left(\text{e}^{-\text{i}\pi/3}\sigma_{i_0}\tau_{i_0} \right) U_{i_0} = -\tau_{i_0} \, . 
\end{equation}
Therefore
\begin{equation}\label{eqn:HNgauge}
\begin{split}
U_{i_0}^\dagger H_N U_{i_0}  =& -\frac{1}{3\sqrt{3}}\sum\limits_{i=1,i\neq i_0}^L  (1+\sigma_i^\dagger\sigma_{i+1}+\sigma_i\sigma_{i+1}^\dagger)-\frac{1}{3\sqrt{3}}\sum\limits_{i=1,i\neq i_0}^L  (1+\tau_i +\tau_i^\dagger) \\
& - \frac{1}{3\sqrt{3}} \left(\tau_{i_0} \sigma_{i_0+1}^\dagger +\tau_{i_0}^\dagger \sigma_{i_0+1}\right) \, .
\end{split}
\end{equation}
The defect Hamiltonian in equation (\ref{eqn:HNgauge}) and a similar version for $H_{N'}$ arise from the $\Z_3$ gauging procedure, which we discuss further in \autoref{sec:KWgauging}. Note that the $N$ and $N'$ defects are characterized by the absence of the transverse field term in addition to a modified nearest neighbor interaction between sites $i_0$ and $i_0+1$. They form the Potts analog of the duality defect in the Ising model, which was analyzed extensively in Refs.~\cite{Grimm2001, Hauru2016,Aasen2016,Aasen2020,Roy2021a,Tan:2022vaz,Li:2023mmw}. 

The lack of the transverse field, together with the nontrivial modification of the nearest-neighbor hopping, leads to several characteristics of the $N, N'$ defects, which are absent in the previously mentioned symmetry defects. First, the ground states for both $N$ and $N'$ defects are two-fold degenerate. The degenerate states differ in their $Z_3$ charge values. Second, the ground states for the $N$ and $N'$ defects have non-zero momenta. In particular, since $N$ and $N'$ are related by charge conjugation, the momenta of states for the two defects have the same norm and carry opposite signs. 

The local unitary operators which shifts the location of the $N$ and $N'$ defects are most conveniently expressed in the TL formalism, in terms of the braiding operators $g_j$ as defined in equation \eqref{eq:g_def}. For example, conjugation by $g^{-1}_{2i_0} g^{-1}_{2i_0+1}$ shifts the $N'$ defect location from between sites $i_0+1$ and $i_0+2$ to between the sites $i_0$ and $i_0+1$ of the corresponding spin chain. The modified lattice translation operators in presence of an $N$ or $N'$ defect are given by
\begin{align}
\label{eq:T_N_N'} 
T_N = g_{2i_0}g_{2i_0 + 1}\uP, \ T_{N'} = g_{2i_0}^{-1}g_{2i_0 + 1}^{-1}\uP,
\end{align}
where as usual, $u$ is the operator which shifts all Potts spin sites by one unit to the right. The DMRG results obtained for the defect Hamiltonian in equation \eqref{eq:H_N} are listed in Table \ref{tab:Ndefect}. Similar results were obtained for the defect Hamiltonian \eqref{eq:H_N'} and are not shown for brevity.
\begin{table}[h]
\renewcommand{\arraystretch}{1.2}

    \centering
    \begin{tabular}{ |P{3cm}| P{2.5cm}|P{2.5cm}|P{2.5cm}|P{2.5cm} | }
    \hline
     \multicolumn{5}{|c|}{States and Conformal dimensions $N$-defect - Up to 60 Potts site - Scaling done with L} \\
     \hline
     State & $h + \bar{h}$ & Theoretical value   & $h - \bar{h}$& Theoretical value  \\ [2ex]
     \hline
    
    $\ket{\frac{1}{15},\frac{1}{40}}$        & 0.0916   & 0.0916      & 0416667 & 0416667          \\
    $\ket{\frac{1}{15},\frac{1}{40}}$        & 0.0916   & 0.0916      & 0416667 & 0416667          \\
    $\ket{0,\frac{1}{8}}$                    & 0.125   & 0.125   & -0.125 & -0.125          \\
    $\ket{\frac{2}{5},\frac{1}{40}}$         & 0.426 & 0.425   & 0.375  & 0.375          \\
    
    $\ket{\frac{1}{15},\frac{21}{40}}$       & 0.593  & 0.592   & -0.4583333   & -0.4583333          \\
    $\ket{\frac{1}{15},\frac{21}{40}}$       & 0.593  & 0.592   & -0.4583333   & -0.4583333          \\
    $\ket{\frac{2}{3},\frac{1}{8}}$          & 0.792   & 0.791667 &  0.5416667 & 0.5416667          \\
    $\ket{\frac{2}{3},\frac{1}{8}}$          & 0.792   & 0.791667 &  0.5416667 & 0.5416667          \\

    $\ket{\frac{2}{5},\frac{21}{40}}$         & 0.922 & 0.925   & -0.125  & -0.125        \\
     \hline
    \end{tabular}    
    \caption{\label{tab:Ndefect}Comparison of the $N$ defect spectrum obtained using DMRG with theoretical values.}
\end{table}

\subsection{The $W$ Defect}
\label{sec: D_W}
The last nontrivial defect Hamiltonian we discuss is the one for the W defect:
\begin{equation}
\label{eq:H_W}
H_{W} = H_{\idTDL} + \frac{\gamma}{\pi\sin\gamma\cos\gamma}\left(e_je_{j+1} + e_{j+1}e_j\right).
\end{equation}
(we keep the expression in terms of the TL generators, since plugging in the three-state spin representation~[Eq.~\eqref{eq:TL_Potts}] does not lead to a simplified expression). In contrast to the other defects, the W-defect is realized by $H_W$ only in the scaling limit when $L \gg 1$. Like for the $N$ and $N'$ defects, the ground state is two-fold degenerate~[see Eq.~\eqref{eq:W_Part}], with the two degenerate states having different values of the $\mathbb{Z}_3$ charge. 

The corresponding translation operator for the $W$-defect Hamiltonian is given by
\begin{equation}
T_W = -\left(1-\frac{e_{j+1}}{\cos\gamma}\right)\left(1-\frac{e_{j+2}}{\cos\gamma}\right)\, u.
\end{equation}
It can be analytically seen that the operator 
\begin{equation}
U[j+1]: =  -\left(1-\frac{e_{j+1}}{\cos\gamma}\right)\left(1-\frac{e_{j+2}}{\cos\gamma}\right) \, , 
\end{equation}
moves the defect by two sites for RSOS Hamiltonian, hence by one site for Potts spin chain.
Results for this  $W$ defect are given in the table below. We clearly identify the low-lying energy-states from the defect Hilbert space $\cH_W$ in (\ref{eq:W_Part}).

\begin{table}[h!]
\renewcommand{\arraystretch}{1.2}

    \centering
\begin{tabular}{ |p{2.8cm}| p{2.8cm}|m{2.8cm}|m{2.8cm}| m{2.8cm} | }
\hline
 \multicolumn{5}{|c|}{ States and Conformal dimensions in $W$ - Up to 200 (13 for $h - \bar{h}$) Potts site - Scaling done with L} \\
 \hline
 State  & $h + \bar{h}$ & Theoretical value & $h - \bar{h}$& Theoretical value \\ [2ex]
 \hline
 $\ket{\frac{1}{15},\frac{1}{15}} $       &  0.1326   &  0.133333    &    0   & 0    \\
 $\ket{\frac{1}{15},\frac{1}{15}} $       &  0.1329    &  0.133333   &    0    & 0    \\

 $\ket{\frac{4}{10},0} $       &   0.3777 & 0.4    &    0.3657    &  0.4    \\
 $\ket{0,\frac{4}{10}} $       &   0.3771    & 0.4    &   -0.3657     &  -0.4    \\
 $\ket{\frac{2}{3},\frac{1}{15}} $       &  0.7353   & 0.7333    &    0.5883      &  0.6    \\
 $\ket{\frac{1}{15},\frac{2}{3}} $       &  0.7375     & 0.7333     &    -0.5883      &  -0.6    \\
 $\ket{\frac{2}{3},\frac{1}{15}} $       &   0.7356     & 0.7333   &    0.5883     &  0.6    \\
 $\ket{\frac{1}{15},\frac{2}{3}} $       &  0.7359    & 0.7333    &    -0.5883     &  -0.6    \\

 $\ket{\frac{4}{10},\frac{4}{10}} $       &  0.8293  &  0.8    &     0 & 0    \\

 \hline
\end{tabular}
\label{(2,1)defRSOStable}
\caption{Comparision of the $W$ defect spectrum obtained using DMRG with theoretical values.}
\end{table}

\section{Topological Defects in the direct channel: Entanglement Entropy }

In the previous section, the energy and momenta of eigenstates of the defect Hamiltonians were computed using DMRG and the numerical results were compared with the CFT predictions after  finite-size scaling analysis. In this section, the Affleck-Ludwig g-functions~\cite{Affleck1991} of the topological defects are investigated. While initially proposed in the context of thermodynamic entropies, at conformal fixed points, the  g-function can be extracted by a finite-size scaling analysis of the entanglement entropy~\cite{Calabrese2004, Calabrese2009}. The latter is computed for a block located symmetrically around the defect when the system is in the ground state. For defect CFTs, this is often referred to in the literature as the symmetric entropy~\cite{Gutperle2015, Brehm2016}, distinguishing it from the so-called interface entropy. In the latter case, one of the boundaries of the subsystem coincides with the location of the defect~\cite{Sakai2008, Brehm2015, Karch:2023evr}. In this work, we present the results only for the symmetric entanglement entropy, which, unlike the interface entropy, is directly related to the g-function associated with the defect. That the symmetric entropy is related to the defect g-function can be easily seen in a folded picture~\cite{Oshikawa1997}. Folding the one-dimensional system {\it at} the defect, after suitable identification of the left and right moving modes, leads to a boundary CFT problem where the bulk central charge is doubled~(for a detailed exposition of this for the Ising case in terms of the orbifolded compact boson, see Ref.~\cite{Oshikawa1997}). Changing defects in the unfolded model now changes the boundary condition of the folded one. Importantly, the symmetric entropy in the unfolded model corresponds to entanglement entropy of a block whose one boundary coincides with the boundary in the folded model. It is this boundary contribution to the entanglement entropy~\cite{Calabrese2004, Cardy2016, Affleck1991} that is measured to obtain the defect g-function. We note that while the g-function could equally well be measured from the thermodynamic entropy, numerical computation of the latter is vastly more challenging than the computation of the ground state entanglement entropy. 

\begin{figure}
    \centering
    \includegraphics[scale = 0.55]{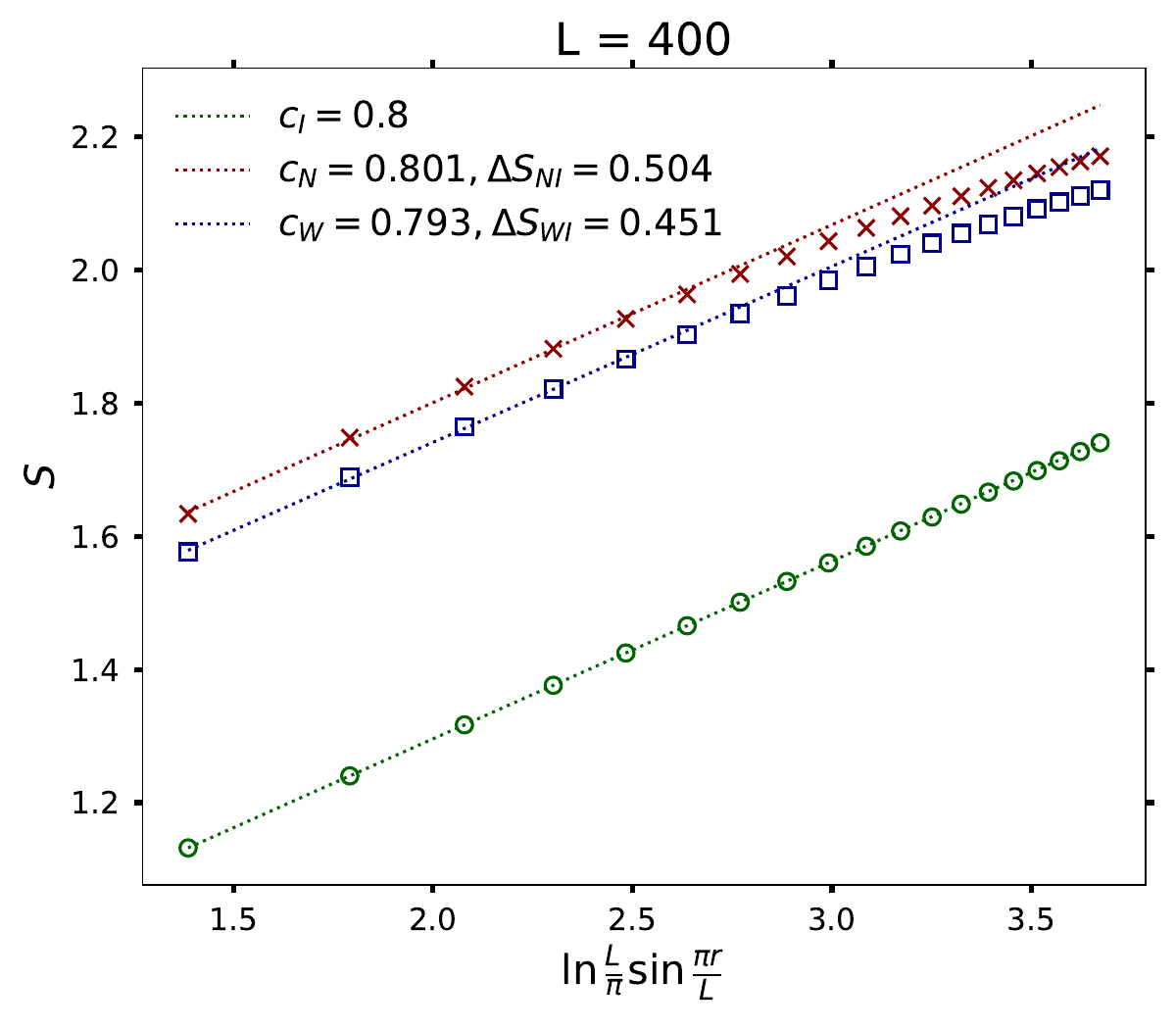}
    \caption{\label{fig:EE_Potts}Results for the Symmetric Entanglement Entropy (S) for a periodic
chain of size 400. For the $N$ and $W$ defects, $\Delta S_{NI}$ and $\Delta S_{WI}$ are numerically calculated in the above graph. Their values can be compared with theoretical ln$(g_{\mathcal{L}})$ values - 0.549306 and 0.481212 for the respective defects.}
    \label{fig:EE_scaling}
\end{figure}

For a periodic system with the identity/no defect, for a block of size~$r$ and system-size~$L$, the entanglement entropy is given by~\cite{Holzhey1994, Calabrese2004, Calabrese2009}
\begin{equation}
\label{eq:S_L}
S (r) = \frac{c}{3}\ln\left[\frac{L}{\pi}\sin\frac{\pi r}{L}\right] + S_0,
\end{equation}
where the subleading term~$S_0$ contains both the non-universal lattice-dependent contributions and entanglement cut boundary entropies. The latter equals the logarithm of the g-function that arise from the boundary conditions at the entanglement cuts~\cite{Cardy2016,Roy2021b,Ohmori:2014eia}. In the presence of a defect, the subleading term gets yet another universal contribution arising from the g-function associated with the defect. Separating the non-universal contribution from the universal entropy contribution in a lattice computation is not possible in general. However, the change in boundary entropy (in the folded picture) can be obtained reliably from the difference of the subleading terms of the entanglement entropies for two different defects~\footnote{Note that subtracting the entanglement entropies for two different defects cancels not only the non-universal lattice contributions but also the boundary entropies associated with the entanglement cuts.}. In particular, choosing the $\idTDL$ defect as the reference defect, one can obtain defect entropies by comparing subleading terms. Notice that the leading logarithmic term of the symmetric entropy is insensitive to the nature of the defect. Intuitively, this can be understood as a consequence of the fact that the entanglement entropy probes the correlations around the boundaries of the subsystem, which, for the symmetric entropy, are far from the defect. However, defects can lead to violation of this well-known logarithmic dependence.\footnote{In the folded model, symmetric entropy corresponds to the ground state entanglement entropy in the setup of CFT on a strip, where the subsystem touches one boundary. Eq. (\ref{eq:S_L}) holds if $r\ll L$. As $r$ increases, the effect of the other boundary condition weighs in leading to the log-dependence violation. At large $r$ where $L-r\ll L$, because of the bi-partition, symmetric entropy approaches the no-defect entropy for subsystem size $L-r$. Analytical expressions for generic cases are not known, except for cases where the two boundary conditions of the strip are the same \cite{Calabrese2004,Calabrese2009,Taddia_2013,Taddia:2016dbm,Estienne_2022} (recently, progress has been made in \cite{Estienne_2023} regarding the Renyi entropies for generic mixed boundary conditions), or for cases of free theories \cite{Fagotti:2010cc,Klich2017,Jafarizadeh_2022,Capizzi:2023vsz, Estienne:2023ekf}.}  This has been shown to occur for the duality defects in the Ising chain when the subsystem size is comparable to the total system size~\cite{RoySaleur2022, Rogerson_2022}. Similar effects occur for the Potts chain in the case of the~$N, N'$ and the~$W$ defects~(see Fig.~\ref{fig:EE_Potts}). However, as long as $r\ll L$, Eq.~\eqref{eq:S_L} can be taken to be true for all defects, thereby allowing determination of the corresponding g-functions.

\begin{figure}
    \centering
    \includegraphics[scale = 0.55]{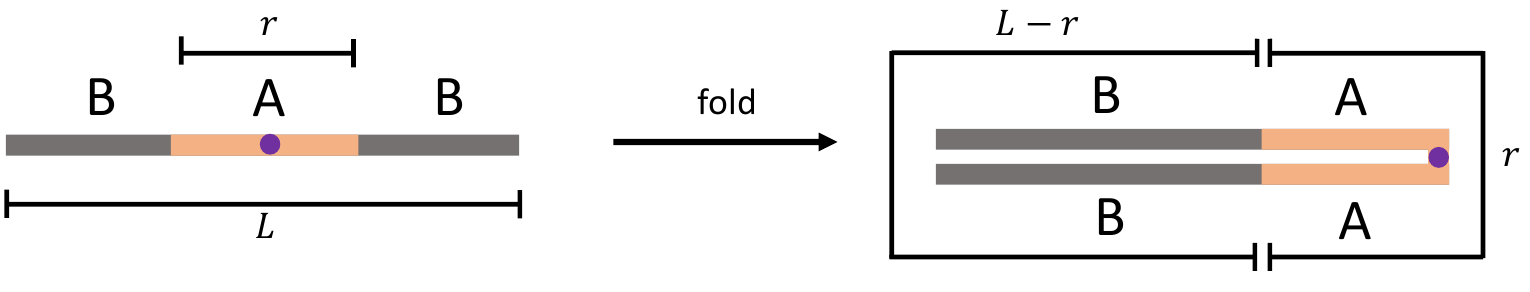}
    \caption{\label{fig:folding}Illustration of the folding procedure to map the CFT with a defect to that with a boundary. The symmetric entanglement entropy in the unfolded picture becomes the entanglement entropy of a block in the folded picture. Importantly, one of the boundaries of the block coincides with the defect leading to usual boundary contribution to the entanglement entropy. In this work, the two edges of the subsystem B are glued together to form a periodic chain with a defect. The subleading term of the entanglement entropy is obtained by fitting to Eq.~\eqref{eq:S_L}. The differences of the g-functions for two different defects are obtained by taking the corresponding differences of the subleading terms. Note that use of Eq.~\eqref{eq:S_L} is appropriate for $N, N'$ and $W$ defects only in the limit~$r\ll L$~(see maintext for further details).}
    \label{fig:EE_scaling}
\end{figure}

In practice, computation of the entanglement entropies of such symmetric blocks is challenging using the matrix product state based DMRG technique. This is because of the very large Hilbert space of the subsystem. Although possible~(see Ref.~\cite{Rogerson_2022} for the Ising case), the system-sizes that can be probed in this way are rather limited. This is particularly important for the defect Hamiltonians since the ground state entanglement entropy for the $N, N'$ and the $W$ defects also exhibit a violation of the well-known logarithmic dependence.  To analyze large systems sizes, the folded model~(see Fig.~\ref{fig:folding}) was simulated as a ladder with couplings at the boundary. As explained above, the symmetric entanglement entropy in this case reduces to the entanglement entropy of a block with boundary at the edge. The latter is obtained naturally in all DMRG computations. While a technical detail, this folding trick is crucial to unambiguously confirm the g-functions of the different defects. 

Fig.~\ref{fig:EE_Potts} shows the results for the symmetric entanglement entropy for the different defects as the subsystem size,~$r$ is varied. For $r\ll L$, the logarithmic dependence is clearly vindicated for all the defects. Notably, the~$N$ and the~$W$ defects have a finite offset compared to the case without defects. The obtained offsets from a fit with Eq.~\eqref{eq:S_L} for the two cases are~$\sim0.504$ and~$\sim0.451$, which are close to the expected values of $(\ln3)/2$ and $[\ln(1+\sqrt{5})/2]/2$. As a sanity check, the central charges from the fit are also shown for the different cases. As expected, the results for the central charges are close to 4/5. 

\section{Conclusion}
To summarize, this work investigated the topological defect lines in the three-state Potts model. In particular, their realizations in (lattice) spin chains were proposed and analyzed numerically using exact diagonalization as well  density matrix renormalization group techniques. Signatures of the defects were obtained in both crossed and direct channels. The corresponding conformal dimensions of the low-lying states were obtained from finite-size scaling analysis of the energy and momentum computed numerically. The g-functions for the different defects were obtained from the computation of entanglement entropies of subsystems for the ground states of the defect Hamiltonians. 

The lattice constructions presented in this work can be combined to investigate fusion of defects on the lattice in both the direct and crossed channels. The latter is of interest particularly for the non-invertible defects,~$N, N',$ and~$W$. In the crossed channel, $N$, $N'$, and $W$ defect correspond to Transfer Matrices with homogeneous   spectral parameters (i.e., equal for  all sites in a row). As the weights of the faces are chosen in such  a way that the Yang-Baxter equation (for face models) is satisfied, the defect operators in the crossed channel commute with each other and also with the identity defect Hamiltonian, as we would also expect in the continuum. The expectation values for the low-lying eigenstates of the Hamiltonian for the $NW$ defect operator are  then just product of expectation values for the individual defect operators. In the direct channel, the corresponding two-defect Hamiltonians can be analyzed as for the single-defect case. Table \ref{tab:WxWdefect:crossed} shows the DMRG results for the low-energy spectra of the two-$W$-defect Hamiltonian and comparison with the direct sum of the $\idTDL$ defect and $W$ defect, in good agreement with CFT expectations.

Before concluding, we note that while presented for the Potts model, the results are clearly general and can be applied for the construction of defects in all RSOS models. Also, while in this paper we mostly emphasized spin chains, our formalism allows for an immediate extension to the construction of  defects for Euclidian lattice models - including cases where the defect lines zig-zag through the lattice, as was studied in \cite{Aasen2020}.  All this will be discussed elsewhere~\cite{Sinha2023a}.

\begin{table}[h!]
\renewcommand{\arraystretch}{1.2}

    \centering
    \begin{tabular}{ |p{3cm}| p{3cm}|m{3cm}| }
\hline
 \multicolumn{3}{|c|}{ States and Conformal dimensions in $W \times W$} \\
 \multicolumn{3}{|c|}{16-60 Potts site in steps of 2 - Scaling done with L} \\

 \hline
 State(Descendant) & $h + \bar{h}$ & Theoretical Value  \\ [2ex]
 \hline
 $\ket{0,0} $   & 0.00058 & 0         \\
 $\ket{\frac{1}{15},\frac{1}{15}} $   &  0.1269 & 0.13333         \\
 $\ket{\frac{1}{15},\frac{1}{15}} $   &  0.1269 & 0.13333         \\
 $\ket{\frac{1}{15},\frac{1}{15}} $   &  0.1464 & 0.13333         \\
 $\ket{\frac{1}{15},\frac{1}{15}} $   &  0.1469 & 0.13333         \\
 $\ket{\frac{2}{5},0} $        & 0.4333     &  0.4  \\
 $\ket{0, \frac{2}{5}} $        &0.4336   &  0.4   \\
 
 $\ket{\frac{2}{3},\frac{1}{15}} $        & 0.7132   &  0.7333 \\
$\ket{\frac{2}{3},\frac{1}{15}} $        & 0.7158    & 0.7333 \\
$\ket{\frac{1}{15},\frac{2}{3}} $        & 0.7160    &  0.7333\\
$\ket{\frac{1}{15},\frac{2}{3}} $        & 0.7161   & 0.7333  \\

 \hline
\end{tabular}

    \caption{DMRG results for the energy spectra of low-lying states for the fusion of two $W$ defects.}
    \label{tab:WxWdefect:crossed}
\end{table}

\noindent {\bf Acknowledgments:} We thank Jonathan Bellet\^ete, Jesper Jacobsen, Azat Gainutdinov, Thiago Silva Tavares, and Aditi Mitra for related collaborations. We also thank Ning Bao, Meng Cheng, Zohar Komargodski, Anatoly Konechny, Robert Konik, Ho Tat Lam, Aditi Mitra, Gregory Moore, Sylvain Ribault, Ingo Runkel, Shu-Heng Shao, Yifan Wang, and Yunqin Zheng for very helpful discussions. We thank Paul Fendley for helpful comments on a draft. AR was supported by a grant from the Simons Foundation (825876, TDN). FY was supported by the DOE grant DE-SC0010008. HS was supported by the French Agence Nationale de la Recherche (ANR) under grant ANR-21-CE40-0003 (project CONFICA). FY thanks the Simons Center for Geometry and Physics for hospitality during the completion of this work. MS acknowledges the hospitality of Institut de physique théorique, CEA Saclay.

\newpage
\begin{appendix}

\section{Three State Potts : Spin Chain and $D_4$ RSOS model }\label{sec:spinchaintorsos}
 In this section we discuss technical aspects of the well known relationship between the two formulations of the  Three State Potts Model: the  Spin Chain  and the $D_4$ RSOS Model. Rather than a general discussion, we content ourselves by considering some simple exmaples.
 
 The Spin Chain formulation has a Hilbert space where each site can be occupied by one of the three states, labelled by say $\ket{ \, \uparrow \, }, \ket{ \searrow }$, and $\ket{ \swarrow }$ with no further restrictions. So, the dimension of $n$-sites Potts spin is $3^{n}$.\footnote{We will consider spin chains with periodic boundary condition here, i.e. the state on the $n + 1^{\text{th}}$ site is the same as the state on the $1^{\text{st}}$ site.}
 
 \bigskip

As we discussed earlier, the basis of the Hilbert space for $D_4$ RSOS model is a random walk over the $D_4$ Dynkin diagram. We can thus decompose the RSOS model into the even and odd sectors. The even sector has $X$ occurring at the first site (and all other odd sites)  and the odd sector has $X$ occurring at the second site (and all other even sites). The even and odd sectors play the role of  the  lattice and dual lattice for the spin chain (this is discussed  in greater detail in section \ref{subsec:duality}). The TL generators and so the Hamiltonian do not connect different sectors. To  map RSOS to the spin chain we use the fact that  the state $X$ always occurs alternatively and so do either 0, 1, or 2. Hence, we think of $X$ as a partition and map the states 0,1, and 2 to $\ket{ \, \uparrow \, }, \ket{ \searrow }$, and $\ket{ \swarrow }$ respectively.\footnote{Note that both the RSOS model and the spin chain at the critical point have an $S_3$ symmetry.}  Let us work in the different sectors and discuss this mapping more concretely. 
 In what follows, we use the following representation of the Potts Algebra 
 \begin{equation}\label{eq:A:Potts_convention_2}
 \sigma =\begin{pmatrix}
1 & 0 & 0\\
0 & \omega & 0\\
0 & 0 & \omega^2
\end{pmatrix}, ~ 
    \tau=\begin{pmatrix}
0 & 0 & 1 \\
1 & 0 & 0 \\
0 & 1 & 0
\end{pmatrix} , ~ \omega=\text{e}^{2\pi \text{i}/3} \, .
 \end{equation} 
 
 It can be shown that the other representation of Potts algebra in equation \eqref{eq:Potts_convention_1} is unitarily equivalent to the one in equation \eqref{eq:A:Potts_convention_2}, i.e. if the matrices in \eqref{eq:Potts_convention_1} are sandwiched between unitary operators $U^{\dagger}$ and $U$, where 
 \begin{equation}\label{eq:unitaryconvention}
U=\frac{1}{\sqrt{3}}\begin{pmatrix}
\omega^{-1} & \omega^{-1} & \omega^{-1}\\
\omega^{-1} & 1 & \omega\\
\omega^{-1} & \omega & 1
\end{pmatrix} \, ,
\end{equation}
then we obtain the matrices in \eqref{eq:A:Potts_convention_2}. Note, the above unitary transformation keeps the charge conjugation matrix, $c_i$, in \eqref{eqn:chargeconj} unchanged.

\subsection{Odd Sector}
In the odd sector, $X$ does not appear at the first site in the states. For example, a product state for an  8 site periodic RSOS model would be $\ket{0,X,0,X,1,X,2,X}$, which can be mapped onto a state of the 4 site periodic spin chain as follows: 
\begin{subequations}
\begin{equation}  
    \ket{0,X,0,X,1,X,2,X} \equiv \ket{ \, \uparrow  \, , \, \uparrow  \, , \searrow ,  \swarrow \, } \, ,
\end{equation}
\begin{equation}
    \ket{x_0, X, x_1, X, x_2, X, x_3, X} \equiv \ket{\bar{x}_0, \bar{x}_1, \bar{x}_2, \bar{x}_3} \, ,  
\end{equation}    
\end{subequations}
where $x_i$ denotes a  height in the RSOS model (0,1, or 2) and $\bar{x}_i$ a spin in   the spin chain  ($\uparrow  \, , \searrow , \text{ or } \swarrow$). This is in general how   the  Hilbert space of the odd sector of  the $2n$ site periodic $D_4$ RSOS is mapped onto  that of the $n$ site Potts spin chain. Now, the operator $e_1$ (or $e_i$, with $i$ odd) acts as
\begin{equation} 
    e_1   \ket{x_0,X,x_1,X,x_2,X,x_3,X} = \sqrt{3} \, \delta_{x_0,x_1} \ket{x_0,X,x_1,X,x_2,X,x_3,X} \, ,
\end{equation}
using \eqref{TLgenPotts}, which can be checked is exactly like the following
\begin{equation}\label{sigmaodd}
    \frac{1}{\sqrt{3}} \left( 1 + \sigma_0^{\dagger} \sigma_1 + \sigma_1^{\dagger} \sigma_0 \right) \ket{\bar{x}_0, \bar{x}_1, \bar{x}_2, \bar{x}_3} \, .
\end{equation}
Similarly, if we consider the action of $e_2$ (or $e_i$, with $i$ even), 
\begin{equation}
      e_2   \ket{x_0,X,x_1,X,x_2,X,x_3,X} = \sum_{a \in \{0,1,2 \}} \frac{1}{\sqrt{3}} \ket{x_0,X,a,X,x_2,X,x_3,X} \, ,
\end{equation}
and this is equivalent to  
\begin{equation}\label{tauodd}
    \frac{1}{\sqrt{3}} \left( 1 + \tau_1 + \tau_1^{\dagger} \right) \ket{\bar{x}_0, \bar{x}_1, \bar{x}_2, \bar{x}_3} \, .
\end{equation}
Obviously this would carry over to chains of arbitrary length. 

\subsection{Even Sector}
In even sector, $X$ now appears at the first RSOS site. An  example of a corresponding state is
\begin{subequations}
\begin{equation}
    \ket{X,0,X,0,X,1,X,2} \equiv \ket{ \, \uparrow \, , \, \uparrow \, , \searrow, \swarrow \, } \, ,
\end{equation}
\begin{equation}
    \ket{X,x_0,X,x_1,X,x_2,X,x_3} \equiv \ket{\bar{x}_0, \bar{x}_1, \bar{x}_2, \bar{x}_3}  \, .
\end{equation}    
\end{subequations}
Now, let us consider the action of $e_1$ on this state 
\begin{equation}
    e_1 \ket{X,x_0,X,x_1,X,x_2,X,x_3} = \frac{1}{\sqrt{3}} \sum_{a \in \{ 0,1,2 \} } \ket{X,a,X,x_1,X,x_2,X,x_3}  \, ,
\end{equation}
which is equivalent to 
\begin{equation}\label{taueven}
        \frac{1}{\sqrt{3}} \left( 1 + \tau_0 + \tau_0^{\dagger} \right) \ket{\bar{x}_0, \bar{x}_1, \bar{x}_2, \bar{x}_3}  \, ,
\end{equation}
and similarly 
\begin{equation}
     e_2 \ket{X,x_0,X,x_1,X,x_2,X,x_3} = \sqrt{3}  \, \delta_{x_0, x_1} \ket{X,x_0,X,x_1,X,x_2,X,x_3}  \, ,
\end{equation}
which is equivalent to 
\begin{equation}\label{sigmaeven}
     \frac{1}{\sqrt{3}} \left( 1 + \sigma_0^{\dagger} \sigma_1 + \sigma_1^{\dagger} \sigma_0 \right)  \ket{\bar{x}_0, \bar{x}_1, \bar{x}_2, \bar{x}_3}  \, .
\end{equation}
Obviously this too would carry over to chains of arbitrary length.

From the above equations, it is clear that 
\begin{equation}
    \sum e_i =     \frac{1}{\sqrt{3}} \sum_{j} \left( 1 + \sigma_j^{\dagger} \sigma_{j + 1} + \sigma_{j+1}^{\dagger} \sigma_j \right)   + \frac{1}{\sqrt{3}} \sum_{j}  \left( 1 + \tau_j + \tau_j^{\dagger} \right) \, ,
\end{equation}
for each sector, allowing us to  recover the Potts spin chain Hamiltonian. As the size of both even and odd sectors of $2n$ sites RSOS model is same as that of $n$ sites spin chain, the size of the full $2n$ sites RSOS model is double that of the spin chain. The Hamiltonians will have the same eigenvalues (when normalized correctly), but the degeneracy will be doubled in RSOS.

\subsection{Charge Conjugation and $\eta$ defect}\label{subsec:C_and_eta_RSOS_Direct}

We now  formulate the charge conjugation and $\eta$ defect Hamiltonian in the RSOS language. In the Potts Spin Chain, the Charge Conjugation defect is given by \eqref{eq:H_C}
\begin{equation}\label{eq:c_defect_direct_potts}
\begin{split}
 H_{C} &=  H_{\idTDL} + \frac{1}{3\sqrt{3}}\left(\sigma_{i_0}^\dagger \sigma_{i_0+1} +{\rm h.c.} \right) - \frac{1}{3\sqrt{3}}\left(\sigma_{i_0}^\dagger \sigma_{i_0+1}^\dagger+{\rm h.c.} \right) \, ,  \\
   & = H_{\idTDL} + \frac{1}{3\sqrt{3}}\left(\sigma_{i_0}^\dagger \sigma_{i_0+1} +{\rm h.c.} \right)  - c_{i_0} \left(  \frac{1}{3\sqrt{3}}\left(\sigma_{i_0}^\dagger \sigma_{i_0+1} +{\rm h.c.} \right) \right) c_{i_0} \ . 
\end{split}
\end{equation}
Recall, $c_{i_0}$ is the local charge conjugation operator for the Potts spin chain \eqref{eqn:chargeconj}. It acts by mapping $\searrow \, \to \, \swarrow $ to $\swarrow  \, \to \, \searrow$  at the site labelled by $\bar{x}_{i_0}$.
For the odd RSOS sector, one can write the equivalent operator as 
\begin{equation}\label{False operator Charge}
  H_{C}^{\text{odd}}= -  \sum_i e_i + e_{2 j + 1}  - C^{\text{RSOS}}_{2j+1} e_{2j+1} C^{\text{RSOS}}_{2j+1} \, , 
\end{equation}
by writing the action of $e_{2j+1}$ on this sector - see \eqref{sigmaodd}, in terms of $\sigma$ operators. Here $C^{\text{RSOS}}_{2j+1} $ acts on the RSOS Hilbert Space at the $2j+1 ^{{\rm th}}$ site (i.e. on the RSOS basis state $\ket{y_0, y_1,\ldots ,y_{2j}, \ldots, y_{2L-1}}$, on the height labelled by $y_{2j}$) by keeping  0 and $X$ fixed, and mapping  1 to 2 and vice versa. We will drop the RSOS superscript from $C^{\text{RSOS}}$ now.

In the even sector, $y_{2j} = X$, hence $C_{2j + 1}$ acts as identity. Therefore the Hamiltonian in Equation \eqref{False operator Charge} acts like the no-defect Hamiltonian for states in even sector. This is remedied by using the following Hamiltonian instead
\begin{equation}\label{eq:c_defect_direct_RSOS}
     H_{C}= -  \sum_i e_i  + e_{2j+1} + e_{2j + 2}  - C_{2j+1} e_{2j+1} C_{2j+1}  - C_{2j+2} e_{2j+2} C_{2j+2} \, . 
\end{equation}
The last term, $- C_{2j+2} e_{2j+2} C_{2j+2}$, does not make any change in the odd sector as the $2j + 2^{{\rm th}}$ site, i.e. $y_{2j + 1}$, is $X$ in the odd sector, hence  $C_{2j+2}$ acts as identity. By mapping either sector of the RSOS model to the Potts spin chain, it can be checked that the Hamiltonian in \eqref{eq:c_defect_direct_RSOS}, in Potts spin language is \eqref{eq:c_defect_direct_potts} with $i_0 = j$.
\bigskip 

The $\eta$ defect Hamiltonian for Potts spin chain is given by \eqref{eq:H_eta}
\begin{equation}\label{eq:eta_direct_potts}
\begin{split}
H_\eta =&  H_{\idTDL} + \frac{1}{3\sqrt{3}}\left(\sigma_{i_0}^\dagger \sigma_{i_0+1} +{\rm h.c.} \right) -\frac{1}{3\sqrt{3}}\left(\omega\sigma_{i_0}^\dagger \sigma_{i_0+1} +{\rm h.c.} \right)    \\
 = & H_{\idTDL} +  \frac{1}{3\sqrt{3}}\left(\sigma_{i_0}^\dagger \sigma_{i_0+1} +{\rm h.c.} \right) - \tau_{i_0}\left( \frac{1}{3\sqrt{3}}\left(\sigma_{i_0}^\dagger \sigma_{i_0+1} +{\rm h.c.} \right) \right)  \tau_{i_0}^{-1} \, .
\end{split}
\end{equation}
The RSOS $\eta$ defect Hamiltonian is obtained by the same technique that was used to construct the RSOS charge conjugation defect Hamiltonian in \eqref{eq:c_defect_direct_RSOS}, instead of conjugating with $C$, we conjugate with $\tau$ operator 
\begin{equation}\label{eq:eta_direct_RSOS}
    H_{\eta} =  -  \sum_i e_i  + e_{2j + 1} + e_{2j + 2}  - \tau_{2j+1} e_{2j+1} \tau_{2j+1}^{-1}  - \tau_{2j+2} e_{2j+2} \tau_{2j+2}^{-1} \, ,
\end{equation}
where $\tau_{i}$ acts on the $i^{{\rm th}}$ site (i.e. height labelled by $y_{i-1}$), keeps $X$ fixed and sends $0,1,$ and $2$ to $2$, $1$, and $0$ respectively.  It can be checked that if we map either the odd or even RSOS sector to Potts spin chain, the Hamiltonian in \eqref{eq:eta_direct_RSOS} is the Hamiltonian in \eqref{eq:eta_direct_potts} with $i_0 = j$.

\bigskip 

One sees numerically that 
\begin{equation}
    \begin{split}
        &[ H_{C}, Y ] = 0  \, , \\
        &[ H_{\eta}, Y ] = 0 \, .
    \end{split}
\end{equation}
Note,  the $\eta$ and $C$ defect in the direct channel cannot be represented by using only elements of  the Affine TL Algebra. \footnote{The $C_{i}$ and $\tau_j$ operators do not lie in Affine TL Algebra}
To summarize, the Charge Conjugation defect is given by  
\begin{equation}
\begin{split}
      &H_{C} =  -  \sum_i e_i  + e_{2j + 1} + e_{2j + 2}  - C_{2j + 1} e_{2j + 1} C_{2j + 1}  - C_{2j+2} e_{2j+2} C_{2j+2} \, ,  \\
 &\text{ where } C_{i} \ket{\ldots x_{i - 2} , 1 / \textcolor{red}{2}, x_{i} , \ldots } = \ket{\ldots x_{i - 2} , 2 / \textcolor{red}{1}, x_{i} , \ldots } \, ,
\end{split}
\end{equation}
and $C_{i}$ acts  like the identity on all other states. The $\eta$ defect is given by 
\begin{equation}
\begin{split}
     & H_{\eta} =  -  \sum_i e_i  + e_{2j + 1} + e_{2j + 2}  - \tau_{2j + 1} e_{2j+ 1} \tau_{2j +1}^{-1}  - \tau_{2j+2} e_{2j+2} \tau_{2j+2}^{-1} \, ,\\
& \text{ where } \tau_{i} \ket{\ldots x_{i - 2} , 0/\textcolor{red}{1}/\textcolor{blue}{2}, x_{i } , \ldots } = \ket{\ldots x_{i - 2}, 2/\textcolor{red}{0}/\textcolor{blue}{1}, x_{i} , \ldots }\, , 
\end{split}
\end{equation}
and $\tau_i$ acts like the  identity on all other states.
\newpage

\section{Kramers-Wannier duality defects via gauging}\label{sec:KWgauging}

For $(1+1)$-d bosonic theories with a non-anomalous $0$-form discrete symmetry, one can construct Kramers-Wannier duality type interfaces, by gauging the symmetry in a half of the space. For theories that are self-dual upon gauging, this procedure then produces a Kramers-Wannier duality type defect. Recently this construction is generalized to higher dimensions, starting with e.g. \cite{Choi:2021kmx, Kaidi:2021xfk}. Here we carry out this procedure in 1d quantum spin chains, producing the defect Hamiltonian for the KW defects. As a warm-up, we first consider the critical transverse-field Ising model\footnote{We thank Ho Tat Lam for previous helpful discussions.}, then move on to the critical three-states Potts model.

\subsection{(1+1)-d critical TFI  model}

We consider the following 1d critical Transverse Field Ising chain with open boundary conditions:
\begin{equation}
H=-\sum\limits_{i=1}^{N-1} \sigma_i^z \sigma_{i+1}^z - \sum\limits_{i=1}^N \sigma_i^x \, .
\end{equation}
This model has a global non-anomalous $\Z_2$ spin-flip symmetry. 
After gauging this $\Z_2$ symmetry, one obtains again a TFI model, but defined on the dual lattice. 

To gauge the $\Z_2$ symmetry we introduce $\Z_2$ gauge fields on the links, denoted by $\tau^x_{i+\frac{1}{2}}$, and thus consider  the following Hamiltonian:
\begin{equation}\label{eqn:Z2gauge}
H' = -\sum\limits_{i=1}^{N-1}\sigma_i^z \tau_{i+\frac{1}{2}}^x \sigma_{i+1}^z -\sum\limits_{i=1}^N \sigma_i^x \, ,
\end{equation}
subject to the following Gauss-law constraints
\begin{equation}\label{eqn:gauss}
\begin{split}
\tau_{i-\frac{1}{2}}^z \sigma_i^x \tau_{i+\frac{1}{2}}^z &= 1, ~ i=2,...,N-1 \, ,\\
\sigma_1^x \tau_{\frac{3}{2}}^z =1,& ~ \tau_{N-\frac{1}{2}}^z \sigma_N^x =1 \, .
\end{split}
\end{equation}

Some remarks are in order:
\begin{itemize}
	\item In the Hamiltonian (\ref{eqn:Z2gauge}), there is a local $\Z_2$ gauge redundance because now one can flip each individial $\sigma_i^z$, where the sign is absorbed into the $\Z_2$ gauge field $\tau^x_{i+\frac{1}{2}}$ living on the links. 
	
	\item We have used the convention that $\tau^x$ is the $\Z_2$ gauge field while $\tau^z$ denotes the $\Z_2$ flux.
	
	\item Away from the open boundaries, (\ref{eqn:gauss}) is understood as the $\Z_2$ Gauss-law constraints at each site $i$, where $\sigma_i^x$ corresponds to the $\Z_2$ charge. (Recall that in the ungauged Ising model, the $\Z_2$ charge is obtained by a string of $\sigma_i^x$.)
	
	\item The open boundary condition can be thought as the analog of the Neumann boundary condition for the gauge field. The $\Z_2$ flux has to be fully absorbed by the $\Z_2 $ charge on the boundary. This is different from the Dirichlet boundary condition, which plays a role in the construction of KW defect.
	 
\end{itemize}

From the tensor product Hilbert space of all sites and links, the Gauss-law constraints (\ref{eqn:gauss}) project onto a subspace of physical states. Locally consider the product Hilbert space $(\mathbb{C}^2)_{i-\frac{1}{2}}  \otimes (\mathbb{C}^2)_i  \otimes  (\mathbb{C}^2)_{i+\frac{1}{2}}$, the Gauss-law constraint projects states into a subspace spanned by
\begin{equation}\label{eqn:basislinksite}
| 0 + 0 \rangle , ~ |0 - 1\rangle, ~ |1 - 0\rangle, ~ | 1 + 1 \rangle \, , 
\end{equation}
where $|0\rangle, |1\rangle$ are eigenstates for the Pauli matrices with the superscript $z$, and $|\pm \rangle$ are eigenstates for the Pauli matrices with the superscript $x$. Moreover one can check that on the subspace of physical states, $\sigma^z_i\tau^x_{i+\frac{1}{2}}\sigma^z_{i+1}$ acts the same as $\tau^x_{i+\frac{1}{2}}$, and $\sigma^x_i$ acts the same as $\tau^z_{i-\frac{1}{2}}\tau^z_{i+\frac{1}{2}}$. The latter is easy to see from the Gauss-law constraints. To check the former statement, notice that in local bases of the form (\ref{eqn:basislinksite}), $\sigma^z_i\tau^x_{i+\frac{1}{2}}\sigma^z_{i+1}$ acts as ``spin flips" ($| 0\rangle \leftrightarrow |1\rangle$ and $|+\rangle \leftrightarrow | -\rangle$) on two neighboring sites. Due to the Gauss-law constraints, within the physical states subspace, this kind of action has the same effect as just acting with $\tau^x_{i+\frac{1}{2}}$. 

Therefore the gauged TFI Hamiltonian is equivalent to
\begin{equation}
H'=-\sum\limits_{i=1}^{N-2}\tau_{i+\frac{1}{2}}^z \tau_{i+\frac{3}{2}}^z -\sum\limits_{i=1}^{N-1}\tau_{i+\frac{1}{2}}^x-\tau_{\frac{3}{2}}^z - \tau_{N-\frac{1}{2}}^z \, .
\end{equation}
This differs from the original open chain TFI Hamiltonian by boundary terms, similar to what happens with the KW duality transform \cite{DeWolfe:2022qof, Li:2023mmw}. 

To construct the KW duality defect, we only gauge the $\Z_2$ symmetry on part of the whole open chain, say to the left of site $i_0+1$. Moreover we impose the Dirichlet boundary condition for the gauge field at link $i_0+\frac{1}{2}$. Namely we consider the following Hamiltonian 
\begin{equation}
H_D = -\sum\limits_{i=1}^{i_0} \sigma_i^z \tau_{i+\frac{1}{2}}^x \sigma_{i+1}^z  -\sum\limits_{i=i_0+1}^{N-1}\sigma_i^z\sigma_{i+1}^z - \sum\limits_{i=1}^N \sigma_i^x \, , 
\end{equation}
subject to the following Gauss-law constraints
\begin{equation}
\begin{split}
\tau_{i-\frac{1}{2}}^z \sigma_i^x \tau_{i+\frac{1}{2}}^z &= 1, ~ i=2,...,i_0 \, , \\
 \sigma_1^x \tau_{\frac{3}{2}}^z & =1 \, .
\end{split}
\end{equation}
Projecting onto the subspace of physical states works out analogously as before, in particular, $\sigma^z_{i_0}\tau^x_{i_0+\frac{1}{2}}\sigma^z_{i_0+1}$ acts the same as $\tau^x_{i_0+\frac{1}{2}}\sigma^z_{i_0+1}$. We then obtain the following Hamiltonian
\begin{equation}
\begin{split}
H_D = &- \sum\limits_{i=2}^{i_0} \tau_{i-\frac{1}{2}}^z \tau_{i+\frac{1}{2}}^z-\sum\limits_{i=1}^{i_0-1}\tau_{i+\frac{1}{2}}^x - \tau_{i_0+\frac{1}{2}}^x \sigma_{i_0+1}^z \\
&-\sum\limits_{i=i_0+1}^{N-1}\sigma_i^z \sigma_{i+1}^z -\sum\limits_{i=i_0+1}^N \sigma_i^x -\tau_{\frac{3}{2}}^z \, .
\end{split}
\end{equation}
Defining $\sigma_i^{x,z}:=\tau_{i+\frac{1}{2}}^{x,z}$ for the chain to the left of site $i_0+1$\footnote{This redefinition uses the isomorphism between the left half chain and its dual.}, we then have
\begin{equation}
H_D=-\sum\limits_{i=1,i\neq i_0}^{N-1} \sigma_i^z\sigma_{i+1}^z -\sum\limits_{i=1,i\neq i_0}^{N}\sigma_i^x -\sigma_{i_0}^x \sigma_{i_0+1}^z -\sigma_1^z \, ,
\end{equation}
which reproduces the open chain duality defect Hamiltonian, again with a subtle boundary term associated with the open boundary condition.

\subsection{(1+1)-d critical three-state Potts model}

The $\Z_3$ gauging construction of the duality defect in the three-state Potts model works out similarly as for the TFI model. Here we consider the duality defect $N$, whose open chain defect Hamiltonian from integrability is (after rescaling and shifting by a constant)
\begin{equation}
H_N' = - \sum\limits_{i=1,i\neq i_0}^{N-1} \sigma_i^\dagger \sigma_{i+1} -\sum\limits_{i=1,i\neq i_0}^N \tau_i +\text{e}^{-\text{i}\frac{\pi}{3}}\sigma_{i_0}\tau_{i_0}\sigma_{i_0+1}^\dagger + \text{h.c.} \, ,
\end{equation}
where our convention here is
\begin{equation}
\sigma=\begin{pmatrix}
0 & 1 & 0\\ 0 & 0 & 1\\ 1 & 0 & 0
\end{pmatrix}, ~ \tau=\begin{pmatrix}
1 & 0 & 0\\ 0 & w & 0\\ 0 & 0 & w^2
\end{pmatrix}, ~ w=\text{e}^{2\pi \text{i}/3} \, .
\end{equation}

This Hamiltonian is not in the natural form resulting from gauging. However we can perform a local unitary conjugation, using
\begin{equation}
U_{i_0} = \frac{1}{\sqrt{3}}\begin{pmatrix}
w & 1 & 1\\ 
1 & w & 1\\
1 & 1 & w
\end{pmatrix} \, ,
\end{equation}
which satisfies
\begin{equation}
U \sigma U^\dagger = \sigma, ~ U \left( \text{e}^{-\text{i}\frac{\pi}{3}}\sigma\tau \right) U^\dagger = -\tau \, . 
\end{equation}
We then have
\begin{equation}\label{eqn:PottsHN}
\begin{split}
H_N : &= U_{i_0} H'_N U_{i_0}^\dagger\\
&= - \sum\limits_{i=1,i\neq i_0}^{N-1} \sigma_i^\dagger \sigma_{i+1} -\sum\limits_{i=1,i\neq i_0}^N \tau_i -\tau_{i_0}\sigma_{i_0+1}^\dagger + \text{h.c.} \, .
\end{split}
\end{equation}

In the following we will demonstrate how $\Z_3$ gauging in the original Potts model produces (\ref{eqn:PottsHN}), again up to subtle boundary terms from the open boundary condition. Our starting point is the following three-state Potts Hamiltonian with open boundary conditions
\begin{equation}
H=-\sum\limits_{i=1}^{N-1}\sigma_i^\dagger \sigma_{i+1} - \sum\limits_{i=1}^N \tau_i +\text{h.c.} \, .
\end{equation}

The model has a non-anomalous $\Z_3$ symmetry generated by string of $\tau_i^\dagger$ operators. Now we gauge the $\Z_3$ symmetry to the left of site $i_0+1$, obtaining
\begin{equation}
H''_N = -\sum\limits_{i=1}^{i_0}\sigma_i^\dagger \tilde{\tau}_{i+\frac{1}{2}}^\dagger\sigma_{i+1} -\sum\limits_{i=i_0+1}^{N-1}\sigma_i^\dagger \sigma_{i+1}-\sum\limits_{i=1}^N \tau_i +\text{h.c.} \, ,
\end{equation}
subject to the following Gauss-law contraints 
\begin{equation}
\begin{split}
\tilde{\sigma}_{i-\frac{1}{2}}^\dagger \tau_i^\dagger \tilde{\sigma}_{i+\frac{1}{2}} & =1, i=2,...,i_0 \, , \\
\tau_1^\dagger \tilde{\sigma}_{\frac{3}{2}} &=1 \, .
\end{split}
\end{equation}
Projecting onto the subspace of physical states gives us
\begin{equation}
\begin{split}
H_N = & -\sum\limits_{i=2}^{i_0} \tilde{\sigma}^\dagger_{i-\frac{1}{2}}\tilde{\sigma}_{i+\frac{1}{2}}-\sum\limits_{i=1}^{i_0-1} \tilde{\tau}_{i+\frac{1}{2}}^\dagger -\tilde{\tau}^\dagger_{i_0+\frac{1}{2}}\sigma_{i_0+1}\\
& - \sum\limits_{i=i_0+1}^{N-1}\sigma_i^\dagger \sigma_{i+1}-\sum\limits_{i=i_0+1}^N \tau_i -\tilde{\sigma}_{\frac{3}{2}}^\dagger + \text{h.c.} \, . 
\end{split}
\end{equation}
Defining $\sigma_i:=\tilde{\sigma}_{i+\frac{1}{2}}, \tau_i:=\tilde{\tau}_{i+\frac{1}{2}}$ to the left of site $i_0+1$, we recover (\ref{eqn:PottsHN}) up to the boundary term associated with the open boundary condition.

\newpage

\section{Correspondence between 2d RCFTs and 3d TQFTs and its application on TDLs in non-diagonal models}\label{sec:2d3d}

In this Appendix, we aim to label the topological lines in the three-state Potts CFT using the Kac labels of $M(6,5)$ minimal model. This labeling is helpful for the general RSOS construction.\footnote{We remark that, the slab construction used here is along the lines of \cite{Kapustin:2010if,Fuchs:2012dt,Carqueville:2017ono,Komargodski:2020mxz}. It is related to, but different from the setup of recent works on generalized symmetries (e.g. \cite{Gaiotto:2020iye,Freed:2022qnc,Lin:2022dhv,Lin:2023uvm}), symmetry TFTs (e.g. \cite{kong2015boundarybulk,Apruzzi:2021nmk,Moradi:2022lqp,Kaidi:2022cpf,Kaidi:2023maf}) and categorical symmetries (e.g. \cite{Ji:2019jhk,Ji:2019eqo}).} 

Rational CFTs (RCFTs) in 2d are closely related to 3d topological field theories (TQFTs). Concretely, the chiral algebra of the RCFT provides data of a modular tensor category (MTC) describing the anyons of an associated 3d TQFT. Such data are often also referred to as the Moore-Seiberg data \cite{Moore:1988qv,Moore:1988ss}. The most notable examples are 2d $G_k$ WZW models with a simply-connected group $G$ at level $k$, which can be constructed from $\hat{\mathfrak{g}}_k$ Chern-Simons (CS) theories on an interval \cite{Witten:1988hf}. 

\insfigsvg{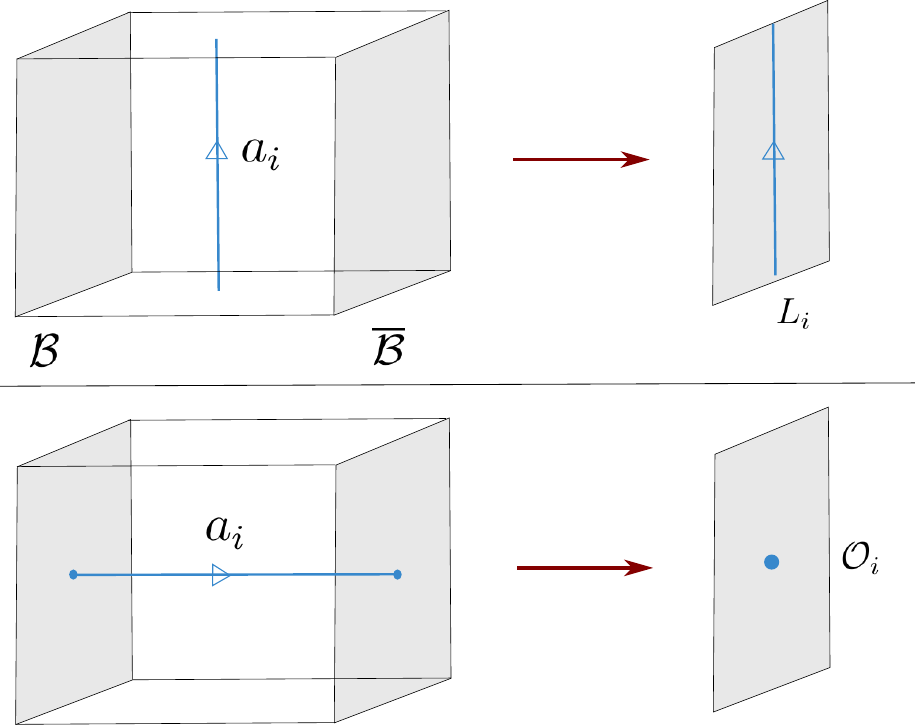}{0.5}{Illustration of 3d Chern-Simons theory on an interval with boundary conditions $\cB$ and $\overline{\cB}$. Wilson lines parallel to the boundary descend to Verlinde lines in the corresponding 2d WZW model, while Wilson lines stretched between the boundaries give rise to local primary fields in 2d. }

The diagonal $G_k$ WZW models have topological lines $L_i$ labeled by integrable representations $i$ of $\hat{\mathfrak{g}}_k$, preserving the left and right $\hat{\mathfrak{g}}_k$ current algebras. Such lines are often denoted as the Verlinde lines \cite{Verlinde:1988sn}. They are in 1-to-1 correspondence with the primary fields $\cO_i$ and obey the same fusion rules. This correspondence can be explained from the picture of 3d CS theory on an interval. Recall that the 3d CS theory has Wilson lines $a_i$ labeled by integrable representations. The primary fields $\cO_i$ in the 2d theory, corresponding to the Verma module $V_i\otimes V_{\bar{i}}$, are obtained by stretching Wilson lines $a_i$ between the two boundaries $\cB$ and $\overline{\cB}$. On the other hands, the Verlinde lines $L_i$ in 2d correspond to Wilson lines $a_i$ running parallel to the 2d spacetime. This is illustrated in \autoref{fig:2d3ddiag}. The 1-to-1 correspondence follows from the fact that, both $\cO_i$ and $L_i$ come from the Wilson lines $a_i$ in the 3d CS theory. The Drinfeld center of the fusion category of 2d Verlinde lines $L_i$ gives rise to data of the MTC describing Wilson lines $a_i$ in the 3d CS theory.\footnote{Compared with a general fusion category, a MTC is also equipped with braiding relations.}

The above statements also hold for more general 2d diagonal RCFTs and 3d TQFTs, with Wilson lines replaced by the anyon lines in the 3d TQFT. For non-diagonal 2d RCFTs, the structure is a bit more complicated. Recall that given a chiral algebra, to specify the full CFT we need to pick a modular invariant which glues together the chiral and anti-chiral parts. This data is encoded into the multiplicities $m_{i,\bar{j}}\in\Z_{\geq 0}$ counting the number of times the Verma module $V_i\otimes V_{\bar{j}}$ appearing in the Hilbert space on $S^1$:
\begin{equation}
\cH = \bigoplus_{i,\bar{j}} m_{i,\bar{j}} V_i\otimes V_{\bar{j}} \, .
\end{equation}
Therefore the classification of RCFT with respect to a given chiral algebra is equivalent to the classification of modular invariants. For example, $\widehat{\mathfrak{su}(2)}_k$ modular invariants admits an ADE classification depending on the level $k$ \cite{Cappelli:1986hf,Cappelli:1987xt,Kato:1987td}. The choice of a modular invariant could also be understood as a generalized gauging of a non-anomalous ``subpart" of the Verlinde lines in the corresponding diagonal model \cite{2001math......1219K,Fuchs:2002cm,Frohlich:2009gb,Komargodski:2020mxz}. 

The modular invariant of 2d RCFTs also has an interpretation in the context of corresponding 3d TQFTs, where it maps to the choice of a topological surface defect $S$ inserted inside the interval in 3d \cite{Kapustin:2010if,Fuchs:2012dt,Carqueville:2017ono}. Concretely, a local primary field $\cO_{i,\bar{j}}$ in the Verma module $V_i\otimes V_{\bar{j}}$ of the 2d RCFT is realized by 3d anyon lines $a_i$ and $a_{\bar{j}}$ connected at the topological surface $S$. The modular invariant $m_{i,\bar{j}}$ counts the interface operators on the surface $S$, connecting the anyon lines $a_i$ and $a_{\bar{j}}$. This is illustrated in \autoref{fig:2d3dlocal}.

A subset of topological lines in the 2d RCFT is constructed by placing 3d anyon lines either to the left or to the right of the surface $S$, as illustrated in \autoref{fig:2d3dline}. We denote such lines in 2d by $L^+_i$ and $L^-_{\bar{j}}$ respectively. More generally, one can insert the 3d anyon lines $a_i$ and $a_{\bar{j}}$ simultaneously to the left and to the right of the surface $S$. The resulting 2d topological line, which we denote as $L_i^+\otimes L_{\bar{j}}^-$, is in general not simple, namely it could be written as a sum of simple topological lines. The fusion category of topological lines in 2d is identified with the fusion category of topological lines on the surface $S$ inside the interval in 3d.

\insfigsvg{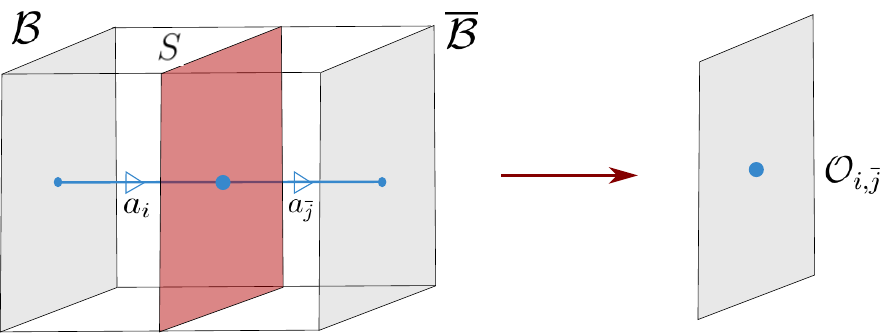}{0.56}{Illustration of a 3d TQFT on an interval with the insertion of a topological surface in between. Anyon lines stretched between the boundaries descend to local primary operators in the corresponding 2d non-diagonal RCFT. }

\insfigsvg{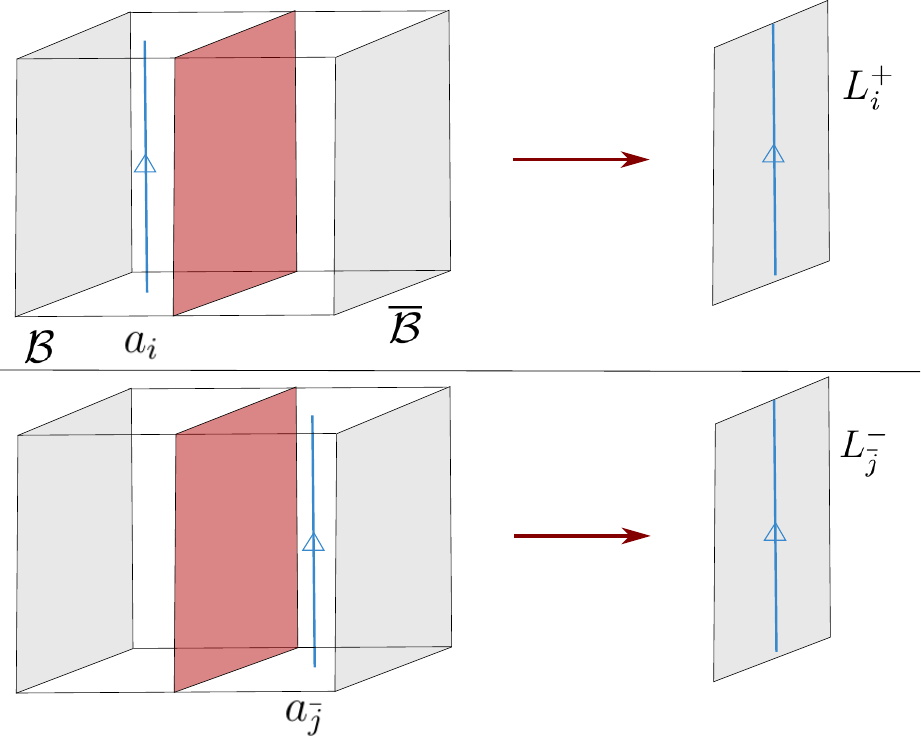}{0.5}{Illustration of a 3d TQFT on an interval with the insertion of a topological surface in between. Anyon lines parallel to the boundaries descend to topological lines in the corresponding 2d non-diagonal RCFT. }

An important piece of data associated with a topological line $L$ is the spectrum of operators that the line can end on. Via radial quantization, such operators correspond to states in the Hilbert space $\cH_L$ on $S^1$, with $L$ inserted at a point on $S^1$. The spectrum of defect Hilbert spaces can also be computed from the 3d TQFT point of view. One way to produce the configuration of a topological line ending on a point operator in 2d is by inserting anyon junctions in the 3d TQFT. This is illustrated in \autoref{fig:defectspectra}, for the case of $L^+_k$, $L^-_{\bar{l}}$ and $L^+_k\otimes L_{\bar{l}}^-$ respectively.

\insfigsvg{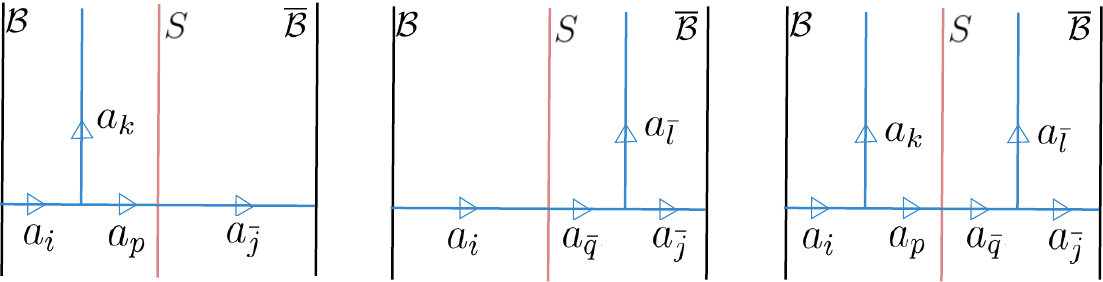}{0.7}{Illustration of a 3d TQFT on an interval with the insertion of a topological surface in between. Certain anyon junction configurations in 3d descend to configurations of a topological line ending on a point operator in 2d.}

Denoting the defect Hilbert space associated with a topological line $L$ as $\cH_L$, we can write down its decomposition into the Virasoro modules as follows:
\begin{align}
\cH_{L^+_k}&=\bigoplus_{i,\bar{j}} \left( \sum\limits_p N^{i}_{k,p} m_{p,\bar{j}} \right) V_i\otimes V_{\bar{j}},\\
\cH_{L^-_{\bar{l}}}&=\bigoplus_{i,\bar{j}} \left( \sum\limits_{\bar{q}} N^{\bar{q}}_{\bar{l},\bar{j}} m_{i,\bar{q}} \right) V_i\otimes V_{\bar{j}},\\
\cH_{L^+_k\otimes L^-_{\bar{l}}}&=\bigoplus_{i,\bar{j}} \left( \sum\limits_{p,\bar{q}} N^i_{k,p} m_{p,\bar{q}}N^{\bar{q}}_{\bar{l},\bar{j}} \right) V_i\otimes V_{\bar{j}},
\end{align}
where $N^{*}_{*,*}$ denotes the fusion coefficient and $m_{*,*}$ is the modular invariant gluing together the chiral and anti-chiral algebras.

We now put the above general statements into the context of three-state Potts model in 2d. The corresponding 3d TQFT has anyon lines $a_i$, carrying Kac labels for the $M(6,5)$ minimal model. Concretely we have\footnote{See for example \cite{DiFrancesco:1997nk} regarding the convention of Kac labels.}
\begin{equation}
i \in \{ (1,1), (1,2), (1,3), (1,4), (1,5), (2,1), (2,2), (2,3), (2,4), (3,1) \} \, .
\end{equation}

The three-state Potts model can be constructed by putting the 3d TQFT on an interval with boundary conditions $\cB$ and $\overline{\cB}$, with a topological surface defect $S$ inserted inside the interval. The topological surface $S$ corresponds to a type-D modular invariant, which is encoded as the multiplicities $m_{i,\bar{j}}$ of the Verma module $V_i\otimes V_{\bar{j}}$ appearing in the Hilbert space on $S^1$:
\begin{equation}
\begin{split}
\cH = & \bigoplus_{i,\bar{j}} m_{i,\bar{j}} V_i \otimes V_{\bar{j}}\\
= & \left(V_{(1,1)}\otimes V_{(1,1)}\right) \oplus \left(V_{(2,1)}\otimes V_{(2,1)}\right) \oplus \left(V_{(3,1)}\otimes V_{(3,1)}\right) \oplus \left(V_{(1,5)}\otimes V_{(1,5)}\right)\\
& \oplus 2 ~ \left(V_{(2,3)}\otimes V_{(2,3)}\right) \oplus 2 ~  \left(V_{(1,3)}\otimes V_{(1,3)}\right) \oplus \left(V_{(1,1)}\otimes V_{(1,5)}\right)\\
& \oplus \left(V_{(1,5)}\otimes V_{(1,1)}\right) \oplus \left(V_{(2,1)}\otimes V_{(3,1)}\right) \oplus \left(V_{(3,1)}\otimes V_{(2,1)}\right) \, .
\end{split}
\end{equation}

By inserting anyons with junctions as shown in \autoref{fig:defectspectra}, we obtain topological lines in the Potts model. In our notation, a generic topological line takes the form of $L_k^+\otimes L_{\bar{l}}^-$. Most such lines are not simple, namely they can be written as a sum of several simple lines from the following set:
\begin{equation}
I, \eta, \bar{\eta}, W, \eta W, \bar{\eta}W, C, \eta C, \bar{\eta}C, WC, \eta WC, \bar{\eta} WC, N, N':=CN, WN, WN' \, .
\end{equation}
Moreover there could be different choices of $k$ and $\bar{l}$ in $L^+_k\otimes L_{\bar{l}}^-$, such that they descend to the same topological line in 2d. In particular, we list below choices which yield the simple lines $N, N', W, WN, WN'$.
\begin{equation}
\begin{split}
N:& ~ L^+_{(1,2)}, L^+_{(1,4)}, L^+_{(1,2)}\otimes L^-_{(1,5)}, L^+_{(1,4)}\otimes L^-_{(1,5)}\\
N':& ~ L^-_{(1,2)}, L^-_{(1,4)}, L^+_{(1,5)}\otimes L^-_{(1,2)}, L^+_{(1,5)}\otimes L^-_{(1,4)}\\
W:& ~ L^+_{(2,1)}, L^+_{(3,1)}, L^-_{(2,1)}, L^-_{(3,1)},L^+_{(2,1)}\otimes L^-_{(1,5)}, L^+_{(3,1)}\otimes L^-_{(1,5)},\\
& ~ L^+_{(1,5)}\otimes L^-_{(2,1)}, L^+_{(1,5)}\otimes L^-_{(3,1)}\\
WN:& ~ L^+_{(2,2)}, L^+_{(2,4)}, L^+_{(2,2)}\otimes L^-_{(1,5)}, L^+_{(2,4)}\otimes L^-_{(1,5)}, L^+_{(1,2)}\otimes L^-_{(2,1)},\\
& ~ L^+_{(1,2)}\otimes L^-_{(3,1)}, L^+_{(1,4)}\otimes L^-_{(2,1)}, L^+_{(1,4)}\otimes L^-_{(3,1)}\\
WN':& ~ L^-_{(2,2)}, L^-_{(2,4)}, L^+_{(1,5)}\otimes L^-_{(2,2)}, L^+_{(1,5)}\otimes L^-_{(2,4)}, L^+_{(2,1)}\otimes L^-_{(1,2)},\\
& ~ L^+_{(2,1)}\otimes L^-_{(1,4)}, L^+_{(3,1)}\otimes L^-_{(1,2)}, L^+_{(3,1)}\otimes L^-_{(1,4)} \, .
\end{split}
\end{equation}

\newpage
\section{Tambara - Yamagami Fusion Category }\label{sec:TYtoRSOS}
The $\mathbb{Z}_N$ Tambara-Yamagami category has $N + 1$ simple objects, $N$ of which are labelled by integers - 0, 1, 2, ..., $N - 1$, and the $(N + 1)^{\rm th}$ is labelled by $X$. The fusion rules are such that 
\begin{subequations}
\begin{equation}
    X \otimes X = \bigoplus_{a = 0}^{N-1} a \, ,
\end{equation}
\begin{equation}
    a \otimes b = (a+b) \text{ mod } N \, ,
\end{equation}
\begin{equation}
    a \otimes X = X \otimes a = X \, .
\end{equation}
\end{subequations}
For an object $a$ in the category, the dual object, $a^{\star}$, is an object such that $a \otimes a^{\star} = 0$. $X$ and $0$ are self-dual, but any other simple object, $a$, is dual to $N - a$. This prohibits us from writing the anyonic chain corresponding  to the three-state Potts model  in the way used e.g. in \cite{Feiguin:2006ydp},\cite{Buican:2017rxc} for $A$ type RSOS models. 

For categories in which  objects are not self dual, we must draw arrows between different heights \cite{Aasen2020},  \cite{Huang:2021nvb}.\footnote{Note, the arrows indicate which way the anyonic tree goes up. This way it is easy to write the correct F-symbols while making change of basis transformations.}

\begin{figure}[h]
\centering
\begin{tikzpicture}[scale = 1.5]
\begin{scope}[very thick, every node/.style={sloped,allow upside down}]
\draw (0,0)-- node {\midarrow} (1,0);
\draw (1,0)-- node {\midarrow} (2,0);
\draw (2,0)-- node {\midarrow} (3,0);
\draw (3,0)-- node {\midarrow} (4,0);
\draw (4,0)-- node {\midarrow} (5,0);
\draw (5,0)-- node {\midarrow} (6,0);
\draw (1,0)-- node {\midarrow} (1,1);
\draw (2,0)-- node {\midarrow} (2,1);
\draw (3,0)-- node {\midarrow} (3,1);
\draw (4,0)-- node {\midarrow} (4,1);
\draw (5,0)-- node {\midarrow} (5,1);
\node[above right] at (.5,.6) {$X$};
\node[above right] at (1.5,0.6) {$X$};
\node[above right] at (2.5,0.6) {$X$};
\node[above right] at (3.5,0.6) {$X$};
\node[above right] at (4.5,0.6) {$X$};
\node[above right] at (0.2,-0.45) {$0/1/2$};
\node[above right] at (1.2,-0.4) {$X$};
\node[above right] at (2.15,-0.45) {$0/1/2$};
\node[above right] at (3.25,-0.4) {$X$};
\node[above right] at (4.15,-0.45) {$0/1/2$};
\node[above right] at (5.2,-0.4) {$X....$};
\end{scope}
\end{tikzpicture}
\caption{Three State Potts Anyonic Chain.}
\label{fig:pottstree}
\end{figure}
Ordinary (A-type) RSOS models can be reformulated as anyonic chains and thus given a categorical interpretation. While it is easy to generalize the corresponding construction to D-type models, we haven't seen the details published anywhere in the literature, and thus provide them here for completeness. 

The formal construction of anyonic chains relies on a fusion category and  a specific special object. For example for the A-type models, the category is $su(2)_k$, and the special object is the object with spin ${1\over 2}$. The local Hamiltonian is then defined as a projection operator onto a particular fusion channel \cite{Feiguin:2006ydp}. For A-type models this channel is chosen to be spin  $0$ (the identity object). It turns out  that the projection operators thus obtained  are the TL generators up to a scale factor (exactly the quantum dimension of $1/2$). Such constructions are discussed in detail in \cite{Aasen2020}, \cite{Shao:2023gho}, \cite{Inamura:2023qzl}, among others.

The story is essentially  the same for the Three-state Potts or $D_4$ RSOS model, but involves instead the 
$Z_3$ Tambara Yamagami category. We have thus  $3$ $\mathbb{Z}_3$ states - $0,1,$ and $2$ and the non-abelian object $X$ which we choose as the special object (instead of spin ${1\over 2}$ for A-type RSOS). The fusion rules dictate that if the first object is either $0,1$, or $2$, the next object must be $X$, then followed by $0/1/2$, as shown in \autoref{fig:pottstree}. We will use the label $a$ or $b$ when the object is not $X$. As for the   channel to project onto,  we choose the simple object $0$.
We will show that  the corresponding Projection operator  is, up to a scale ($d_X = \sqrt{3}$), the TL generator for the 3-state Potts model.

Now, let us define the operator $E_j := d_X P_{j}^{0}$, where $d_X = \sqrt{3}$ is the quantum dimension. If one tries to calculate $E_{j}\ket{....h_{j-1}, h_j, h_{j+1}....}$, there can be two cases, 
\begin{enumerate}
    \item $h_{j-1}, h_{j+1}  = a,b$ and $h_j = X$, 
    \item $h_{j-1}, h_{j+1} = X$ and $h_j = a$.
\end{enumerate}

To see how the projection operator acts, we must use $F-$ transformations, to go into the correct channel, as was done for  $su(2)_k$  chains in \cite{Buican:2017rxc}.  We first consider case 1, the fusion tree for which is shown in \autoref{Case1Ej}, do a change of basis transformation as shown in \autoref{F-Movecase1}. We then apply the projection operator and then again do a change of basis to get back to the standard form, as shown in \autoref{Projection Operator}. In  \autoref{F-Movecase1} and \autoref{Projection Operator}, we show the action of the projection operator if $a = b$. If $a \neq b$, then instead of 0, we would have obtained $(a-b)$ mod 3 in \autoref{F-Movecase1}, but the projection operator would have killed this term. Hence, we obtain
\begin{equation}
    P_{j}^{(0)}\ket{..a,X,b..} = \delta(a,b) \ket{..a,X,b..} \, .
\end{equation}
 Hence $E_j \ket{..a,X,b..} = \sqrt{3} \ \delta(a,b)\ket{..a,X,b..}$  

\begin{figure}[h]
\centering
\begin{tikzpicture}[scale = 1.5]
\begin{scope}[very thick, every node/.style={sloped,allow upside down}]
\draw (2,0)-- node {\midarrow} (3,0);
\draw (3,0)-- node {\midarrow} (4,0);
\draw (4,0)-- node {\midarrow} (5,0);
\draw (3,0)-- node {\midarrow} (3,1);
\draw (4,0)-- node {\midarrow} (4,1);
\node[above right] at (2.5,0.6) {$X$};
\node[above right] at (3.5,0.6) {$X$};
\node[above right] at (2.15,-0.45) {$a$};
\node[above right] at (3.25,-0.45) {$X$};
\node[above right] at (4.15,-0.45) {$b$};
\end{scope}
\end{tikzpicture}
\caption{State for case 1.}
\label{Case1Ej}
\end{figure}

\begin{figure}[h!]
\centering
\begin{tikzpicture}[scale = 1.4]
    \begin{scope}[very thick, every node/.style={sloped,allow upside down}]
    \draw (0,0)-- node {\midarrow} (1,0);
    \draw (1,0)-- node {\midarrow} (2,0);
    \draw (2,0)-- node {\midarrow} (3,0);
    \draw (0,0) -- (1,0);
    \draw (1,0) -- node {\midarrow}  (1,1);
    \draw (2,0) -- node {\midarrow}  (2,1);
    \draw (3.5,0.6) -- (4,0.6);
    \draw (3.5,0.4) -- (4,0.4);
    \draw (6,0)-- node {\midarrow} (7.5,0);
    \draw (7.5,0)-- node {\midarrow} (9,0);
    \draw (7.5,0)-- node {\midarrow} (7.5,1);
    \draw (7.5,1) --node {\midarrow} (6.5,2);
    \draw (7.5,1) --node {\midarrow} (8.5,2);
    \node[above right] at (.5,.6) {$X$};
    \node[above right] at (1.5,0.6) {$X$};
    \node[above right] at (0.2,-0.5) {$a$};
    \node[above right] at (1.2,-0.5) {$X$};
    \node[above right] at (2.2,-0.5) {$a$};
    \node[above right] at (6.5,-0.5) {$a$};
    \node[above right] at (8,-0.5) {$a$};
    \node[above right] at (7.5,0.4) {$0$};
    \node[above right] at (7,1.6) {$X$};
    \node[above right] at (7.6,1.6) {$X$};
    \node[scale = 1.2, above] at (5,0.3) {$\alpha$};
    \end{scope}
\end{tikzpicture}
\caption{There is only one tree in the RHS as the rest are 0 due to fusion rules, $a$ is fixed $\in \{ 0,1,2 \}$. }
\label{F-Movecase1}
\end{figure}

\begin{figure}[h!]
    \centering
    \begin{tikzpicture}[scale = 0.8]
    \begin{scope}[very thick, every node/.style={sloped,allow upside down}]

    \draw (0,0) --node {\midarrow} (1.5,0);
    \draw (1.5,0) --node {\midarrow} (3,0);
    \draw (1.5,0) --node {\midarrow} (1.5,1); 
    \draw (4.5,0.6) -- (5,0.6);
    \draw (4.5,0.4) -- (5,0.4);
    \draw (7,0) --node {\midarrow} (8.5,0);
    \draw (8.5,0) --node {\midarrow} (10,0);
    \draw (8.5,0) --node {\midarrow} (8.5,1);
    \draw (8.5,1) --node {\midarrow} (7.5,2);
    \draw (8.5,1) --node {\midarrow} (9.5,2);
    \draw (1.5,1) --node {\midarrow} (0.5,2);
    \draw (1.5,1) --node {\midarrow} (2.5,2);
    \node[above right] at (1.5,0.4) {$0 $};
    \node[above right] at (0.5,-0.7) {$a$};
    \node[above right] at (1.8,-0.7) {$a$};
    \node[above right] at (0.7,1.6) {$X$};
    \node[above right] at (1.6,1.6) {$X$};
    \node[scale = 1.2, above] at (-2,0.1) {$P_{j}^{(0)}$};
    
    \node[above right] at (7.5,-0.7) {$a$};
    \node[above right] at (9,-0.7) {$a$};
    \node[above right] at (8.5,0.4) {$0$};
    \node[above right] at (7.6,1.6) {$X$};
    \node[above right] at (8.4,1.6) {$X$};
    \node[scale = 1.2, above] at (6,0.1){$ \, \, \alpha $};
    \node[scale = 1.2, above] at (-0.5,0.1) {$\alpha $};
    \path[out=135,in=225] (-.5,-1) edge (-.5,2);
    \path[in= 315, out=45] (3.6,-1) edge (3.6,2);
    
    \draw (10.5,0.6) -- (11,0.6);
    \draw (10.5,0.4) -- (11,0.4);

    \draw (12,0) --node {\midarrow} (13.5,0);
    \draw (13.5,0) --node {\midarrow} (15,0);
    \draw (15,0) --node {\midarrow} (16.5,0);
    \draw (13.5,0) --node {\midarrow} (13.5,1.2);
    \draw (15,0) --node {\midarrow} (15,1.2);
    \node[above right] at (12.5, 0.6) {$X$};
    \node[above right] at (14, 0.6) {$X$};
    \node[above right] at (12.5, -0.7) {$a$};
    \node[above right] at (14, -0.7) {$X$};
    \node[above right] at (15.5, -0.7) {$a$};
    
    \end{scope}
    \end{tikzpicture}
    \caption{Action of the Projection Operator.}
    \label{Projection Operator}
\end{figure}

\bigskip

Now, let us go to case 2, the fusion tree for which is shown in \autoref{Case2Ej}. Again, let us do the basis transformation and then apply the projection operator.

\begin{figure}[h!]
\centering
\begin{tikzpicture}[scale = 1.5]
\begin{scope}[very thick, every node/.style={sloped,allow upside down}]
\draw (2,0)-- node {\midarrow} (3,0);
\draw (3,0)-- node {\midarrow} (4,0);
\draw (4,0)-- node {\midarrow} (5,0);
\draw (3,0)-- node {\midarrow} (3,1);
\draw (4,0)-- node {\midarrow} (4,1);
\node[above right] at (2.5,0.6) {$X$};
\node[above right] at (3.5,0.6) {$X$};
\node[above right] at (2.15,-0.45) {$X$};
\node[above right] at (3.25,-0.4) {$a$};
\node[above right] at (4.15,-0.45) {$X$};
\end{scope}
\end{tikzpicture}
\caption{State for case 2.}
\label{Case2Ej}
\end{figure}
\begin{figure}[h!]
\centering
\begin{tikzpicture}[scale = 1.5]
    \begin{scope}[very thick, every node/.style={sloped,allow upside down}]
    \draw (0,0)-- node {\midarrow} (1,0);
    \draw (1,0)-- node {\midarrow} (2,0);
    \draw (2,0)-- node {\midarrow} (3,0);
    \draw (0,0) -- (1,0);
    \draw (1,0) -- node {\midarrow}  (1,1);
    \draw (2,0) -- node {\midarrow}  (2,1);
    \draw (3.5,0.6) -- (4,0.6);
    \draw (3.5,0.4) -- (4,0.4);
    \draw (7,0)-- node {\midarrow} (8.5,0);
    \draw (8.5,0)-- node {\midarrow} (10,0);
    \draw (8.5,0)-- node {\midarrow} (8.5,1);
    \draw (8.5,1) --node {\midarrow} (7.5,2);
    \draw (8.5,1) --node {\midarrow} (9.5,2);
    \node[above right] at (.6,.6) {$X$};
    \node[above right] at (1.6,0.6) {$X$};
    \node[above right] at (0.2,-0.4) {$X$};
    \node[above right] at (1.2,-0.4) {$a$};
    \node[above right] at (2.2,-0.4) {$X$};
    \node[above right] at (7.5,-0.4) {$X$};
    \node[above right] at (9,-0.4) {$X$};
    \node[above right] at (8.5,0.4) {$\tilde{x}_1$};
    \node[above right] at (8,1.4) {$X$};
    \node[above right] at (8.6,1.4) {$X$};
    \node[scale = 1.2, above] at (6,0.1) {$\sum\limits_{\tilde{x}_1 \in \{ 0,1,2\}}\left({F}_{X}^{X X X }\right)^{-1}_{a \tilde{x}_1}$};
    \end{scope}
\end{tikzpicture}
\caption{$F$-move for case 2 ; $\tilde{x}_1  = X$ is prohibited by the fusion rules.}
\label{F-Movecase2}

\end{figure}

\begin{figure}[h!]
    \centering
    \begin{tikzpicture}[scale = 1]
    \begin{scope}[very thick, every node/.style={sloped,allow upside down}]

    \draw (0,0) --node {\midarrow} (1.5,0);
    \draw (1.5,0) --node {\midarrow} (3,0);
    \draw (1.5,1) -- (1.5,0); 
    \draw (4.3,0.6) -- (4.8,0.6);
    \draw (4.3,0.4) -- (4.8,0.4);
    \draw (7,0) --node {\midarrow} (8.5,0);
    \draw (8.5,0) --node {\midarrow} (10,0);
    \draw (8.5,0) --node {\midarrow} (8.5,1);
    \draw (8.5,1) --node {\midarrow} (7.5,2);
    \draw (8.5,1) --node {\midarrow} (9.5,2);
    \draw (1.5,1) --node {\midarrow} (0.5,2);
    \draw (1.5,1) --node {\midarrow} (2.5,2);
    \node[above right] at (1.5,0.4) {$\tilde{x}_1$};
    \node[above right] at (0.5,-0.6) {$X$};
    \node[above right] at (1.8,-0.6) {$X$};
    \node[above right] at (0.8,1.5) {$X$};
    \node[above right] at (1.7,1.5) {$X$};
    \node[scale = 1.2, above] at (-5,0.1) {$P_{j}^{(0)}$};
    
    \node[above right] at (7.5,-0.6) {$X$};
    \node[above right] at (9,-0.6) {$X$};
    \node[above right] at (8.5,0.4) {$0$};
    \node[above right] at (7.9,1.5) {$X$};
    \node[above right] at (8.5,1.5) {$X$};
    \node[scale = 1.2, above] at (6,0.1){$\left({F}_{X}^{X X X }\right)^{-1}_{a  0}$};
    \node[scale = 1.2, above] at (-2,0.1) {$\sum\limits_{\tilde{x}_1 \in \{0,1,2 \}}\left({F}_{X}^{X X X }\right)^{-1}_{a \tilde{x}_1}$};
    \path[out=135,in=225] (-3.6,-1) edge (-3.6,2);
    \path[in= 315, out=45] (3.5,-1) edge (3.5,2);
    \end{scope}
    \end{tikzpicture}
    \caption{Action of the Projection Operator}
    \label{Projection Operator - Case 2}
\end{figure}

\begin{figure}
    \centering
    \begin{tikzpicture}[scale = 1.2]
    \begin{scope}[very thick, every node/.style={sloped,allow upside down}]

    \draw (-2,0) --node {\midarrow} (-0.5,0);
    \draw (-0.5,0) --node {\midarrow} (1,0);

    \draw (-0.5,0) --node {\midarrow} (-0.5,1); 
    \draw (0.8,0.6) -- (1.3,0.6);
    \draw (0.8,0.4) -- (1.3,0.4);
    \draw (7,0) --node {\midarrow} (8,0);
    \draw (8,0) --node {\midarrow} (9,0);
    \draw (9,0) --node {\midarrow} (10,0);
    \draw (8,0) --node {\midarrow} (8,1);
    \draw (9,0) --node {\midarrow}(9,1);
    \draw (-0.5,1) --node {\midarrow} (-1.5,2);
    \draw (-0.5,1) --node {\midarrow} (0.5,2);
    \node[above right] at (-0.5,0.4) {$0$};
    \node[above right] at (-1.5,-0.5) {$X$};
    \node[above right] at (0.1,-0.5) {$X$};
    \node[above right] at (-1.1,1.5) {$X$};
    \node[above right] at (-0.4,1.5) {$X$};
    
    \node[above right] at (7.2,-0.5) {$X$};
    \node[above right] at (8.2,-0.6) {$\tilde{y}$};
    \node[above right] at (9.2,-0.5) {$X$};
    \node[above right] at (7.5,0.4) {$X$};
    \node[above right] at (8.5,0.4) {$X$};
    \node[scale = 1.2, above] at (-2,0.1) {$\left({F}_{X}^{X X X }\right)^{-1}_{a  0}$};
    \node[scale = 1.2, above] at (4.5,0.1) {$ \sum\limits_{\tilde{y} \in \{0,1,2 \}} \left({F}_{X}^{X X X }\right)^{-1}_{a  0}({F}_{X}^{X X X })_{0 \tilde{y}}$};
    \end{scope}
    \end{tikzpicture}
    \caption{Inverse $F$ transformation. }
    \label{InverseF2}
\end{figure}
For all $a$, $(F^{XXX}_{X})_{0a} = \frac{1}{\sqrt{3}}$  - \cite{Aasen2020} and by taking the inverse, we get $({F}_{X}^{X X X })^{-1}_{b 0} = \frac{1}{\sqrt{3}}$. Now, using the action of projection operator and changing basis using F -symbols, as explicitly shown in \autoref{F-Movecase2}, \autoref{Projection Operator - Case 2}, and \autoref{InverseF2}, we see that 
\begin{equation}
\begin{split}
    P_j^{(0)} \ket{..X,a,X..}  = &  \sum\limits_{\tilde{y} \in \{0,1,2 \}} \left({F}_{X}^{X X X }\right)^{-1}_{a  0}({F}_{X}^{X X X })_{0 \tilde{y}} \ket{..X,\tilde{y},X..} \\ 
    = & \sum\limits_{\tilde{y} \in \{ 0,1,2\}} \frac{1}{3} \ket{ .. X, \tilde{y}, X .. } \, .
\end{split}
\end{equation}

So, $E_j \ket{..X,a,X..} = \frac{1}{\sqrt{3}}\sum_{b = 0}^{2} \ket{..X,b,X..}$.

\bigskip 

Summarizing the action of $E_j$ 
\begin{subequations}
    \begin{equation}
            E_j \ket{..a,X,b..} = \sqrt{3} \ \delta(a,b)\ket{..a,X,b..} \, ,  
    \end{equation} 
    \begin{equation}
            E_j \ket{..X,a,X..} = \frac{1}{\sqrt{3}}\sum_{b = 0}^{2} \ket{..X,b,X..}  \, , 
    \end{equation}
\end{subequations}
which is the same as Equation \ref{TLgenPotts}.

\end{appendix}

\bigskip

\newpage

\bibliographystyle{utphys}

\bibliography{library_1}

\end{document}